\documentclass[12pt]{article}
\usepackage{amsmath}
\usepackage{graphicx}
\usepackage{epsfig}
\usepackage{amsthm}
\usepackage[small,sl]{caption}
\usepackage{amsfonts}
\usepackage{amssymb}
\usepackage{color}

\def\dsp{\def\baselinestretch{1.1}\large\normalsize}
\dsp
\def\beq{\begin{equation}}

\title{{\bf Structure of neutrino mass matrix\\ and CP violation}}

\author{{Michele Frigerio} \footnote{\tt frigerio@he.sissa.it}\\
{\small\it  INFN, Section of Trieste {\rm and} International  School}\\ 
{\small\it for Advanced Studies (SISSA), Via Beirut
4, I-34014 Trieste, Italy.}\\
{Alexei Yu. Smirnov} \footnote{\tt smirnov@ictp.trieste.it}\\
{\small\it The Abdus Salam International  Center for Theoretical Physics
(ICTP), I-34100 Trieste, Italy}\\
{\small and \it Institute for  Nuclear Research, Russian Academy of
Sciences, Moscow, Russia.} }

\date{}

\begin{document}
\maketitle
\begin{abstract}
We reconstruct the neutrino mass matrix
in the flavor basis, using all available experimental data on neutrino
oscillations. Majorana nature of neutrinos, normal mass
hierarchy (ordering)
and validity of the LMA MSW solution of the solar neutrino problem
are assumed.
We study  dependences of the mass matrix
elements, $m_{\alpha\beta}$, on the CP violating Dirac, $\delta$, and
Majorana, $\rho$,
$\sigma$,  phases, for  different values of the
mixing angle $\theta_{13}$ and of the  absolute mass
scale, $m_1$.
The contours of constant mass in the  $\rho-\sigma$ plane
have been constructed for all $m_{\alpha\beta}$.
These  $\rho-\sigma$ plots allow to systematically scan all possible 
structures of the mass matrix. 
We identify regions of parameters in which the matrix has
(i) a structure with the dominant  $\mu\tau$-block, (ii) various
 hierarchical structures, (iii) flavor alignment, 
(iv) structures with special ordering or
equalities of elements, (v) the democratic form.
In certain cases the matrix can be parameterized by powers of a unique
expansion (ordering) parameter $\lambda \approx 0.2-0.3$
($\lambda_{ord}\approx 0.6 - 0.7$). 
Perspectives to further restrict the
structure of  mass matrix in future experiments, in particular in the
$\beta\beta_{0\nu}$-decay searches, are discussed.

\vspace*{.3cm}
\noindent Ref.SISSA 17/2002/EP
 \hfill
hep-ph/0202247
\end{abstract}
\begin{center}
\end{center}

\newpage

\section{Introduction}

Significant amount of  information about neutrino masses and mixing has
already been
obtained from  experiments on the atmospheric \cite{superk} and solar
neutrinos \cite{SKsolar,sno}, from laboratory experiments, in particular  
the reactor experiments \cite{chooz} and
neutrinoless double beta decay searches \cite{kk}, from astrophysics and
cosmology.  New substantial results  are expected soon.

What are implications of these  results for the fundamental
theory, for  the mechanism of neutrino mass generation, 
origin of large lepton  mixing, 
relation between the quark and the lepton masses?
The neutrino masses and mixing appear from diagonalization of
the neutrino mass matrix. 
In a sense, the  mass matrix {\it unifies} the information which  is
contained in the masses  and mixing angles 
(which appear as independent physical observables). 
So, the questions we are asking should be
considered in terms of properties of the mass matrix.

It is expected that the  structure of the mass matrix can be  explained by
certain  (broken) symmetry realized in certain basis at some  high
mass scale \cite{fronie}.  We will call this basis the {\it symmetry
basis}. Thus, to approach the fundamental
theory, one should find the mass matrix in the symmetry basis and
at the corresponding {\it symmetry scale}.
Both   abelian ({\it e.g.}, \cite{abelian,altfer})
and non abelian  ({\it e.g.}, \cite{nonabe})  symmetries, broken
(spontaneously) at the
various symmetry scales,   have been widely considered (see also the
reviews \cite{frirev,barr}).
Also, the possibilities have been studied to
identify the flavor symmetry scale
with other known scales in  the theory
like the Grand Unification scale or the string scale.

The first step to the fundamental theory
is the reconstruction of the matrix in the {\it flavor basis}
using all available experimental data.
The flavor basis, formed by $\nu_e, \nu_{\mu}, \nu_{\tau}$,
is determined
as the basis in which the mass matrix of charge leptons is diagonal.
However, the symmetry basis may not coincide with the flavor basis,
while the structure of the mass matrix depends on the  basis substantially.
Furthermore, using the existing  experimental information we can
reconstruct the mass matrix at the low (electroweak) scale.
The scale at which possible flavor symmetry is realized (broken)
is unknown.
So, the bottom-up approach would consist of determination of the
structure of the mass matrix at low scale, selection of the appropriate
symmetry basis and selection of the correct symmetry scale.

This picture should be taken with some caution: 
{\it a priory} it is not 
clear whether  certain symmetry is behind properties of the mass matrix.  
This should be established by studying possible regularities of
the mass matrix. 

There is a number of attempts to reconstruct neutrino mass matrix in the
flavor basis using the available experimental results
\cite{barb,vissani,akh}. Most of the studies
have been performed in the context of three Majorana neutrinos and the data
on  atmospheric and solar neutrinos  as well as from the CHOOZ reactor
experiment \cite{superk,SKsolar,sno,chooz} have been
used as an input.  Clearly this information  is not enough to reconstruct
the  mass matrix completely.
Apart from the oscillation parameters (mass squared differences
and mixing angles), the mass matrix
depends  on  non-oscillation parameters:  the absolute mass scale, $m_1$,
and the CP violating Majorana phases.
Furthermore, even not all the oscillation parameters are known.
In particular,
there is  only an upper bound on the mixing angle  $\theta_{13}$ and there
is no information about the value  of the Dirac phase $\delta$.
Also the type of mass hierarchy
(ordering) of the states is unknown.

The studies performed so far  were concentrated, mainly, on identification
of the dominant structures of the  mass matrix and possible zeros of
certain matrix elements.
It was realized that in the case of spectrum with normal mass hierarchy,
$m_1 \ll m_2 \ll m_3$, the mass matrix has   structure with the dominant
$\mu \tau$-block, formed by $M_{\mu \mu}, M_{\mu \tau}, M_{\tau \tau}$
elements, and small elements of the  $e-$row ($M_{ee}, M_{e\mu}, M_{e\tau}$)
\cite{vissani,yana}.
In the case  of inverted mass hierarchy, the dominant structure can be
formed by elements of the $e-$row:  $M_{e\mu}$ and
$M_{e\tau}$ \cite{altfer,barb}.  These structures may be related to
an underlying $L_e - L_{\mu} - L_{\tau}$  symmetry.

In the case of degenerate mass spectrum  new dominant structures appear
depending on the CP-parities of the mass eigenstates
(see, {\it e.g.}, \cite{altfer,barb,FSV}).
In particular, it has been found that the diagonal elements,  being equal
to each other, can form the dominant
structure for equal CP-parities of all three neutrinos.
Another interesting possibility is the dominant
structure formed by the $ee-$, $\mu\tau-$ and $\tau\mu-$elements (moreover, 
$|M_{ee}| \approx |M_{\mu \tau}|$), which could imply,
{\it e.g.},  $SO(3)$ flavor symmetry or $U(1)$ symmetry,
with charge prescription  $(0, 1,  -1)$ and an additional permutation
symmetry.  Recently,  the possibility  that some  matrix elements
equal exactly  zero has been considered \cite{FGM}.

It was shown that experimental data can be explained
in models with universal Yukawa couplings \cite{frix}, which
lead to  ``democratic" mass matrices with all
mass matrix elements having  the same modulus but different phases.

Completely different approach is based on
``Anarchy" of the mass matrix \cite{HMW}.  It has been proposed that
the  elements of the mass matrix appear as random numbers
from certain interval and
there is no special structure of the mass matrix
dictated by certain symmetry.
It was estimated  how  frequently neutrino oscillation
data can be reproduced in this way. 
Random values of the complex phases of the 
mass matrix elements $M_{\alpha \beta}$ have  also been  considered
\cite{vissani}.\\

It was realized that the structure of the mass matrix 
depends  strongly on the unknown CP violating phases, especially in the
case  of degenerate spectrum. In general, in the system of three
Majorana neutrinos there are three CP violating phases: the
Dirac phase, $\delta$, the unique phase in the mixing matrix
relevant for oscillations, and two
Majorana phases, which are  relative phases of the three mass eigenstates.

In most of previous studies, the CP violating phases
were neglected and CP-parities have been discussed mainly
(see, however, \cite{FGM} and also \cite{savoy}, where 
the role of phases in the generation of large solar mixing is considered).\\

In this paper  we perform a systematic and comprehensive study  of
dependence of the neutrino mass matrix structure  on the CP violating
phases. We concentrate on  the first step  in the
``bottom - up" approach: reconstruction of  the mass matrix in the flavor
basis. A short discussion  of basis dependence
(which deserves a separate study) will be presented in sect. 6.7.
We  suggest a  way to analyze  all possible structures of the mass matrix
which are allowed by experimental data.

The paper is organized as follows. In section \ref{matrix} we describe our
approach and summarize physical inputs from neutrino oscillation
experiments. In sections \ref{plots} and \ref{mass1}, we study the
dependence of the mass matrix elements on CP violating phases.
We consider spectra with mass
hierarchy  (section \ref{angles}), partial degeneracy (section
\ref{secpd}) and complete degeneracy (section \ref{degs}).
In section \ref{rosi} we introduce and describe the $(\rho-\sigma)$ plots.
In section \ref{theo}  we consider implications of CP
violating phases for the structure of the mass matrix.
In section \ref{seven}  we discuss our result and draw
conclusions.

\section{Reconstructing $\nu$ mass matrix\label{matrix}}

The reconstruction of the mass matrix
in the flavor basis is the first step of the bottom-up approach. 
The next step -- selection of the  symmetry basis -- requires
additional 
assumptions and therefore is more ambiguous. We will shortly discuss this
issue in section \ref{symmetry}.  However, already in 
the flavor basis one  can 

- search for  regularities in the mass matrix,  

- study correlations of different matrix elements, 

- study  correlations between the neutrino mass matrix 
and charged lepton masses, that is  study of  possible flavor alignment.   

The flavor basis is
convenient for searches of symmetries associated with the lepton numbers
$L_e$, $L_{\mu}$, $L_{\tau}$. Last but not least, it is not excluded that
flavor basis is not much different from the symmetry basis.

\subsection{Mass matrix in flavor basis. Parameterization}

The neutrino mass matrix in flavor basis, $M$, can be written as
\begin{equation}
M = U^*M^{diag}U^{\dag}~,
\label{matr}
\end{equation}
where
\begin{equation}
M^{diag} \equiv Diag(m_1e^{-2i\rho},~m_2,~m_3e^{-2i\sigma})~.
\label{mmm}
\end{equation}
Here $m_i$ are the moduli of neutrino mass eigenvalues
and $\rho$ and $\sigma$ are the two CP violating Majorana phases,
varying between $0$ and $\pi$.  
The neutrino mixing matrix $U$ is defined by
\beq
\nu_{\alpha L}=U_{\alpha i}\nu_{iL}\;, \;\;\;\; \alpha =e,\mu,\tau\;,
\;\;\;\; i=1,2,3\;,
\nonumber
\end{equation}
where $\nu_\alpha$ are the flavor neutrino states,
and $\nu_i$ are the mass eigenstates. We use
the standard parameterization for $U$:
\begin{equation}
U =
\left( \begin{array}{ccc}
 c_{13}c_{12} & s_{12}c_{13} & s_{13} e^{-i\delta}
\\ -s_{12}c_{23}-s_{23}s_{13}c_{12} e^{i\delta} &
    c_{23}c_{12}-s_{23}s_{13}s_{12} e^{i\delta}
& s_{23}c_{13} \\ s_{23}s_{12}-s_{13}c_{23}c_{12} e^{i\delta}
 & -s_{23}c_{12}-s_{13}s_{12}c_{23} e^{i\delta} & c_{23}c_{13}
\end{array} \right)\;,
\label{U}
\end{equation}
where $c_{ij} \equiv \cos \theta_{ij}$, $s_{ij} \equiv \sin \theta_{ij}$
and $\delta$ is the CP violating Dirac phase. The mixing angles
vary between $0$ and $\pi/2$ and $\delta$ varies between $0$ and $2\pi$.

The matrix $M$ is symmetric and, therefore,  defined by six elements\footnote
{Notice that, in general, elements  are determined by 6 absolute
values and 6 phases. Three phases can be eliminated by renormalization
of the neutrino wave functions, so that only 9 quantities have physical 
meaning. This matches  9 parameters (3 phases, 3 masses and 3 angles)
which appear in our parameterization.}.
According to Eqs.(\ref{matr},\ref{mmm}), they can be written explicitly as
\beq
M_{\alpha\beta}= (U_{\alpha 1}^*U_{\beta 1}^*)
\;m_1e^{-2i\rho} + (U_{\alpha 2}^*U_{\beta 2}^*)
\; m_2 + (U_{\alpha 3}^*U_{\beta 3}^*)
\;m_3e^{-2i\sigma}\;,\;\;\;\; \alpha,\beta=e,\mu,\tau\;.
\label{ABC}
\end{equation}
The expression for $M_{\alpha\beta}$, in terms
of $m_i,\theta_{ij},\delta,\rho,\sigma$, is given in the appendix.

The mass matrix elements, as  functions of  CP violating phases, 
depend on the parameterization. Our choice  of  parameterization
has the following motivations. 

In contrast to previous works, {\it e.g.} \cite{FSV,mina}, 
we ascribe the Majorana phases to the first and the third mass
eigenstates (\ref{mmm}), so that in the limit of strong mass hierarchy
the dependence on  the  phase $\rho$ disappears.
Furthermore, in the limit of degeneracy  the interplay  of the
two Majorana phases (due to mixing) is weaker if $\sigma$
is attached to the third mass eigenstate.

We use the standard parameterization (Eq.(\ref{U})) of the mixing matrix
for two reasons: (1) it is the most often used parameterization,
in particular in studies of the CP violation in neutrino oscillations;
(2) the Dirac phase is associated to $s_{13}$. So,  the influence
of $\delta$ on structure of the matrix is suppressed,  and moreover,  with
improvements of bound on $s_{13}$ the effect of the phase will decrease.

Notice that in our parameterization $m_{ee}$ element depends on all three
phases. In particular, the phase $\delta$ enters in combination 
$\sigma - \delta$.  The dependence of $m_{ee}$ on $\sigma -
\delta$ is very weak being suppressed by $s_{13}^2$. 
Another  parameterization of the mixing matrix has been used, {\it e.g.}, in
\cite{FSV}, in which $m_{ee}$ does not depend on $\delta$.
We find that our parameterization is more convenient 
when all elements of the mass matrix 
(and not only $m_{ee}$) are analyzed.  
Moreover, transition to our  parameterization in $m_{ee}$  
is reduced to the simple  shift $\sigma \rightarrow \sigma  - \delta$.

In what follows we will, mainly, analyze the absolute values of mass
matrix
elements, $m_{\alpha\beta}$:  
\begin{equation}
M_{\alpha\beta} = m_{\alpha\beta} e^{i\phi_{\alpha\beta}}
\;,\;\;\;\; \alpha,\beta=e,\mu,\tau\; , 
\label{elem}
\end{equation}
since the absolute values may give more straightforward information on
possible underlying symmetry. 
The phases of the mass  matrix elements, $\phi_{\alpha\beta}$, are
also important for theory.  The phases  are  
 known functions of 9 physical parameters:  
$\phi_{\alpha\beta} = \phi_{\alpha\beta}(m_i, \theta_{ij},
\rho,\sigma,\delta)$. They  can be found if
these  9 parameters are measured. We comment on the phases 
and include separate discussion in section \ref{MEphases}.

Notice that, in flavor basis, $m_{\alpha\beta}$ are physical quantities,
that is, they can be {\it directly} measured in physical processes.
In particular, the rate of the neutrinoless $2\beta$-decay
is proportional to $m_{ee}^2$.
Other entries are in principle measurable in processes
with $\Delta L=2$, like the  decay $K^{+}\rightarrow \pi^-\mu^+\mu^+$ or
the scattering 
$e^-p\rightarrow\nu_el^{\pm }{l'}^{\pm }X$ (for a review see \cite{zuber}).
The rates of these processes are proportional
to $m_{\alpha\beta}^2$, where $\alpha$ and $\beta$
are the flavors of the two produced  leptons in the final state 
or leptons in the initial and final states. 

The present bounds on the elements $m_{\alpha \beta}$ other than
$m_{ee}$ are many orders of magnitude weaker than indirect limits. For
instance the bound on $m_{e \mu}^2$ is 16 orders of magnitude above the
limit obtained from oscillations \cite{rode}. Clearly, the possibility to
improve direct limits deserves further studies.

We introduce the dimensionless quantities
$$
\tilde{M}_{\alpha\beta}\equiv
\frac{M_{\alpha\beta}}{m_3}\;,  \qquad \tilde{m}_{\alpha\beta}\equiv
\frac{m_{\alpha\beta}}{m_3}\;,
$$
where $m_3$ is the largest mass eigenvalue.

\subsection{Conservation of the sum of masses squared \label{sumrule}}

According to Eq. (\ref{matr}), the mixing  matrix distributes the masses
from $M^{diag}$ to the elements of the flavor
mass matrix $M$:
$$
m_1, m_2, m_3 \rightarrow m_{\alpha\beta}.
$$
The following sum rule is useful for analysis of the flavor mass matrix:
\begin{equation}
S_0 \equiv \sum_{i = 1,2,3} m^2_i = \sum_{\alpha,\beta=e,\mu,\tau}
m_{\alpha\beta}^2 \;.
\label{sum}
\end{equation}
That is, the sum of moduli squared of all the elements of the mass matrix
is invariant under basis transformation (rotation). The equality
(\ref{sum}) is the straightforward consequence of the unitarity of
transformation.
Indeed, denoting $M_i\equiv M_{ii}^{diag}$  ($|M_i|\equiv m_i$), we can
write,  using Eqs.(\ref{matr},\ref{mmm}):
$$
\sum_{\alpha,\beta} m_{\alpha\beta}^2 =
\sum_{\alpha,\beta} \left|\sum_{i} U^*_{\alpha i}U^*_{\beta i}M_i
\right|^2 =
$$
\begin{equation}
\sum_{i} \sum_{\alpha,\beta} \left|U^*_{\alpha i}U^*_{\beta i}M_i
\right|^2 +
\sum_{i > j} \sum_{\alpha,\beta}
\left[U^*_{\alpha i}U^*_{\beta i}  U_{\alpha j} U_{\beta j} M_i M_j^*  +
h.c.
\right] .
\label{sum1}
\nonumber
\end{equation}
The first term is immediately reduced to
$\sum_{i} \left|M_i \right|^2 = \sum_{i} m_i^2$, whereas
the second term is zero due to orthogonality: $\sum_{\alpha} U^*_{\alpha i}
U_{\alpha j}=  0$ for $i \neq j$.

\subsection{Experimental input}

In what follows we will find
$m_{\alpha\beta}=m_{\alpha\beta}(m_i,\theta_{ij},\delta,\rho,\sigma)$,
using all available neutrino data.
We will restrict our analysis to the case
of normal mass hierarchy (ordering): $m_1\ll m_2\ll m_3$ ($m_1<m_2<m_3$).
The inverted hierarchy (ordering) is disfavored
by supernova SN1987A data \cite{supernova} (see, however, \cite{barger}).
Normal hierarchy
implies that $m_2^2-m_1^2 \equiv \Delta m^2_{sol}$ and
$m_3^2-m_2^2 \simeq m_3^2-m_1^2 \equiv \Delta m^2_{atm}$.
We will also restrict ourself to the LMA MSW solution
of the solar neutrino problem, which gives the best
global fit of the solar neutrino data \cite{globalfits}.
This solution looks especially plausible after
SNO data \cite{sno} and it can be tested
in the already operating KamLAND experiment \cite{kamland}.
We accept interpretation of the atmospheric neutrino results
\cite{superk} in terms of $\nu_{\mu}\rightarrow \nu_{\tau}$
oscillations as the dominant mode.

The following experimental information is used.
\begin{itemize}
\item
The best fit point for the LMA solution \cite{globalfits}:
\beq
\tan^2 \theta_{12}=  0.41, ~~~~~ \Delta m^2_{sol} = 6.2 \cdot 10^{-5} {\rm
eV}^2\;.
\label{sol}
\end{equation}
At $99\%$ C.L. the following intervals are allowed:
\beq
\tan^2 \theta_{12}= 0.27 - 0.75, ~~~~
\Delta m^2_{sol} = (2.7 -  25)\cdot 10^{-5}{\rm eV}^2\;.
\label{solar}
\end{equation}
\item
From atmospheric neutrino
analysis we take, at the  $99\%$ C.L. \cite{superk}:
\beq
\tan \theta_{23} = 1 ^{+ 0.54}_{-0.35}, ~~~~
\Delta m^2_{atm}=(2.5^{+2.5}_{-1.3}) \cdot 10^{-3} {\rm eV}^2\;.
\label{atmo}
\end{equation}
\item
We use the CHOOZ bound on $\theta_{13}$
at the $90\%$ C.L. \cite{chooz}: 
\beq
\sin \theta_{13} \lesssim 0.22  ~~~(\Delta m^2_{atm}\gtrsim 2\cdot
10^{-3}{\rm eV}^2).
\label{choozexp}
\end{equation}
\end{itemize}
In our discussion we will take into account
the upper limit on the Majorana neutrino mass
from neutrino-less $2\beta$ decay \cite{kk,igex}:
\beq
m_{ee} < 0.35 {\rm ~eV}\qquad(90\% {\rm ~C.L.})~, 
\label{meebeta}
\end{equation}
and  $m_{ee}\lesssim 1$ eV, if uncertainties in the nuclear matrix
elements are taken into account. 
We think, it is premature to  include in the analysis the recent
result on  $2\beta_{0\nu}$-decay \cite{evidence} 
which has a controversial interpretation (see discussion in
\cite{FSV,antievi}).
We consider the direct kinematic bound on the mass of electron
neutrino \cite{mainz},
\beq
m_e < 2.2{\rm ~eV}\qquad(95\% {\rm ~C.L.})\;.
\label{mainz}
\end{equation}

The unknown CP violating phases $\delta,\rho,\sigma$,  as
well as the absolute mass scale, $m_1$, and the angle $\theta_{13}$
are treated as free parameters.  

Let us emphasize that the experimental input
(\ref{sol}) - (\ref{choozexp}) does not depend on the Majorana phases,
$\rho$ and $\sigma$, and on $m_1$, because only the differences
of $m_i^2$ enter the oscillation probabilities.
The input does not depend also on the Dirac phase, $\delta$.
This can be explicitly seen from the parameterization (\ref{U}):
at the level of present experimental accuracy,
the solar and atmospheric neutrino results are determined
by $U_{e1},U_{e2}$ and $U_{\mu 3},U_{\tau 3}$ respectively.
CHOOZ gives the bound on $|U_{e3}|$. All these quantities
do not depend on $\delta$.

\subsection{$\mu\tau$-block and $e$-row elements}

In view of large 2-3 mixing, it is convenient  to split
the six independent elements  of the mass matrix into two groups:

- elements of the $\mu \tau$-block: $m_{\mu \mu}$, $m_{\mu \tau}$,
$m_{\tau \tau}$, with zero electron lepton number, $L_e = 0$;

- elements of the $e$-row: 
$m_{e \mu}$, $m_{e \tau}$  with $L_e = 1$, and 
$m_{ee}$ with $L_e = 2$. 
 
As we will see later, these groups of elements have
different dependences on CP violating phases.
Moreover, such a split can be motivated by phenomenology.

\subsection{Small parameters, mass ratios  and limits}

There are several small parameters in the problem: 

1)  The ratio of masses squared differences:
\beq
r_{\Delta} = 
\sqrt{\frac{\Delta m^2_{sol}}{\Delta m^2_{atm}}}~= 
0.16~^{+ 0.29}_{- 0.09}~, 
\label{rdelta}
\end{equation}
where the central value corresponds to the  best fit values of mass
squared
differences (see
(\ref{sol}),(\ref{atmo})), and the interval is obtained   
varying $\Delta m^2_{atm}$ and $\Delta m^2_{sol}$
in the ranges given in (\ref{atmo}) and (\ref{solar}). 

2) The 1-3 mixing: $s_{13}\lesssim 0.2$ (see (\ref{choozexp}));

3) The deviation of 2-3 mixing from maximal value
(see (\ref{atmo})), which can be described by
\beq
\xi\equiv\cos{2\theta_{23}} \sim (- 0.4  \div  0.4) .
\label{xidef}
\end{equation}
Future experimental studies will further restrict all these three
parameters.\\

Let us introduce  dimensionless
parameters - the  ratios of mass eigenvalues:
\beq
k \equiv \frac{m_1}{m_2}, ~~~~~~r\equiv\frac{m_2}{m_3}.  
\label{kandr}
\end{equation}
In the case of strong mass hierarchy, $k\approx 0$ and 
$r = r_{\Delta}$.
Clearly, we may have $k \sim 1$ and  $r \ll 1$. If
$r \sim 1$, then $k \approx 1$.\\

Let us consider the mass matrix in various limits.

1) $r=s_{13}=\xi=0$.
In this case  we arrive at a matrix with
zero $e$-row elements
and $\mu\tau$-block elements equal to $m_3/2$.
Obviously no dependence on CP-phases appears.

2) $r = s_{13} = 0$, $\xi \neq 0$ (see (\ref{mmu})-(\ref{mtt})). We get
\beq
\tilde{m}_{\mu\mu}= \frac{1}{2}(1-\xi)\;,\qquad
\tilde{m}_{\tau\tau}= \frac{1}{2}(1+\xi)\;,\qquad
\tilde{m}_{\mu\tau}= \frac{1}{2}\sqrt{1-\xi^2}\;.
\label{spread}
\end{equation}
The element $m_{\mu\tau}$ is almost
unchanged with respect to maximal $\theta_{23}$,
while $m_{\mu\mu}$ and $m_{\tau\tau}$ vary
with $\xi$ significantly and  in opposite directions.
The determinant of the $\mu\tau$-block  is zero. Again, there is
no dependence on CP violating phases.

3) $s_{13}=\xi=m_1=0$, but $r \neq 0$. We have
\beq
M = \frac{m_3}{2}\left(
\begin{array}{ccc}
2s_{12}^2r & \sqrt{2}s_{12}c_{12}r & -\sqrt{2}s_{12}c_{12}r \\
\dots & e^{-2i\sigma}+c_{12}^2r & e^{-2i\sigma}-c_{12}^2r\\
\dots & \dots & e^{-2i\sigma}+c_{12}^2r~
\end{array}
\right)~.
\label{rlimit}
\end{equation}
Now dependence on the Majorana phase appears in the
$\mu \tau$-block, but there is no phase dependence of the $e$-row
elements. 

The influence of the CP violating phases on the
matrix structure is very weak in the limit of
strong mass hierarchy and small $s_{13}$. 
Indeed, for $r \rightarrow  1$, the effect of phase 
$\sigma$ disappears,  
dependence of the elements on the Dirac phase is
associated with $s_{13}$, so the effect of $\delta$
decreases with $s_{13}$,  the dependence   on the phase
$\rho$ is associated to the mass $m_1$ and it is negligible
when $r \ll  1$. 

\subsection{Analytic expressions and phase diagrams}

Exact analytic expressions for the mass matrix elements in terms
of mass eigenvalues, $m_i$, mixing angles and phases are
given in the Appendix. We present the matrix elements as
sums of three contributions corresponding to
three different mass eigenvalues in (\ref{mee} - \ref{mtt}) and  as series
in powers of $s_{13}$ in (\ref{big}). Representation of 
$m_{\alpha\beta}$ as the sums of three terms with different phases  
is given in  (\ref{mutau-m}, \ref{em-m}, \ref{ee-m}). 
We will use various approximate expressions for $m_{\alpha\beta}$ which
can be obtained from Eqs.(\ref{mee} - \ref{ee-m}).
 
\begin{figure}
[t]
\begin{center}
\epsfig{file=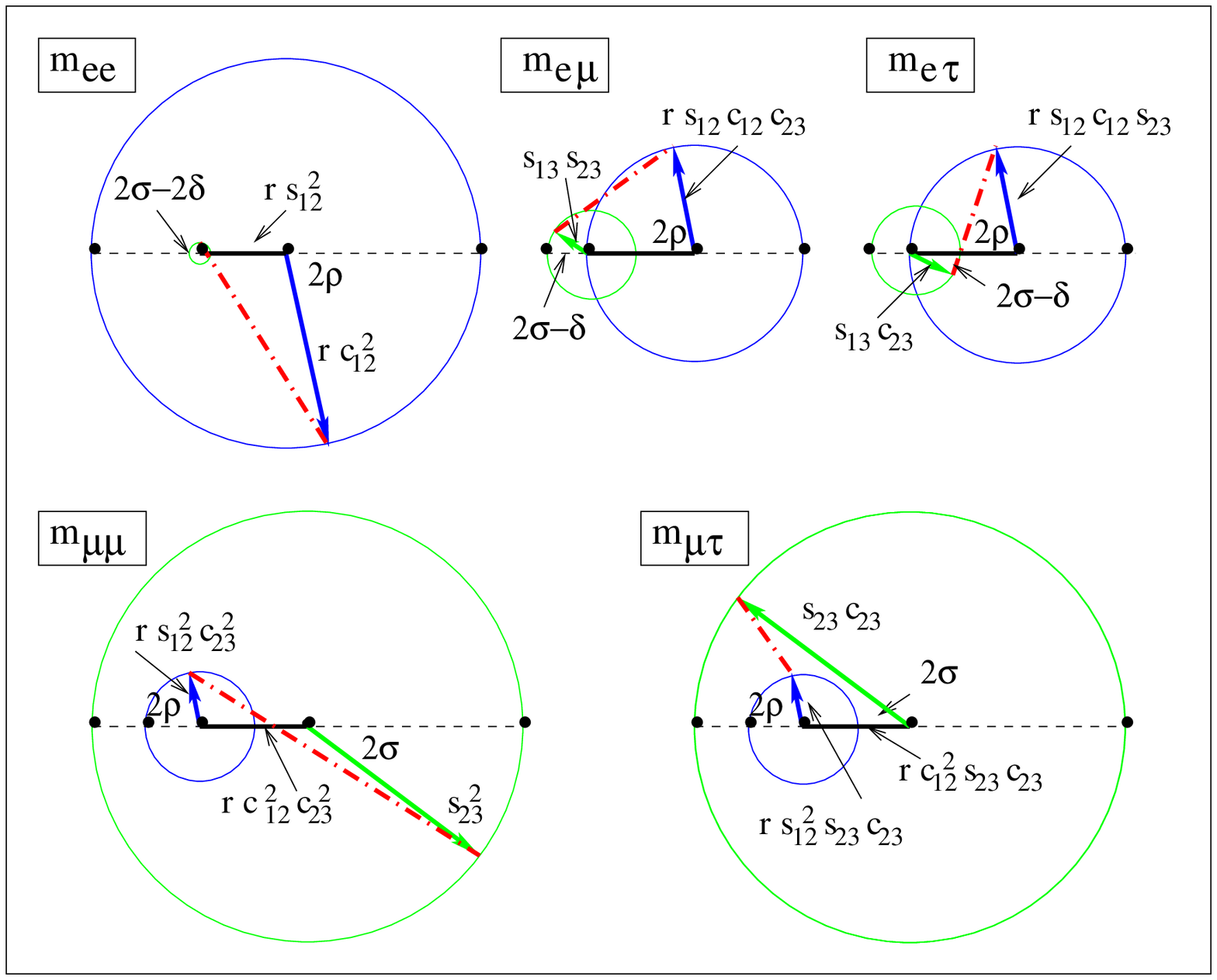,width=450pt,bbllx=0, bblly=0, bburx=510, bbury=420}
\end{center}
\caption{Phase diagrams. Shown is  a graphic representation of the mass
matrix elements  $m_{\alpha\beta}$ in the complex plane. Thick lines
represent the  contributions of the three leading terms in expressions
(\ref{mutau-m},\ref{em-m},\ref{ee-m}).  The diagram for $m_{\tau\tau}$ is
obtained from the $m_{\mu\mu}$ diagram  with the substitution
$s_{23}\leftrightarrow c_{23}$. The length  of the dash-dotted line gives
the value of the matrix element. The diagrams  correspond to the spectrum
with partial degeneracy ($k\approx 1$, $r\lesssim 1$).}
\label{phase}
\end{figure}

For small $s_{13}$, one can draw simple graphic representation
of the mass matrix elements in the complex plane
(Fig.\ref{phase}). 
Neglecting  terms of order $s_{13}$  in the brackets of
Eqs.(\ref{mem} -  \ref{mtt}) or, equivalently, $\epsilon$ terms 
in Eqs.(\ref{mutau-m}, \ref{em-m}), we find that each mass $m_{\alpha\beta}$
turns out to be the sum of three terms with phase factors
which depend on
certain combinations of the phases $\delta$, $\rho$, $\sigma$.
So, in the complex plane the masses $m_{\alpha\beta}$ can be represented 
as sums of three vectors (corresponding to the three terms).
The lengths of these vectors are determined by mass eigenvalues
(ratios $k$ and $r$) and mixing angles. The angles between vectors are
given by combinations of the phases $\delta$, $\rho$, $\sigma$.

We will call this graphic representation, 
used for $m_{ee}$ in \cite{smikla} and mentioned  
in \cite{viss,NMF}, the {\it phase diagram}. 
The phase  diagrams allow one easily to find minimal and maximal values 
as well as phases of the
matrix elements,  and  possible correlations between them.
In Fig.\ref{phase} we show phase diagrams for the case
of partial degeneracy: $k\approx 1\;,\;\;r\lesssim 1$.

The mass matrix elements are periodic functions of the
CP violating phases. In the next section  we will analyze  these
dependences by 
quantifying for each phase the amplitude of variations, the period and
the average value of the element. The latter  we define as
the average between maximal and minimal possible value of the element.
We will also  consider the relative phases of variations of  different
elements and correlations between them.

\section{CP phases  in the case of hierarchical mass spectrum
\label{plots}}

In  the limit  of strong mass hierarchy,
when $m_1\approx 0$, $m_2^2 \approx \Delta m^2_{sol}$
and $m_3^2 \approx \Delta m^2_{atm}$, 
$k\approx 0$,  only one Majorana phase, $\sigma$,  is relevant. 
If also $s_{13}\approx 0$, we have, for the matrix of the moduli:
\beq
\tilde{m} = \left(
\begin{array}{ccc}
s_{12}^2r & c_{23}s_{12}c_{12}r & s_{23}s_{12}c_{12}r \\
\dots & \left| c_{12}^2 c_{23}^2
r + s_{23}^2 e^{-2i\sigma} \right| & 
s_{23} c_{23} \left| c_{12}^2 r - e^{-2i\sigma}
\right|\\
\dots & \dots & \left| c_{12}^2 s_{23}^2 r + c_{23}^2
e^{-2i\sigma}\right|
\end{array}
\right)~.
\label{hmatrix}
\end{equation}
Notice that the $e$-row  elements are real, $\phi_{e\alpha} =
0$, whereas for the phases of 
$\mu\tau$-block elements we have $\phi_{\alpha\beta} \approx - 2\sigma$. 
The corrections are proportional to  $r$. 

\subsection{Dependence of $m_{\alpha\beta}$ on CP violating phases
\label{angles}}

\begin{figure}
[t]
\begin{center}
\epsfig{file=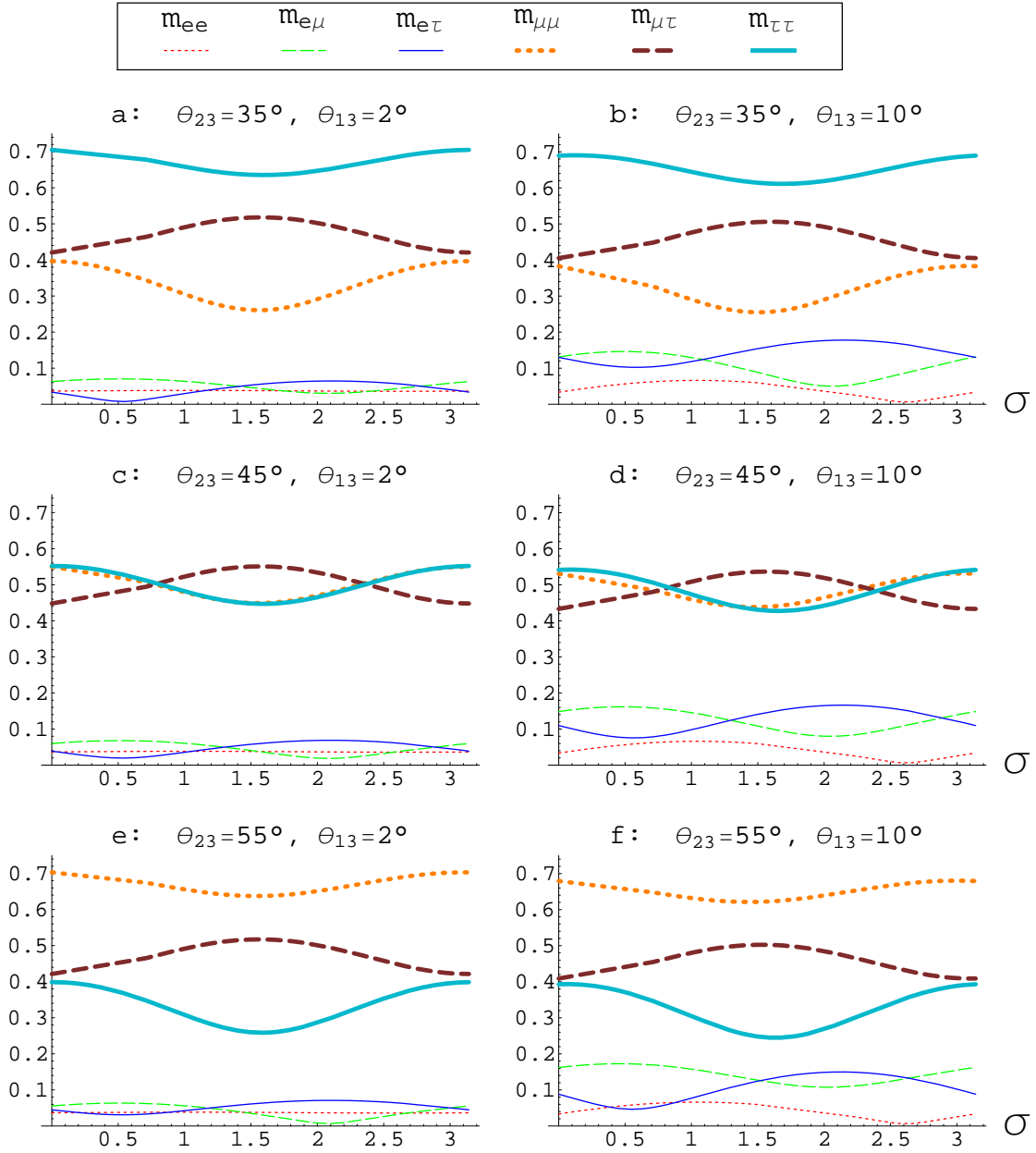,width=450pt,bbllx=70, bblly=230, bburx=550, bbury=770}
\end{center}
\caption{Dependence of the absolute value of  neutrino mass matrix
elements (in units $\sqrt{\Delta m_{atm}^2}=0.05$ eV)  on $\sigma$, for
different values of $\theta_{23}$ and $\theta_{13}$.  We take
$\tan^2\theta_{12}=0.36$, $\delta=\pi/3$, $m_2=0.14~m_3$, $m_1=0$.}
\label{sigma}
\end{figure}

In Figs.~\ref{sigma}, \ref{sigmam2}  we show the six mass matrix
elements, $\tilde{m}_{\alpha\beta}, \; \alpha,\beta=e,\mu,\tau$, 
as functions of the phase  $\sigma$, for different values of the mixing
angles $\theta_{23}$ and $\theta_{13}$, from  the allowed regions given
in (\ref{atmo}) and (\ref{choozexp}).
Main features of the dependences can be well understood taking
the lowest order  terms in $r$ and $s_{13}$ from
Eqs.(\ref{mutau-m},\ref{em-m},\ref{ee-m}).

According to  (\ref{hmatrix}),  the dependence of 
$\mu\tau$-block elements on $\sigma$ is a 
result of the interplay of the main, ${\cal O}(1)$, term and of the
${\cal O}(r)$ term. In the lowest  order, the $\mu\tau$-block elements do
not  depend on $s_{13}$ (see Figs.\ref{sigma},\ref{sigmam2});
$m_{\mu\tau}$ has an
opposite phase with respect  to the  two  other elements. The relative
amplitudes of variations equal 
\beq
\frac{\Delta m^{\sigma}}{m} \approx
c_{12}^2 r \times
\left\{
\begin{array}{ll}
\cot^2{\theta_{23}}, &  ~~~m_{\mu\mu}\\
\tan^2{\theta_{23}}, &  ~~~m_{\tau\tau}\\
1 ,                  &  ~~~m_{\mu\tau}
\end{array}
\right. .
\label{blocks}
\end{equation}
For the best fit values of the parameters  the amplitudes  are of order
$10\%$. In the
case of non-maximal 2-3 mixing, the amplitudes  can reach $\sim 25\%$.
The corrections $\sim r s_{13}$ (\ref{corr})
lead to small phase shift and small change of  the amplitude of
variations.

Neglecting terms of the order $r s_{13}s_{12}^2$,
we get  from (\ref{em-m}) expressions for $m_{e\mu}$ and  $m_{e\tau}$:
\beq
\tilde m_{e\mu} \approx \left| r
s_{12}c_{12}c_{23}
+s_{13}s_{23}e^{i(\delta-2\sigma)} \right|\;,
\label{memd}
\end{equation}
\beq
\tilde m_{e\tau} \approx \left| r
s_{12}c_{12}s_{23} - s_{13}c_{23}e^{i(\delta-2\sigma)} \right|\;.
\label{metd}
\end{equation}
So,  the elements  $m_{e\mu}$ and $m_{e\tau}$
depend on phases in the combination
$(\delta-2\sigma)$, they change  with $(\delta-2\sigma)$ in
opposite phases,  their  values are determined by the
interplay of
the order $r$ and order $s_{13}$ terms, which can have  comparable  sizes.
Maximal values of $m_{e\mu}$ and $m_{e\tau}$ increase with
$s_{13}$.

The relative amplitude of variations of $m_{e\mu}$ with $(\delta - 2\sigma)$
is maximal when the  two terms in (\ref{memd}) have the same modulus:
\beq
s_{13}=s_{13}^0 \equiv \frac 12 r \sin{2\theta_{12}}
\cot \theta_{23} \approx 0.07\;.
\label{s130}
\end{equation}
If $s_{13}=s_{13}^0$,
\beq
m_{e\mu}=0 \qquad \textrm{for} \qquad \delta-2\sigma=\pi \;.
\nonumber
\end{equation}
Maximal value of $m_{e\mu}$ equals  $m_{e\mu}^{max}  = 2 \bar{m}_{e\mu} =
\sin{2\theta_{12}} c_{23}r$.
For
$s_{13}<s_{13}^0$, (Fig.\ref{sigma}a,\ref{sigma}c,\ref{sigma}e),
the average value of $m_{e\mu}$ is determined  by the
first term in  (\ref{memd}), whereas the amplitude of variations is given
by $s_{13}/s_{13}^0$.
For $s_{13}>s_{13}^0$, the second term   in (\ref{memd}) dominates. It
determines the average  value of $m_{e\mu}$, around  which variations
occur. The relative amplitude of variations  is given  by the factor
$s_{13}^0/s_{13}$ (Fig.\ref{sigma}b,\ref{sigma}d,\ref{sigma}f).

Behavior of the element   $m_{e\tau}$ is similar: the two 
terms in (\ref{metd}) are equal and can cancel each other 
if 
\beq
s_{13}=\bar{s}_{13}^0 \equiv \frac 12 r \sin{2\theta_{12}}
\tan\theta_{23}
\end{equation}
($\bar{s}_{13}^0=s_{13}^0$  for maximal 2-3 mixing),  so 
\beq
m_{e\tau}=0 \qquad \textrm{for} \qquad \delta-2\sigma=0 \;.
\nonumber
\end{equation}
Corrections of order  $rs_{13}s_{12}^2$, neglected in
(\ref{memd}) and (\ref{metd}), produce a
small relative  shift of phases  of $\tilde{m}_{e\mu}$ and
$\tilde{m}_{e\tau}$ (see Fig.\ref{sigma}).

For the $ee$-element  we have:
\beq
\tilde m_{ee} \approx   \left|c_{13}^2 s_{12}^2 r + s_{13}^2
e^{2i(\delta-\sigma)} \right|\;.
\label{me}
\end{equation}
It   depends on the combination of phases
$2(\delta-\sigma)$.
Due to the factor  $r s_{12}^2$ in the first term, both contributions  in
(\ref{me}) can be comparable  in spite of the
$s_{13}^2$-order of the second term.  Two  terms are equal at
$\tan{\theta_{13}} =  s_{12}\sqrt{r}\approx0.21$,  that is, near the upper
limit for $s_{13}$. In this case the amplitude of variation  can be
maximal and
\beq
m_{ee}=0 \qquad {\rm for}~~~~\delta-\sigma=\pi/2,~3\pi/2\;.
\nonumber
\end{equation}
Such a situation  is approximately realized in
Fig.\ref{sigma}b,\ref{sigma}d,\ref{sigma}f.
For small values of $s_{13}$ ($s_{13} \ll 0.2$), the dependence of
$m_{ee}$ on phases is negligible  
(Fig.\ref{sigma}a,\ref{sigma}c,\ref{sigma}e).  The
relative amplitude  of variations is determined by the
ratio $\tan^2{\theta_{13}}/(rs_{12}^2)$   and the average value  equals
$\overline{m}_{ee} \approx s_{12}^2 r$.

Let us analyze the dependence of matrix elements on the Dirac phase $\delta$.
The elements of $\mu\tau$-block  depend on  $\delta$  very weakly,
via order $r s_{13}$ corrections (see (\ref{corr})).
E.g., for $s_{13} \approx 0.14$
(Fig.\ref{sigma}b,\ref{sigma}d,\ref{sigma}f),  
we find 
$\Delta m_{\mu\mu}^{\delta}/\bar{m}_{\mu\mu}\approx 
\Delta m_{\tau\tau}^{\delta}/\bar{m}_{\tau\tau} \sim 0.02$. The
dependence of $m_{\mu\tau}$ on $\delta$ is further suppressed by  the
factor  $\xi\equiv \cos{2\theta_{23}}$.

The elements of $e$-row have much stronger  relative dependence on
$\delta$.  As we pointed out, the
elements $m_{e\mu}$ and $m_{e\tau}$  depend on phases in the combination
$(\delta - 2 \sigma)$ (this feature is weakly violated  by corrections
$\sim r s_{13}$,  which depend on the phase $\delta$ only).
So, up to corrections of order $s_{13}r$, one can extract the
information on the $\delta$ dependence of the elements from the
Fig.\ref{sigma} (or Fig.\ref{sigmam2} for large $r$) immediately.
The change  of $\delta$ by amount $\Delta\delta$ is
equivalent to  horizontally shift the lines which correspond
to $m_{e\mu}$  and $m_{e\tau}$ along with $\sigma$-axis by  $\Delta\delta/2$
and  the $m_{ee}$ line by  $\Delta\delta$,
with respect to the  lines of $\mu\tau$-block,  which are almost
unchanged.  The phase $\delta$ can be selected
in such a way that certain features of the $m_{ee}$  line and  other
$e$-row lines will occur at the same value of $\sigma$.
For instance, according to Fig.\ref{sigma}b, one can get
$\tilde{m}_{ee} \ll \tilde{m}_{e\mu} \ll \tilde{m}_{e\tau}$.

All  the elements have the same period of variation
with $\sigma$, although  the phases of variations are different.
There is a phase shift by $\pi$ within different groups:
\beq
\phi(\tilde{m}_{\mu\mu}) =
\phi(\tilde{m}_{\tau\tau}) = \phi(\tilde{m}_{\mu\tau}) + \pi = - 2\sigma\;;
\label{shiftmt}
\end{equation}
\beq
\phi(\tilde{m}_{e\mu}) =\phi(\tilde{m}_{e\tau}) + \pi = - 2\sigma + \delta \;.
\label{shifter}
\end{equation}
There is a relative shift of phase between $\mu\tau$-block and $e$-row
elements which is determined by  $\delta$:
\beq
\phi(\tilde{m}_{e\mu}) - \phi(\tilde{m}_{\mu\mu})
=  \delta~, ~~~~
\phi(\tilde{m}_{ee})  - \phi(\tilde{m}_{e\mu}) =  \delta.
\end{equation}
These relations are weakly broken by corrections of order $r s_{13}$.

\subsection{Dependence of masses  on $\theta_{12}$, $r$ and $k$
\label{mass}}

Variations of $\theta_{12}$ within  the allowed LMA region, given in
(\ref{solar}), do not produce substantial changes of results shown in
Fig.\ref{sigma}. With increase  of $\theta_{12}$, the
amplitudes of variations of $\mu\tau$-block elements  with $\sigma$ (see
(\ref{blocks})) decrease as $c_{12}^2$.  For maximal 1-2 mixing we get
$\sim 30\%$ decrease in comparison with  the best fit value of
$\theta_{12}$. In contrast, the amplitude of  variations of these elements
with $\delta$ increases as $\sin{2\theta_{12}}$  (see
(\ref{corr})). The dependence on $\delta$ remains weak, because 
the increase of the amplitude can be only $10\%$. 
For the  $e$-row elements, the critical
value $s_{13}^0$ is proportional to  $\sin{2\theta_{12}}$ (see (\ref{s130})).
The $ee$-element $m_{ee}$  can be 
two times larger for almost maximal  solar mixing angle than for 
the best fit value (see (\ref{me})).

Changes  of $\Delta m^2_{sol}$ and $\Delta m^2_{atm}$ within the 
allowed regions, (\ref{solar}) and (\ref{atmo}),
produce  strong effect on the  structure of  mass
matrix. In Fig.\ref{sigmam2} we  show the dependence of mass  matrix
elements on $\sigma$  for $r=0.3$,  corresponding,
{\it e.g.}, to $\Delta m^2_{sol}\approx 2\cdot 10^{-4}{\rm eV}^2$  and
$\Delta m^2_{atm}\approx 2\cdot 10^{-3}{\rm eV}^2$.
For the $\mu\tau$-block elements, the amplitudes  increase
linearly with $r$ (see (\ref{blocks})) and  for $r\approx 0.3$
they can be   larger than   $30\%$. For   the $e$-row elements the
critical value $s_{13}^0$ (see (\ref{s130})) also  increases linearly with
$r$;  for $r\approx 0.3$, we get
$s_{13}^0\approx 0.13$. For $s_{13} < s_{13}^0$,  the average values of
elements increase as $m_{e\mu} \sim m_{e\tau} \sim r$,  but the
amplitude of variations with $(\delta-2\sigma)$ does not
change (compare Fig.\ref{sigmam2},
panels a,c,e, with corresponding panels in Fig.\ref{sigma}).
For $s_{13}\gtrsim s_{13}^0$,  the average values of $m_{e\mu}$ and
$m_{e\tau}$ do not depend on $r$,  while their amplitudes can be maximal
(Fig.\ref{sigmam2}, panels b,d,f).
The average value of  $ee$-element
increases with $r$: $m_{ee}\sim r$;  the amplitude of variations
with $2(\delta-\sigma)$ does not change.

Till now,  we have considered
the case $m_1=0$. A strong normal  hierarchy among mass eigenvalues, $m_1
\ll m_2 \ll m_3$, holds for $m_1$  up to approximately
$0.002 {\rm ~eV}$ ($k<0.3$).
Notice that, for $m_1\neq 0$,  both Majorana phases become relevant (see
(\ref{ABC})).
We have checked that varying  $m_1$ between $0$ and $0.002 {\rm eV}$, the
dependence of $m_{\alpha\beta}$  on angles and CP phases, showed in
Figs.\ref{sigma},\ref{sigmam2},  is qualitatively the same as for $m_1=0$,
except for the dependence of $m_{ee}$. The $ee$-element  can be about
two times larger.  Indeed, neglecting terms of order  $s_{13}^2$,  we get:
\beq
\tilde{m}_{ee} \approx  r s_{12}^2
\left(1+k \cot^2\theta_{12}\cos{2\rho}\right)\;.
\nonumber
\end{equation}
The second term in the brackets is of  order one for, {\it e.g.},
$m_1 = 0.002$ eV, $m_2 = 0.006$ eV,  $\tan^2\theta_{12} = 0.35$
and $\rho = 0, \pi$.
Depending on $\rho$,  the ratio of   $m_{ee}$ and the other $e$-row
elements can significantly change.

\subsection{Structure of the mass matrix in the hierarchical case}

\begin{figure}
[t]
\begin{center}
\epsfig{file=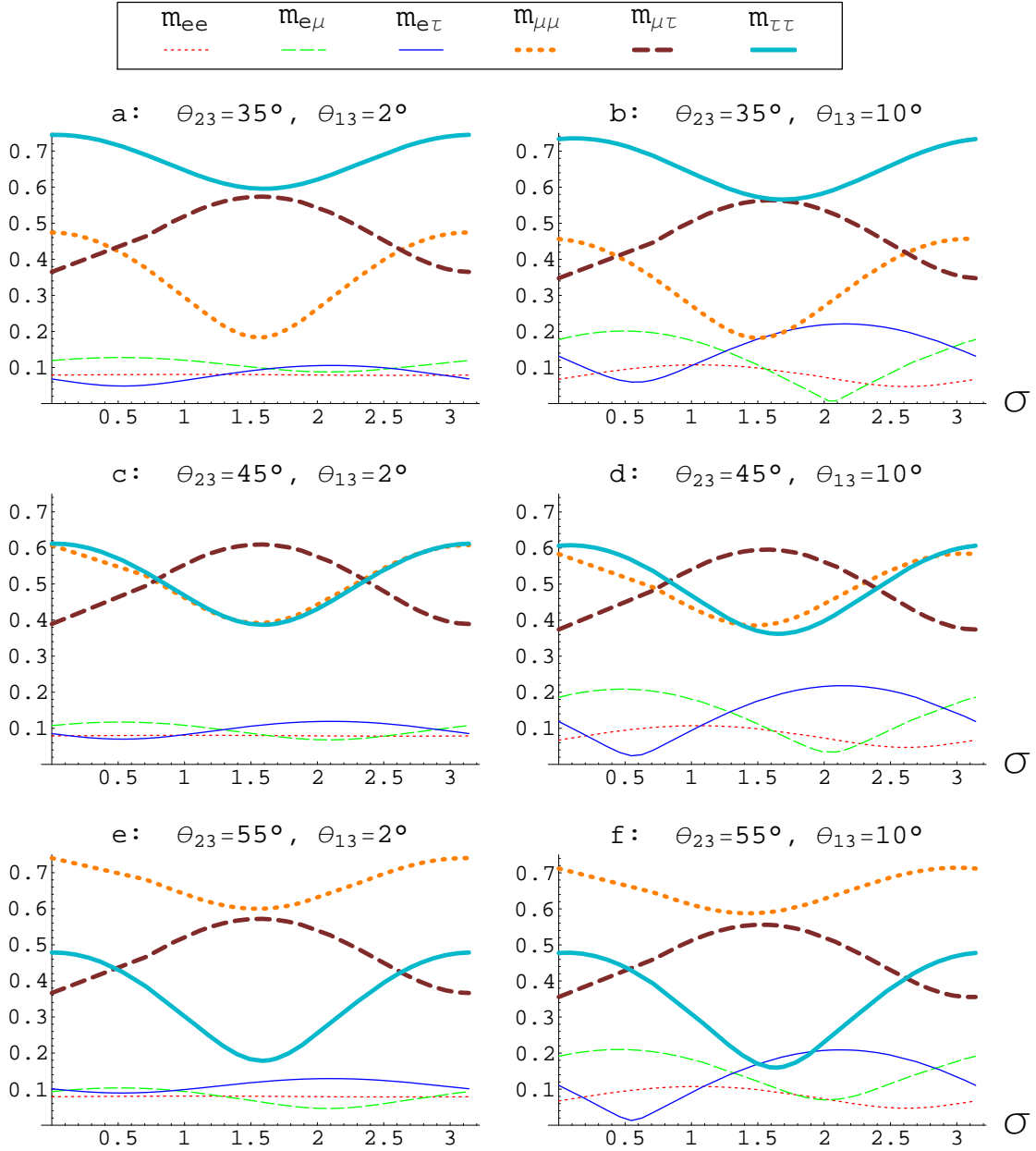,width=450pt,bbllx=70, bblly=230, bburx=550, bbury=770}
\end{center}
\caption{The same as in Fig.\ref{sigma}, but for $m_2=0.3~m_3$.}
\label{sigmam2}
\end{figure}

1) As follows from Figs.\ref{sigma},\ref{sigmam2},  the sharp structure
with the 
dominant $\mu\tau$-block and  sub\-dominant $e$-row  appears for small
$s_{13}$, small $r$ and near maximal 2-3 mixing. In this case
$$
m_{ee}\sim m_{e\mu}\sim m_{e\tau}  \ll 
m_{\mu\mu}\sim m_{\mu\tau} \sim m_{\tau\tau},
$$
\beq
\frac{m(e-{\rm row})}{m(\mu\tau-{\rm block})} \sim [s_{13}, r] \sim 0.1 ,
\label{ra}
\end{equation}
where $m(e-{\rm row})$ and $m(\mu\tau-{\rm block})$ refer to typical
masses of the $e-$row
and $\mu\tau$-block elements.
Improvements of the upper bound on  $s_{13}$  and on  
$\xi$, as well  as  establishing $\Delta
m^2_{sol}$ near its
present best fit value, will confirm this structure in assumption of
mass hierarchy. In the limit of sharp structure
[$\mu\tau$-block]-[$e$-row], the elements of  dominant block
depend very weakly   on $\delta$ and have about $10\%$ variations
(determined by $r$) due to the phase $\sigma$. The elements $m_{e\mu}$ and
$m_{e\tau}$  depend significantly  on the combination $(\delta-2\sigma)$,
unless very strong upper bound on $s_{13}$ will be established.
The $ee-$element varies with $2(\delta-\sigma)$, with amplitude
$\sim s_{13}^2$.
Thus, uncertainties  in the structure of the mass matrix due to
unknown CP violating phases can be substantially reduced by further
measurements of mixing angles and mass squared differences.

According to  Figs.~\ref{sigma},\ref{sigmam2},  for a large part of the
parameter space $(\theta_{23},\theta_{13},r,\delta,\sigma)$,  the
structure [$\mu\tau$-block]-[$e$-row] is less profound or even disappears.
Indeed, in the case of large $\xi$ or/and large  $r$,
the split between  masses within  the $\mu\tau$-block can be larger than
the gap
between $m(e-{\rm row})$   and $m(\mu\tau-{\rm block})$, depending on
$\sigma$. 
Separation of the elements in two
groups loses any sense. For the extreme case of  large values of $r$,  the
elements $m_{e\mu}$ and $m_{e\tau}$ can be even larger than  $m_{\mu\mu}$
or $m_{\tau\tau}$.

2) Dependence  of the gap between $\mu\tau$-block and  $e$-row elements on
$s_{13}$ and $r$ can be seen  comparing left and right panels in
Figs.\ref{sigma},\ref{sigmam2} and
 Fig.\ref{sigma}  with Fig.\ref{sigmam2}, respectively.
The deviation of $\theta_{23}$  from $45^{\circ}$, 
leading to a spread among the $\mu\tau$-block elements
(see (\ref{spread})), can strongly decrease the gap.

Let us quantify the size of gap. Taking  
only leading terms in $\xi$, $r$ and $s_{13}$, 
one has for the $\mu\tau$-block elements: 
\beq
m(\mu\tau-{\rm block})  \geq  m(\mu\tau-{\rm block})^{min}
\equiv \frac{m_3}{2}(1 - |\xi| -r c_{12}^2)\;,
\nonumber
\end{equation}
where $m(\mu\tau-{\rm block})^{min}$ is the value of $m_{\mu\mu}$ or
$m_{\tau\tau}$,
for $\sigma=\pi/2$. The upper bound on the $e$-row elements is given by
\beq
m(e-{\rm row})  \leq  m(e-{\rm row})^{max} \equiv
\frac{m_3}{\sqrt{2}}(s_{13}+rc_{12}s_{12})  \;,
\nonumber
\end{equation}
where $m(e-{\rm row})^{max}$ is the value of $m_{e\mu}$ or $m_{e\tau}$,
for $\delta-2\sigma=0$ or $\pi$, respectively. Therefore,
minimal value of the   gap equals 
\beq
m(\mu\tau-{\rm block})^{min} - m(e-{\rm row})^{max} =  \frac{m_3}{2}
\left[1 -
|\xi| - r (c_{12}^2+\sqrt{2}c_{12}s_{12})
- \sqrt{2}s_{13}\right]\;.
\end{equation}

One can also characterize the split of the elements by
the ratio of mean values of the $e$-row and $\mu\tau$-block elements.
Up to terms quadratic in $\xi$,  $r$ and $s_{13}$,
$\overline{m}(\mu\tau-{\rm block}) \approx
m_3/2$, while for the $e$-row
we can take
\beq
\overline{m}(e-{\rm row}) \equiv \sqrt{\frac{m_{e\mu}^2+m_{e\tau}^2}{2}}
\approx \frac{m_3}{\sqrt{2}} (s_{13}^2 + r^2 c_{12}^2 s_{12}^2)^{1/2}\;.
\nonumber
\end{equation}
Then
\beq
\frac{\bar{m}(e-{\rm row})}{\bar{m}(\mu\tau-{\rm block})}
\approx  \sqrt{2(s_{13}^2 + r^2 c_{12}^2 s_{12}^2)}\;,
\label{mean}
\end{equation}
in accordance with (\ref{ra}).
The ratio (\ref{mean}) does not depend on CP phases and on $\theta_{23}$.

3) Apart from special choice  of phases, the $ee$-element is typically of
the order of the other $e$-row elements.

4) The CP violating  phases can change significantly the structure of 
$e$-row. As 
follows from  Figs.~\ref{sigma},\ref{sigmam2}, one can get, {\it e.g.}, 
\beq
\begin{array}{cc}
m_{ee}\ll m_{e\mu} \ll m_{e\tau} &\textrm{(small $r$, large $s_{13}$)}\;;\\
m_{ee}\ll m_{e\mu}\approx m_{e\tau} &\textrm{(small $r$, large $s_{13}$)}\;;\\
m_{ee}\approx  m_{e\mu}\approx m_{e\tau}& \textrm{(large $r$, small
$s_{13}$)}\;;\\
m_{e\mu} \ll m_{ee}\ll m_{e\tau} &\textrm{(large $r$, large $s_{13}$)}\;.
\end{array}
\end{equation}
Any  element of the $e$-row can be the smallest one. 
All  possible orderings of $e$-row elements can be realized by
appropriate choice of the phases.

5) Depending on phases, one can find a configuration with almost uniform splits among the six mass matrix elements and structure with the dominant $\mu\tau$-block
disappears. Still the average value of the $e$-row elements  is smaller
than  the average value of the $\mu\tau$-block elements (see (\ref{mean})).
Thus one can get flavor alignment (correlation of the neutrino masses and
masses of charge leptons).

\section{CP phases  in the case of non-hierarchical\\ mass spectrum 
\label{mass1}}

With respect to the hierarchical case, the structure of the mass
matrix depends on two additional parameters:
the mass ration $k$ and the phase $\rho$. These parameters enter the mass
matrix elements
in the combinations (see (\ref{big}))
\beq
X\equiv s_{12}^2 k e^{-2i\rho} + c_{12}^2\;,
\quad Y\equiv s_{12}c_{12}(1 - k e^{-2i\rho})\;,
\quad Z\equiv c_{12}^2 k e^{-2i \rho} + s_{12}^2 
\label{XYZ}
\end{equation}
(in the hierarchical case,  $k\approx 0\;,\;X\approx c_{12}^2\;,\;Y\approx
s_{12}c_{12}$ and $Z\approx s_{12}^2$).
We will  use the following parameterization:
\beq
X \equiv x e^{i\phi_X}, ~~  Y \equiv y e^{i\phi_Y}, ~~
Z \equiv  z e^{i\phi_Z},
\label{xyz}
\end{equation}
where $x \equiv |X|$,  $\phi_X$, $y \equiv |Y|$,  $\phi_Y$,
$z \equiv |Z|$, $\phi_Z$ are
functions of $\theta_{12}$, $k$ and $\rho$.

In the limit of very small $s_{13}$ 
using  (\ref{big}) and the notation (\ref{xyz}), we have:
\beq
\tilde{m}=\left(
\begin{array}{ccc}
rz & c_{23}r y & s_{23}r y\\
\dots & |c_{23}^2 r x + s_{23}^2 e^{-2i\sigma_X}| &
s_{23} c_{23} |- r x + e^{-2i\sigma_X}|\\
\dots & \dots & |s_{23}^2 r x + c_{23}^2 e^{-2i\sigma_X}|
\end{array}
\right) \;,
\label{0s13}
\end{equation}
where $\sigma_X \equiv \sigma+\phi_X/2$. One can first analyze 
matrix (\ref{0s13}) and then consider corrections of the order  
$s_{13}$.  
Notice that now the elements of the $e$-row have non-zero  phases which
depend on $\rho$. The dependences of the  absolute values and  phases 
of these elements are
correlated. The phases of $\mu\tau$-block elements depend 
mainly on $\sigma$, with corrections which are functions of $\rho$.

\subsection{Non-degeneracy case
\label{nd}}

For $m_1 \lesssim \sqrt{\Delta m^2_{sol}}$, we have $k \lesssim 1$, 
and $r \ll 1$. 
The largest mass is  given by $m_3 \approx \sqrt{\Delta m^2_{atm}}$.
The contributions  of $m_1$ to the $\mu\tau$-block elements
appear  as small corrections,
but they can be of order $1$ for the $e$-row elements.

Neglecting terms of  order $s_{13}$, we can use for
the $\mu\tau$-block elements the expressions from (\ref{0s13}).
Comparing with the hierarchical case (see Eq.(\ref{hmatrix})), 
we find that the
effect of $m_1$ is
reduced to renormalization of the mass ratio $r$ and shift 
of the phase $\sigma$:
\beq
r \rightarrow r_X \equiv r \frac{x}{c_{12}^2}, ~~~~~
\sigma \rightarrow \sigma_X \equiv \sigma + \frac{1}{2} \phi_X ~.
\label{rx}
\end{equation}
That is,   dependence of the elements on phases can be found from
Figs.\ref{sigma},\ref{sigmam2}, by appropriate change of $r$ and $\sigma$.

Depending on the phase $\rho$,  the contribution related to $m_1$ 
can suppress or
enhance the amplitude of variations of $\mu\tau$-block elements  with
$\sigma$ (see (\ref{blocks})). The extreme modifications are determined by
\beq
r_X= \left\{
\begin{array}{ll}
r (1+k \tan^2{\theta_{12}})\;, & {\rm for}\quad \rho = 0 \\
r (1-k \tan^2{\theta_{12}})\;, & {\rm for}\quad \rho = \pi/2
\end{array}
\right. \;.
\nonumber
\end{equation}
For $k \lesssim 1$ and $\tan^2\theta_{12} \lesssim 0.5$, the relative
effect of $m_1$ is below $50\%$. For $\rho=0,\pi/2$, we have 
$\phi_X=0$ and no phase shift occurs.
In general,  the phase $\phi_X$ is in the interval
$(- \phi_X^{max}  \div \phi_X^{max})$, where $\sin \phi_X^{max}
= k \tan^2\theta_{12}$.
This maximal phase corresponds to
$r_X =  r \sqrt{1 - k^2 \tan^4 \theta_{12}}$.

For the elements of $e$-row, the $s_{13}$ corrections  should be taken
into account (see (\ref{big})):
\beq
\begin{array}{l}
\tilde{m}_{e\mu} \approx
\left| c_{23} r y
 + s_{13}s_{23}e^{i(\delta- 2 \sigma_Y)}\right| ,\\
\tilde{m}_{e\tau} \approx
\left| s_{23} r y - s_{13} c_{23}e^{i(\delta-2\sigma_Y)}
\right| \;.
\end{array}
\label{mernd}
\end{equation}
Again, the effect of $m_1$
is reduced to  renormalization  of $r$ and a  shift of phase
(compare with Eqs.(\ref{memd},\ref{metd})):
\beq
r \rightarrow r_Y \equiv r \frac{y}{s_{12}c_{12}}, ~~~~
\sigma \rightarrow \sigma_Y \equiv \sigma + \frac{1}{2}\phi_Y\;.
\end{equation}
Minimal and maximal  values of $r_Y$ are given by:
\beq
r_Y=\left\{
\begin{array}{ll}
r (1-k) \;, & {\rm for}\quad \rho = 0 \\
r (1+k) \;, & {\rm for}\quad \rho = \pi/2
\end{array}
\right.\;.
\nonumber
\end{equation}
In these extreme cases there is no phase shift.
In general, for arbitrary values of $\rho$,
$\phi_Y$ is in the interval $(-\phi_Y^{max} \div \phi_Y^{max})$, where
$\sin \phi_Y^{max} = k$ and
this maximal value corresponds to $r_Y = r\sqrt{1 - k^2}$.

Notice that, for $m_{e\mu}$ and $m_{e\tau}$,  modifications of $r$ can be
larger than for the elements of $\mu\tau$-block; moreover,  $r_Y$ and
$r_X$ are changing with $\rho$ in opposite phases. 
The phases of variations of $\mu\tau$-block elements
are correlated as in the hierarchical case. No
phase shift among these elements
is induced by $m_1$ contribution: in (\ref{shiftmt}) one should
substitute $\sigma \rightarrow \sigma_X$.
Similar conclusion is valid for $e$-row elements:
in (\ref{shifter}) $\sigma$ should be substituted by $\sigma_Y$.

For the ${ee}$-element,   similarly to the previous cases,
we get (including $s_{13}$ corrections):
\beq
\tilde{m}_{ee} = \left|c_{13}^2 s_{12}^2 r_Z + s_{13}^2 e^{2i(\delta -
\sigma_Z)}\right|,
\end{equation}
where
\beq
r_Z \equiv r \frac{ z}{s_{12}^2},  ~~~~\sigma_Z \equiv \sigma +
\frac{1}{2}\phi_Z.
\end{equation}
Now the difference between $r$ and $r_Z$ can be substantially larger:
\beq
r_Z=\left\{
\begin{array}{ll}
r (1+k \cot^2{\theta_{12}}) \;, & {\rm for}\quad \rho = 0 \\
r |1-k \cot^2{\theta_{12}}| \;, & {\rm for}\quad \rho = \pi/2
\end{array}
\right. \; ,
\nonumber
\end{equation}
and $r_Z$ changes with $\rho$ in phase with $r_X$.
Notice that, for $k < \tan^2\theta_{12}$, the shift $\phi_Z$ is restricted
to the interval  $(-\phi_Z^{max} \div \phi_Z^{max})$, where
$\sin \phi_Z^{max} = k \cot^2 \theta_{12}$.
For $k > \tan^2\theta_{12}$, the shift is unrestricted.
The $ee-$element is zero  for
\beq
\tan{\theta_{13}} = \sqrt{r z}=s_{12} \sqrt{r_Z},~~~~~
(\delta - \sigma_Z)  = \frac{\pi}{2},\;\frac{3\pi}{2}\;.
\nonumber
\end{equation}
Since $r_Z$ can be smaller than  $r$, or even zero,
the equality $m_{ee} = 0$ can be realized for 
smaller values of $s_{13}$ than in the hierarchical case. 
Now the strongly hierarchical structure of the  $e$-row,
\beq
m_{ee}\ll m_{e\mu}\ll m_{e\tau}\;,
\nonumber
\end{equation}
can be easily achieved. Maximal value of $m_{ee}$ equals approximately
$\tilde{m}_{ee}^{max} \approx r(s_{12}^2 + k c_{12}^2)$.

The dependences of the mass matrix elements on phases can be
deduced from Fig.\ref{sigma} and Fig.\ref{sigmam2}.
Since, now, the ``effective'' value of $r$ is different for the
$\mu\tau-$block elements ($r_X$) and the $e$-row elements
($r_Y$,  $r_Z$), one should  take, {\it e.g.}, lines which correspond to
the $\mu\tau$-block from  Fig.\ref{sigma} and lines which correspond to
the $e$-row from Fig.\ref{sigmam2} or {\it vice versa}.

Let us analyze the dependence  of the elements on the phase $\rho$.
The relative amplitudes of variations of the $\mu\tau-$block elements
are suppressed by a factor  $ s_{12}^2 r k $ (see (\ref{mutau-m})):
\beq
\frac{\Delta m^{\rho}}{m} \approx
s_{12}^2 r k \times
\left\{
\begin{array}{ll}
\cot^2{\theta_{23}}, &  ~~~m_{\mu\mu}\\
\tan^2{\theta_{23}}, &  ~~~m_{\tau\tau}\\
1,                   &  ~~~m_{\mu\tau}
\end{array}
\right. .
\label{blocks-rho}
\end{equation}
The influence of  $\rho$ on the $e-$row elements is much stronger.
If $s_{13}\approx 0$, we have
$$
\tilde{m}_{e\mu} \approx c_{23} s_{12} c_{12} r |1 - k e^{-2i\rho}|
$$
(for $\tilde{m}_{e\tau}$ one should substitute $c_{23} \rightarrow s_{23}$)
and the relative  amplitude of variations is given  by $k$.
 
The amplitude of $ee$-element  can be maximal if
$k \ge \tan^2\theta_{12}$.

\subsection{Partial degeneracy \label{secpd}}

For  $\sqrt{\Delta m^2_{sol}} \ll m_1 \lesssim \sqrt{\Delta m^2_{atm}}$,
we get the spectrum  with partial degeneracy $m_1 \approx m_2 \lesssim
m_3$. The ratios of masses are
\beq
r \approx \frac{m_1}{\sqrt{m_1^2 + \Delta m^2_{atm}}}\lesssim 1 \;,\qquad 
\qquad k  \approx 1-\frac{\Delta
m^2_{sol}}{2m_1^2}\;.
\label{soldev}
\end{equation}
For $m_1>2 \cdot 10^{-2} {\rm ~eV}$,  the deviation of $k$ from $1$ is
smaller than $5\%$ and we can   neglect it  in comparison with
other corrections (related to possible large  deviations from maximal 2-3
mixing and to $s_{13}\gtrsim 0.1$). Now the scale of masses is determined by
$m_3 \approx \sqrt{m_1^2 + \Delta m^2_{atm}} \sim
(1 \div 2) \sqrt{\Delta m^2_{atm}}$.
The sum (\ref{sum}) of all the matrix elements squared equals 
$$
S_0\approx m_3^2 (1+2r^2)\;.
$$

\begin{figure}
[t]
\begin{center}
\epsfig{file=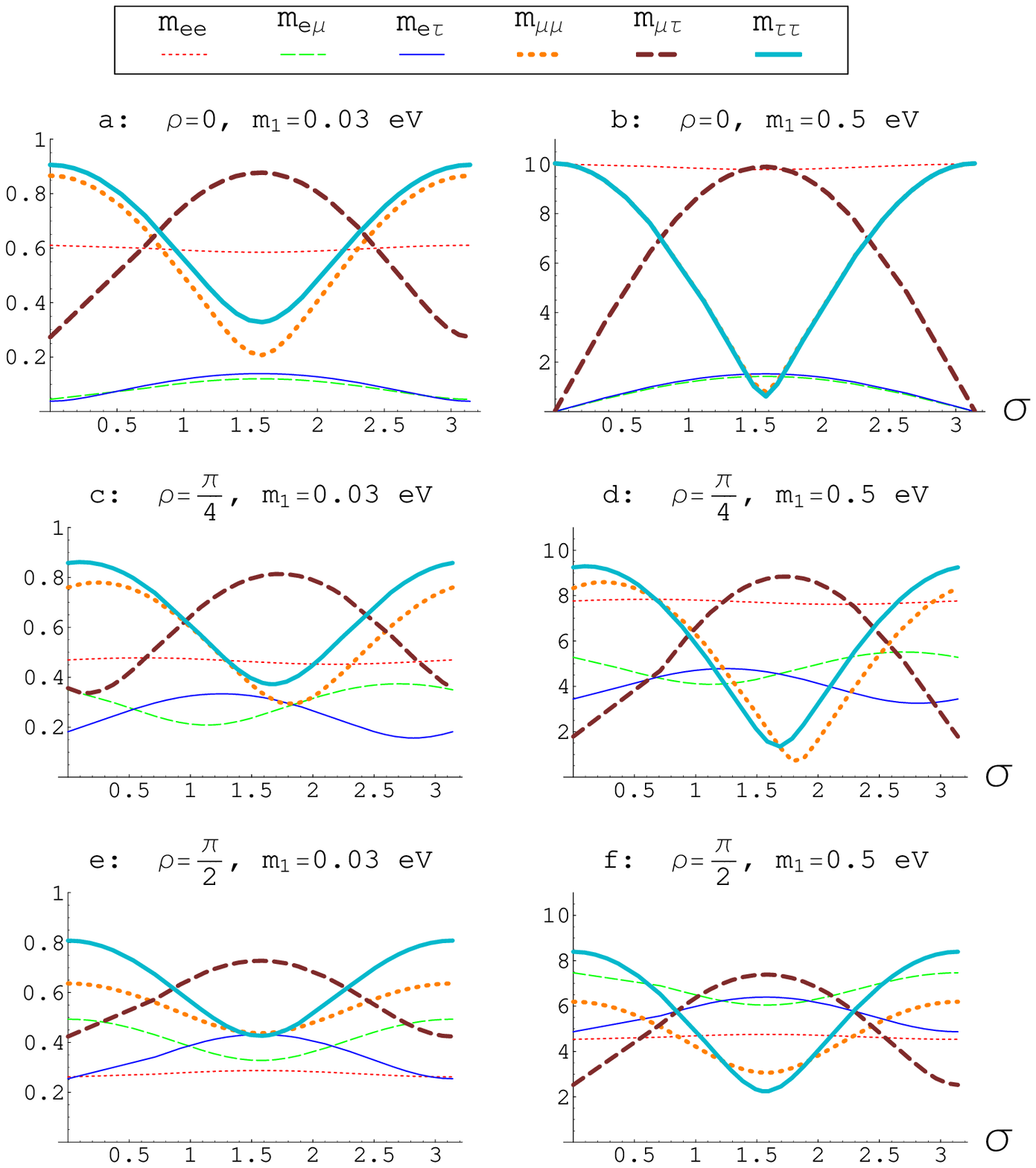,width=450pt,bbllx=90, bblly=275, bburx=525, bbury=760}
\end{center}
\caption{Dependence of the absolute value of mass  matrix elements (in
units $\sqrt{\Delta m_{atm}^2}$) on $\sigma$, for  partially degenerate
spectrum (panels a,c,e) and completely degenerate  spectrum (panels
b,d,f). We show dependences for different values  of the phase $\rho$. We
take $\Delta m^2_{sol}=5 \cdot 10^{-5} {\rm eV}^2$,  $\Delta m^2_{atm}=
2.5 \cdot 10^{-3} {\rm eV}^2$ and 
$\tan^2\theta_{12}=0.36$, $\tan\theta_{23}=0.93$, $s_{13}=0.1$, $\delta=0$.}
\label{m1sigma}
\end{figure}

Let us consider first the dependence of the masses on phase $\sigma$ 
(see Fig.\ref{m1sigma} panels a,c,e and phase diagrams in
Fig.\ref{phase}). In the limit of small $s_{13}$, we get
\beq
\begin{array}{l}
\tilde{m}_{\mu\mu} \approx \left| s_{23}^2 e^{-2i\sigma}
+ r c_{23}^2 X_1(\rho)\right|    \;,\\
\tilde{m}_{\tau\tau}
\approx \left| c_{23}^2 e^{-2i\sigma}
+ r s_{23}^2 X_1(\rho) \right| \;,\\
\tilde{m}_{\mu\tau} \approx s_{23}c_{23}
\left|e^{-2i\sigma} -  r X_1(\rho) \right|\; ,
\end{array}
\label{mutau-mpd}
\end{equation}
where
\beq
X_1(\rho) \equiv X(k = 1) =  c_{12}^2  + s_{12}^2 e^{-2i\rho}.
\label{x1}
\end{equation}
The mass $\tilde{m}_{\mu\mu}$ oscillates with $\sigma$ around $s_{23}^2$;
the amplitude of variations depends on the phase $\rho$.
Maximal amplitude is for $\rho = 0$, which corresponds to
$X_1 = 1$. 
The mass $\tilde{m}_{\tau\tau}$ oscillates in phase with
$\tilde{m}_{\mu\mu}$ around the average value $c_{23}^2$;  
the mass $\tilde{m}_{\mu\tau}$ varies in opposite phase. 
The amplitudes of variations of all $\mu\tau$-block elements decrease 
with increase of the phase $\rho$ and it is minimal for $\rho = \pi/2$.

The corrections of order $s_{13}$ change the amplitudes of variations
and produce a phase shift. 
The largest influence of these corrections is for $\rho = \pi/2$.
From (\ref{mutau-m}) we find
\beq
\begin{array}{l}
\tilde{m}_{\mu\mu} \approx
\left|
s_{23}^2 e^{-2i\sigma}
+ r c_{23}^2 (\cos2\theta_{12} -  2\epsilon_{\mu\mu})
\right|  \;,\\
\tilde{m}_{\tau\tau}
\approx \left|
c_{23}^2 e^{-2i\sigma} +
r s_{23}^2 (\cos2\theta_{12}  + 2 \epsilon_{\tau\tau})
\right| \;,\\
\tilde{m}_{\mu\tau} \approx
s_{23}c_{23}
\left| e^{-2i\sigma} -
r(\cos2\theta_{12}  + 2\epsilon_{\mu\tau})
\right|\;.
\end{array}
\label{mutau-part}
\end{equation}
For  $\delta = 0$ ($\epsilon > 0$),
the corrections suppress the amplitude of variations of
$\tilde{m}_{\mu\mu}$ and enhance the amplitude of
$\tilde{m}_{\tau\tau}$ and $\tilde{m}_{\mu\tau}$ variations
(see Fig.\ref{m1sigma}e).  For $\delta = \pi/2$ ($\epsilon < 0$) the
situation is opposite:   variations of $\tilde{m}_{\mu\mu}$  are enhanced.

In the approximation (\ref{mutau-mpd}), all the elements of 
$\mu\tau$-block depend on the phase $\rho$ in the same
way. So, there is no relative shift 
and the relative phases are determined as in (\ref{shiftmt}).
The phase shift seen in Fig.\ref{m1sigma}c is due to  the interplay of
$\epsilon$ corrections and phase $\rho$.

The dependence of elements of the $e$-row  on  $\sigma$
 (as well as $\delta$) appears due to terms of the order $s_{13}$
(see (\ref{em-m}) and Fig.\ref{m1sigma} panels a,c,e).
Neglecting corrections $\sim r s_{13}$, we
get (for $\rho$ being not to close to 0):
\beq
\tilde{m}_{e\mu} \approx \left|
r  c_{23} Y_1(\rho) + s_{13}s_{23} e^{i(\delta - 2\sigma)} \right|, ~~~
\tilde m_{e\tau} \approx \left|
r s_{23}  Y_1(\rho) -  s_{13}c_{23}e^{i(\delta-2\sigma)} \right|,
\label{em-mpd}
\end{equation}
where
\beq
Y_1(\rho) \equiv   Y(k = 1) =  s_{12} c_{12}\left(1 -
e^{-2i\rho}\right)\;.
\label{y1}
\end{equation}
The masses $\tilde{m}_{e\mu}$ and $\tilde{m}_{e\tau}$ vary with
$(\delta-2\sigma)$
in opposite phase (small phase shift may appear due to
interplay of order $r s_{13}$ corrections and phase $\rho$). 
The amplitude
of variations is proportional to  $s_{13}$. The average values of the
elements  increase with  $\rho$,  and they reach  maxima,
$\tilde{m}_{e\mu}^{max} = r \sin 2\theta_{12} c_{23}$ and
$\tilde{m}_{e\tau}^{max} = r \sin 2\theta_{12} s_{23}$,
at $\rho = \pi/2$.

Configuration with  $\rho = 0$ or $\rho \approx 0$  
is the special one (see Fig.\ref{m1sigma}a).
In this case the main terms in (\ref{em-m}) vanish and
dependence on phases appears due to
$\epsilon$ corrections, defined in (\ref{epsilon}):
\beq
\begin{array}{l}
\tilde{m}_{e\mu} \approx \left| \frac 12
r \sin 2\theta_{12} c_{23}
\left(\epsilon_{e\mu} + \epsilon_{e\mu}' \right)
- s_{13}s_{23} e^{i(\delta - 2\sigma)} \right|\;,  \\
\tilde m_{e\tau} \approx \left|
\frac 12 r \sin 2\theta_{12} s_{23}
\left(\epsilon_{e\tau} + \epsilon_{e\tau}'\right)
- s_{13}c_{23}e^{i(\delta-2\sigma)} \right|\;.
\end{array}
\label{em-mz}
\end{equation}
Notice that elements $\tilde{m}_{e\mu}$ and $\tilde{m}_{e\tau}$
vary in phase; both the average value and the
amplitude are proportional to $s_{13}$.

Changing  the  phase $\delta$ 
by $\Delta \delta$ one shifts lines  which correspond to 
$\tilde{m}_{e\mu}$ and  $\tilde{m}_{e\tau}$,
with respect to the lines of $\mu \tau$-block elements by
$\Delta \sigma =\Delta \delta/2$.  For instance, according to
Fig.\ref{m1sigma}c, one can get equalities
$\tilde{m}_{\mu\mu} = \tilde{m}_{\tau\tau} =  \tilde{m}_{\mu\tau}$ and
$\tilde{m}_{e\mu} = \tilde{m}_{e\tau}$  simultaneously.

Variations of  the $ee$-element (\ref{ee-m})
with $\sigma$ as well as with $\delta$
are strongly suppressed by the factor $s_{13}^2$, so that 
\beq
\tilde{m}_{ee} \approx  r \left|Z_1(\rho)\right|,
\label{ee-mr}
\end{equation}
where
\beq
Z_1(\rho) \equiv   Z(k = 1) =
s_{12}^2 + c_{12}^2 e^{-2i\rho}.
\label{z1}
\end{equation}
The average value  decreases with increase of the  phase $\rho$.
It varies from  $\tilde{m}_{ee}^{max} \approx r$  for
$\rho \approx 0, \pi$ down to $\tilde{m}_{ee}^{min}
\approx r\cos 2\theta_{12}$ for $\rho \approx \pi/2$.
Variations  of  $\tilde{m}_{ee}$ with $\rho$
are in opposite phase  with respect to
$\tilde{m}_{e\mu}$ and $\tilde{m}_{e\tau}$.

Let us analyze the dependence of masses on the phase
$\rho$ (Fig.\ref{m1rho} panels a,c,e).
The amplitudes of variations of the $\mu\tau$-block elements
with $\rho$, $\Delta m^{\rho} \propto r s_{12}^2$, are smaller
(for non-maximal solar mixing) than  the
amplitudes of $\sigma$ variations. The average values of
$\tilde{m}_{\mu\mu}$ and   $\tilde{m}_{\tau\tau}$ decrease
whereas  the average of $\tilde{m}_{\mu\tau}$ increases
with increase of $\sigma$  from 0 to $\pi/2$.
Strong split of masses in the $\mu\tau$-block
(see Fig.\ref{m1rho}a,\ref{m1rho}e) is
due to cancellation of contributions related to
$m_3$ (first term in (\ref{mutau-mpd})) and to
$m_1$ and $m_2$ (second term). For large $s_{13}$, the terms of order
$r s_{13}$ can enhance variations with  $\rho$.

According to (\ref{em-mpd}), variations of the $e$-row elements 
with $\rho$  are strong: the amplitude can be close to
maximal one. For large  values of $s_{13}$, 
the phase $(2\sigma - \delta)$ changes significantly the
average values of the elements 
$m_{e\mu}$ and  $m_{e\tau}$ and also modifies the amplitudes
of variations with $\rho$.\\ 

The matrix elements are all correlated.  This can be seen in the limit of
very small $s_{13}$. For partially degenerate spectrum ($k= 1$), we get: 
\beq
\begin{array}{c}
x = z =\sqrt{1-\sin^2{2\theta_{12}}\sin^2{\rho}}\;,\\
y = \sin{2\theta_{12}} \sin{\rho}\;,
\end{array}
\label{k1xyz}
\end{equation}
so  that $y = \sqrt{1- x^2}$. Moreover,
\beq
\tan{\phi_X}=-\frac{\sin{2\rho}}{\cot^2{\theta_{12}} +\cos{2\rho}}\;.
\end{equation}
The mass matrix can be written as 
\beq
\tilde{m}=\left(
\begin{array}{ccc}
rx & c_{23}r \sqrt{1-x^2} & s_{23}r \sqrt{1-x^2}\\
\dots & |c_{23}^2 r x + s_{23}^2 e^{-2i\sigma_X}| &
s_{23} c_{23} |- r x + e^{-2i\sigma_X}|\\
\dots & \dots & |s_{23}^2 r x + c_{23}^2 e^{-2i\sigma_X}|
\end{array}
\right) \;.
\label{pdmatrix}
\end{equation}
From (\ref{pdmatrix}), we  find  the following relations among
the elements:
\beq
\tilde{m}_{ee}^2+\tilde{m}_{e\mu}^2+\tilde{m}_{e\tau}^2=r^2\;;
\end{equation}
\beq
\frac{m_{e\mu}}{m_{e\tau}}=\tan{\theta_{23}}\;;
\label{memta}
\end{equation}
\beq
\tilde{m}_{\tau\tau}^2-\tilde{m}_{\mu\mu}^2=(1-r^2 x^2)\cos{2\theta_{23}}\;.
\label{mumta}
\end{equation}
Notice that $m_{\tau\tau}=m_{\mu\mu}$ either 
for $\theta_{23}=45^{\circ}$ or for $rx=1$. 
The latter corresponds to completely degenerate spectrum 
and $\rho=0$. In this case
$$
\tilde{m}_{\tau\tau}=\tilde{m}_{\mu\mu} 
= \sqrt{1-\sin^2{2\theta_{23}}\sin^2{\sigma}}\;,
\qquad \tilde{m}_{\mu\tau}=\sin{2\theta_{23}}\sin{\sigma}\;.
$$
Furthermore, we find, for the sum of the $\mu\tau$-block elements,
$$
\tilde{m}_{\mu\mu}^2+\tilde{m}_{\tau\tau}^2+2\tilde{m}_{\mu\tau}^2
= 1+r^2x^2, 
$$ 
and consequently:
$$
\tilde{m}_{\mu\mu}^2+\tilde{m}_{\tau\tau}^2+2\tilde{m}_{\mu\tau}^2
-\tilde{m}_{ee}^2=1\;.
$$

For a given $r$ the mass matrix (\ref{pdmatrix}) is determined by 
 $\sigma_X$ and $x$. In general, $\sigma_X$ and $x$ 
can be treated as two independent parameters. Depending on $\rho$, 
$x$ changes from the minimal value $x^{min}\equiv\cos{2\theta_{12}}$, 
for $\rho=\pi/2$, to $x^{\max}\equiv 1$, for $\rho=0$. 
The phase $\phi_X$ varies in a rather narrow interval, 
which decreases with $\theta_{12}$:
$$
\sin\phi_X \sim (-\tan^2\theta_{12} \div \tan^2\theta_{12})\;.
$$

\subsection{Degenerate spectrum
\label{degs}}

For    $m_1 \gg \sqrt{\Delta m^2_{atm}}$,
we have  $m_1\approx m_2\approx
m_3$,  and  the ratio of masses is given by
\beq
r \approx 1-\frac{\Delta m^2_{atm}}{2m_1^2}\;.
\nonumber
\end{equation}
For $m_1=0.5$ eV, the deviation  of $r$ from $1$ is smaller than $1\%$ and
we can neglect it in comparison  with other small parameters, $s_{13}$ and
$\xi$.
The $e$-row elements and the  $\mu\tau$-block elements are given by
(\ref{mutau-m},\ref{em-m},\ref{ee-m}),  with $k=r=1$. Notice that, in the
approximation $\Delta m^2_{atm} \approx 0$,  the structure of the mass
matrix for normal and inverted hierarchy is the same.

\begin{figure}
[t]
\begin{center}
\epsfig{file=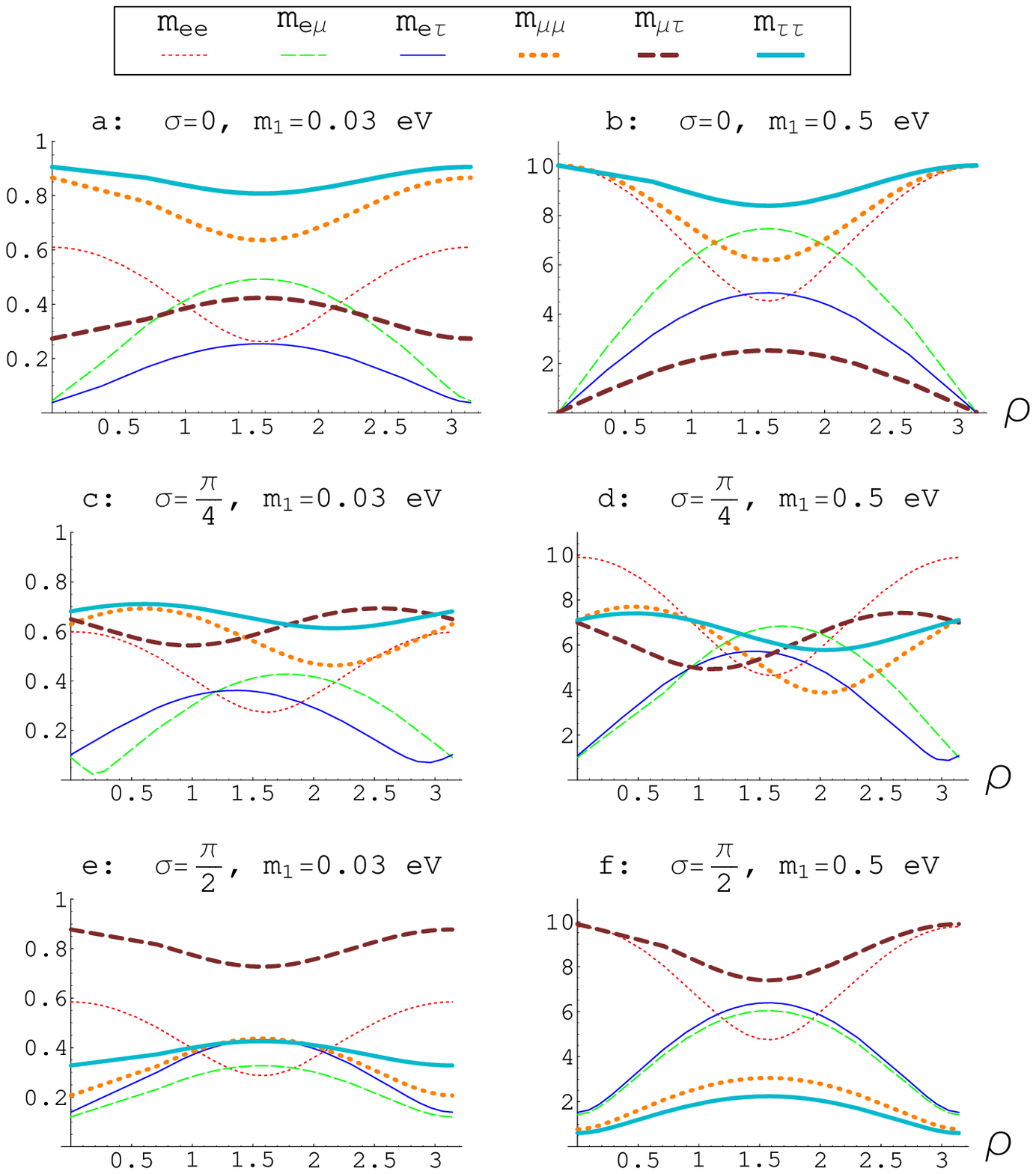,width=450pt,bbllx=95, bblly=290, bburx=505, bbury=760}
\end{center}
\caption{Dependence of  the absolute value of mass matrix elements (in
units $\sqrt{\Delta m_{atm}^2}$)  on $\rho$, for partially degenerate
spectrum (panels a,c,e) and  completely degenerate spectrum (panels
b,d,f). We show dependences  for different values of the phase
$\sigma$. We take $\Delta m^2_{sol}=5 \cdot 10^{-5} {\rm eV}^2$,  $\Delta
m^2_{atm}= 2.5 \cdot 10^{-3} {\rm eV}^2$ and 
$\tan^2\theta_{12}=0.36$, $\tan\theta_{23}=0.93$, $s_{13}=0.1$, $\delta=0$.}
\label{m1rho}
\end{figure}

Transition to the degeneracy case does not produce qualitative changes in
dependences of the matrix elements on phases in comparison with
partial degeneracy case (see Fig.\ref{m1sigma} b,d,f and
Fig.\ref{m1rho} b,d,f).
Amplitudes of variations of the $\mu\tau$-block elements increase and
 can reach maximal size for specific values of phases.
This leads to zero (small) values of certain matrix  elements
and therefore to the appearance of a hierarchical
structure of the mass matrix.
For example, in the case of maximal 2-3 mixing and $\rho=0$, 
we find from (\ref{pdmatrix}):
\beq
\tilde{m}\approx\left(
\begin{array}{ccc}
1 & 0 & 0\\
\dots & \frac 12|1 + e^{-2i\sigma}| &
 \frac 12|- 1 + e^{-2i\sigma}|\\
\dots & \dots & \frac 12|1 +  e^{-2i\sigma}|
\end{array}
\right) \;.
\end{equation}
Therefore, $\tilde{m}_{\mu\mu} = \tilde{m}_{\tau\tau} = 0$ for
$\sigma = \pi/2$ (Fig.\ref{m1sigma}b).
Moreover,  $\epsilon$ corrections  cancel in (\ref{mutau-m}), for $k=1$
and $\rho=0$.
For non-maximal  2-3 mixing,
using (\ref{mutau-mpd}) we find that
$\tilde{m}_{\mu\mu} = 0$  for
\beq
\sin^2\rho  =  \frac{\cos 2\theta_{23}}{c_{23}^4 \sin^2
2\theta_{12}},~~~~
\cos 2\sigma = -
\frac{c_{23}^4\cos2\theta_{12} + s^4_{23}}{2c_{12}^2 s_{23}^2 c_{23}^2}.
\nonumber
\end{equation}
In this case, however, $\tilde{m}_{\tau\tau}$  differs from zero.
Such a configuration is realized approximately in Fig.\ref{m1sigma}d.

The average values  of $\tilde{m}_{e\mu}$ and $\tilde{m}_{e\tau}$
 increase with respect to the partial  degeneracy case, whereas  the
amplitudes of variations with
$\sigma$ and $\rho$ do not change.
Average value of the $ee$-element increases with $r$ and
can reach 1 for $\rho = 0$ (Fig.\ref{m1sigma}b).

The amplitudes of variations with $\rho$ (Fig.\ref{m1rho} b,d,f) 
increase and, for $\rho\approx 0, \pi$, 
hierarchical structure of the mass matrix
appears (Fig.\ref{m1rho}b,\ref{m1rho}f). For
some values of phases  all the elements become approximately equal to each
other (see, {\it e.g.}, Fig.\ref{m1rho}d at  $\rho = 1.3 \pi$).

\subsection{From hierarchy to degeneracy}

\begin{figure}
[t]
\begin{center}
\epsfig{file=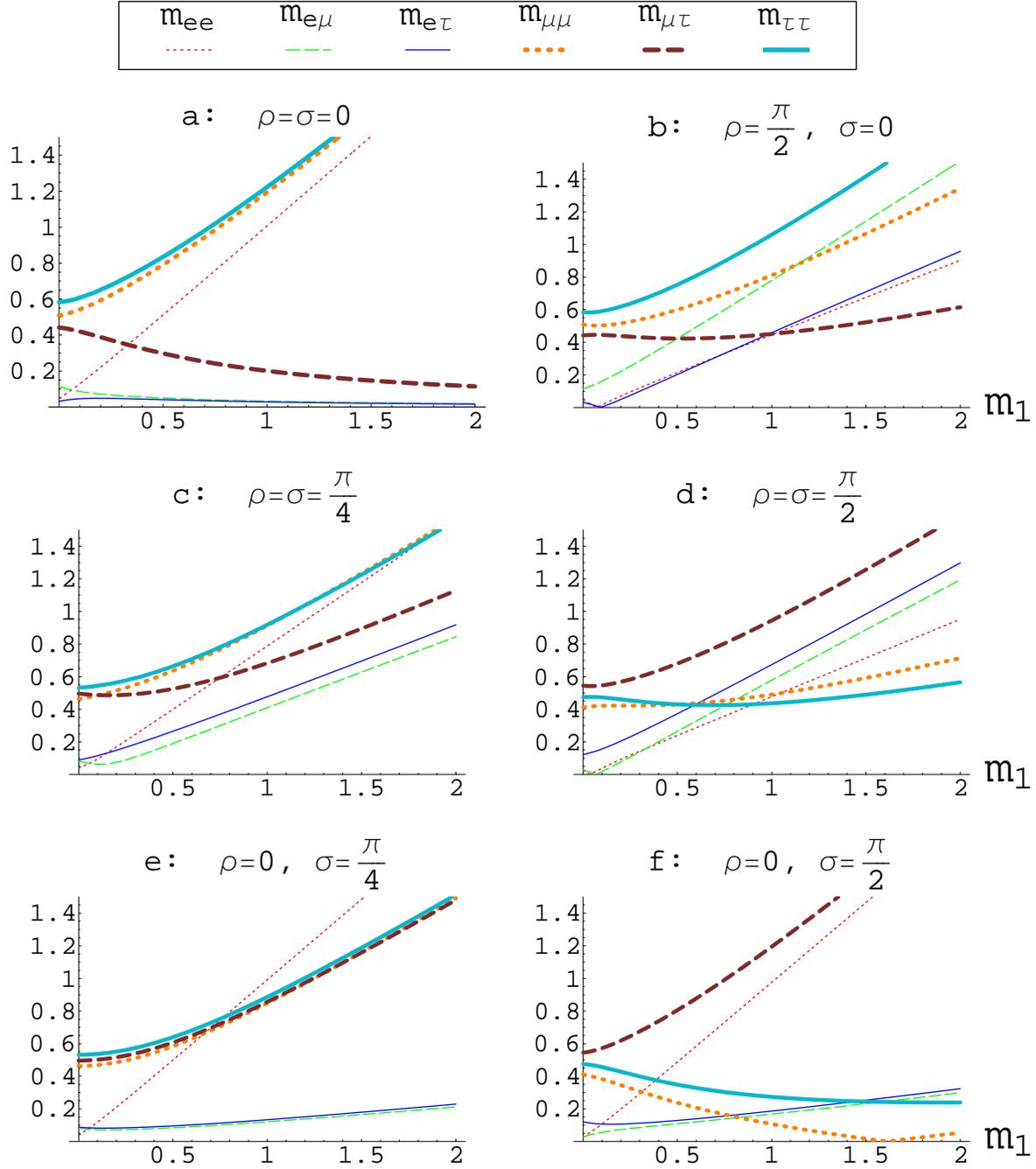,width=450pt,bbllx=85, bblly=310, bburx=480, bbury=760}
\end{center}
\caption{Dependence of the absolute value of  neutrino mass matrix
elements (in units $\sqrt{\Delta m_{atm}^2}$)  on $m_1$. We show
dependences for different values of the phases  $\rho$ and $\sigma$. We
take $\Delta m^2_{sol}=5 \cdot 10^{-5} {\rm eV}^2$,  $\Delta m^2_{atm}=
2.5 \cdot 10^{-3} {\rm eV}^2$ and 
$\tan^2\theta_{12}=0.36$, $\tan\theta_{23}=0.93$, $s_{13}=0.1$, $\delta=0$.}
\label{masszero}
\end{figure}

In Fig.\ref{masszero}, we show the dependence  of $m_{\alpha\beta}$ on
$m_1$  for different values of the Majorana  phases $\sigma$ and
$\rho$.
As follows from the figure, the hierarchical structure
with the dominant $\mu\tau$-block and small  $e$-row elements
exists, independently on phases, for $m_1/\sqrt{\Delta m_{atm}^2} \lesssim
0.1$ ($m_1 \lesssim 0.005$ eV). This
interval of $m_1$ corresponds to hierarchical or non-degenerate
spectra. The structure with  dominant  $\mu\tau-$block
disappears for $m_1/\sqrt{\Delta m_{atm}^2} \sim  0.3 \div 0.5$
($m_1 < (0.02 \div 0.03)$ eV), that is for partially degenerate  spectrum.
For  $m_1 \gtrsim \sqrt{\Delta m_{atm}^2} \approx 0.05$ eV, 
the spectrum converges
to the degenerate one.
In this last case, the structure of the mass matrix
depends substantially on the Majorana phases. Notice that, in general,
the pairs of elements $\tilde{m}_{\mu\mu}$ and $\tilde{m}_{\tau\tau}$ ,
as well as $\tilde{m}_{e\mu}$ and $\tilde{m}_{e\tau}$, have similar
dependences on $m_1$.

For large part of the phase parameter space,  all  elements
of the mass matrix increase with $m_1$ being of the same order.
Some accidental equalities among them may appear.
Particular structures are  realized  for specific
values of phases,
$\rho, \sigma  \approx   0, \pi/4,  \pi/2$, shown in Fig.\ref{masszero}.

\section{$\rho-\sigma$ plots and neutrino-less $2\beta$ decay \label{rosi}}

\subsection{$\rho-\sigma$ plots}

\begin{figure}
[t]
\begin{center}
\epsfig{file=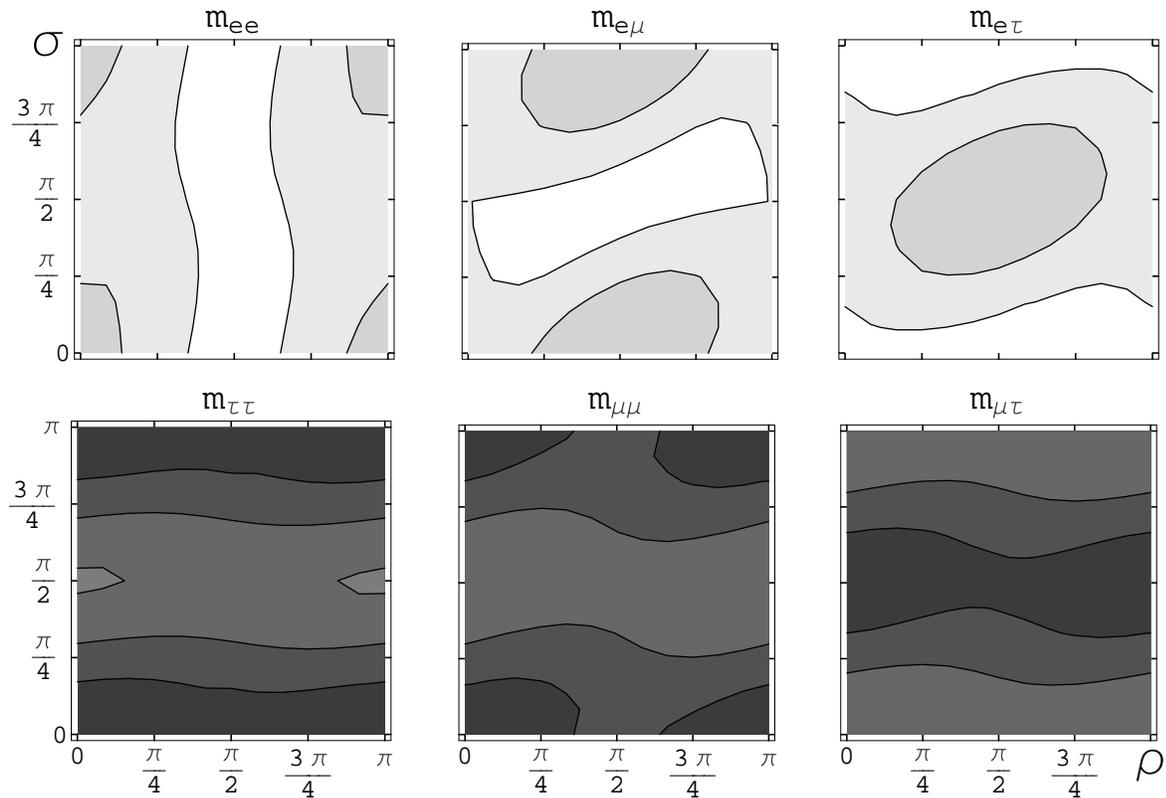,width=450pt,bbllx=70, bblly=455, bburx=520, bbury=765}
\end{center}
\caption{The $\rho-\sigma$ plots for  non-degenerate spectrum, with
$m_1=0.005$ eV. Shown are contours of  constant mass
(iso-mass) $m=(0.1,\,0.2,\,\dots,\,0.9)m^{max}$,  where $m^{max}=0.03$ eV
is the maximal value that the matrix elements can  have, so that the white
regions correspond to the mass interval ($0-0.003$)  eV and the darkest
ones to ($0.027-0.030$) eV. We take  $\Delta m^2_{sol}=5 \cdot 10^{-5}
{\rm eV}^2$, $\Delta m^2_{atm}= 2.5 \cdot 10^{-3} {\rm eV}^2$ and 
$\tan^2\theta_{12}=0.36$, $\tan\theta_{23}=1$, $s_{13}=0.1$, $\delta=0$.}
\label{rs1}
\end{figure}

\begin{figure}
[t]
\begin{center}
\epsfig{file=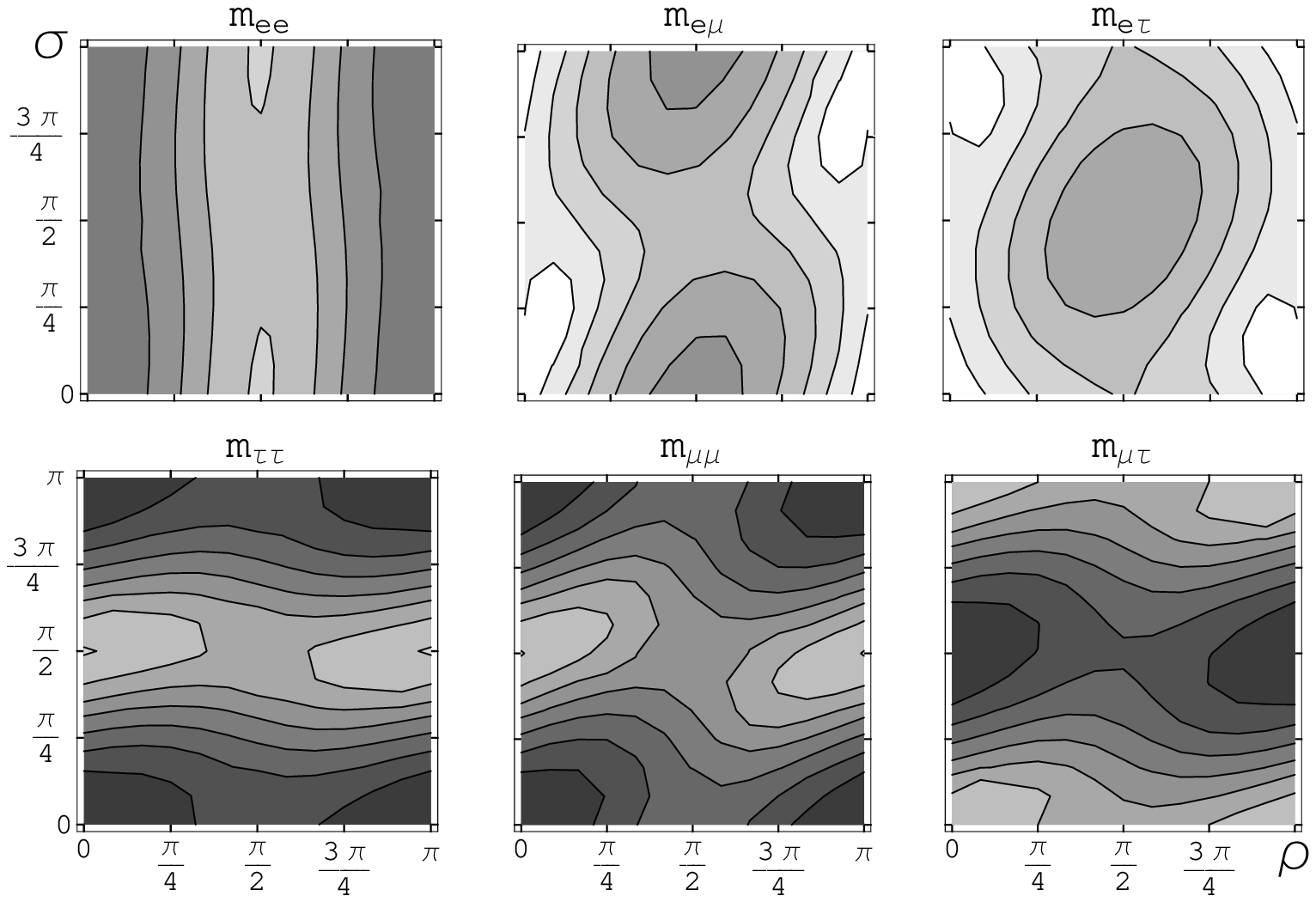,width=450pt,bbllx=70, bblly=455, bburx=520, bbury=765}
\end{center}
\caption{The $\rho-\sigma$ plots  for partially degenerate spectrum, with
$m_1=0.03$ eV. Shown are contours  of constant mass
(iso-mass) $m=(0.1,\,0.2,\,\dots,\,0.9)m^{max}$,  where $m^{max}=0.045$
eV, so that the white regions correspond to the  mass interval
($0-0.0045$) eV and the darkest ones to ($0.0405-0.045$) eV.  We take
$\Delta m^2_{sol}=5 \cdot 10^{-5} {\rm eV}^2$,  $\Delta m^2_{atm}= 2.5
\cdot 10^{-3} {\rm eV}^2$ and 
$\tan^2\theta_{12}=0.36$, $\tan\theta_{23}=1$, $s_{13}=0.1$, $\delta=0$.}
\label{rs2}
\end{figure}

\begin{figure}
[t]
\begin{center}
\epsfig{file=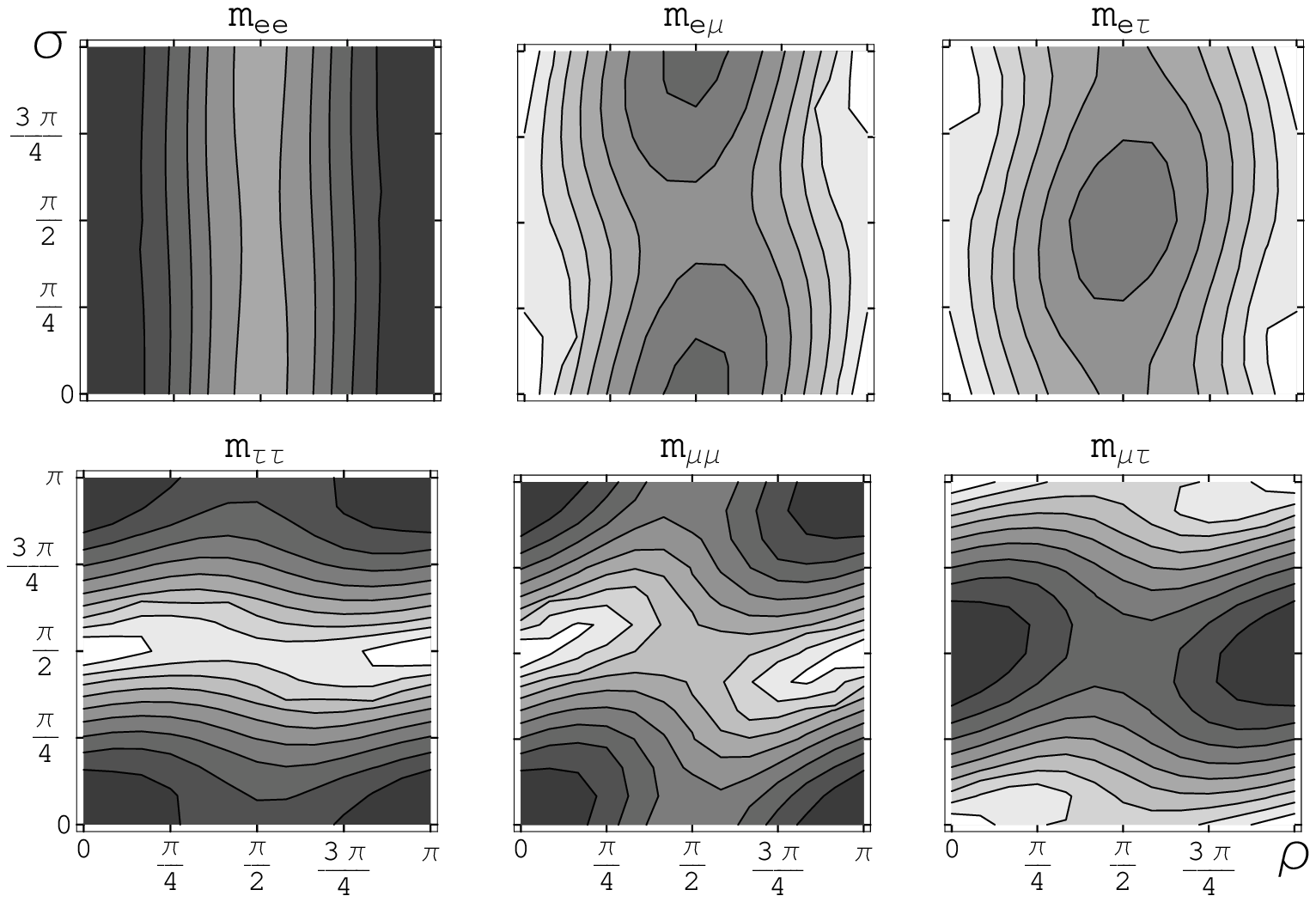,width=450pt,bbllx=70, bblly=455, bburx=520, bbury=765}
\end{center}
\caption{The $\rho-\sigma$ plots for completely  degenerate spectrum, with
$m_1=0.5$ eV. Shown are contours of constant mass  
(iso-mass) $m=(0.1,\,0.2,\,\dots,\,0.9)m^{max}$,  where $m^{max}=0.5$ eV,
so that the white regions correspond to the mass  interval ($0-0.05$) eV
and the darkest ones to ($0.45-0.5$) eV. We take  $\Delta m^2_{sol}=5
\cdot 10^{-5} {\rm eV}^2$,  $\Delta m^2_{atm}= 2.5 \cdot 10^{-3} {\rm
eV}^2$ and  $\tan^2\theta_{12}=0.36$, $\tan\theta_{23}=1$, $s_{13}=0.1$,
$\delta=0$.}
\label{rs3}
\end{figure}

\begin{figure}
[!t]
\begin{center}
\epsfig{file=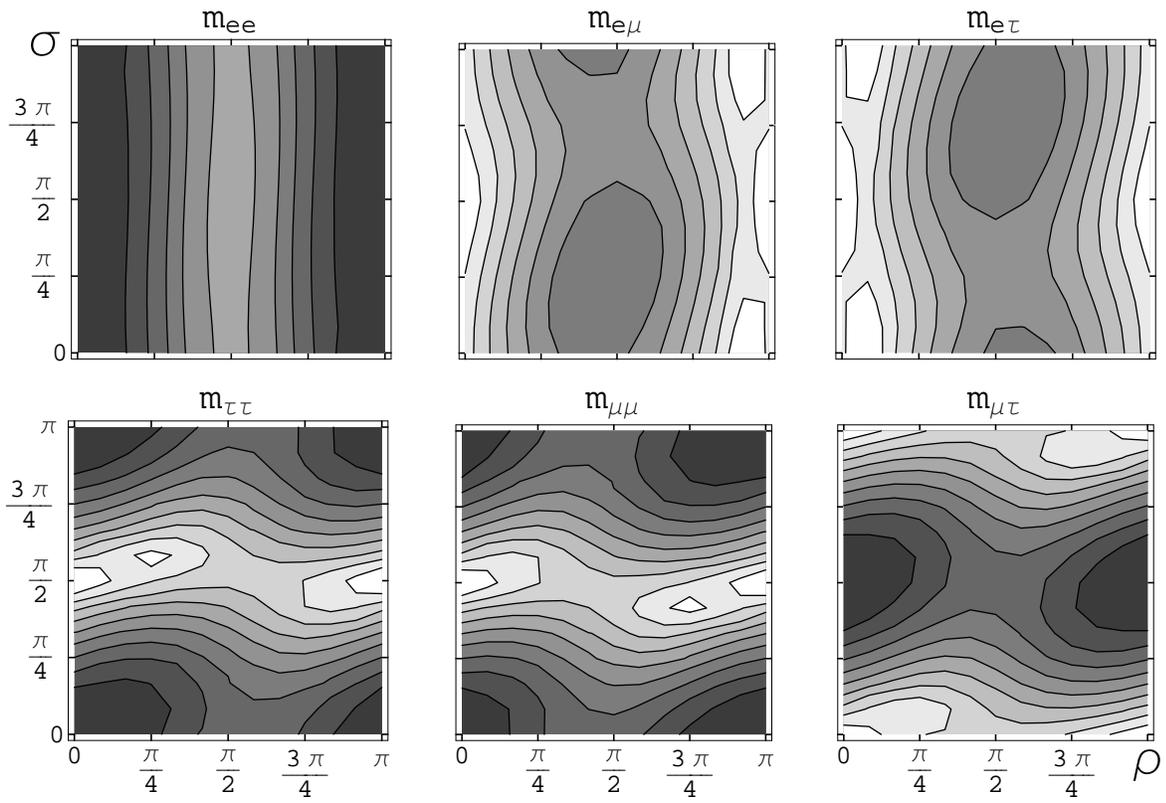,width=450pt,bbllx=70, bblly=455, bburx=520, bbury=765}
\end{center}
\caption{The same as in Fig.\ref{rs3}, but  for non-zero Dirac
phase: $\delta=\pi/2$.}
\label{rs5}
\end{figure}

\begin{figure}
[!b]
\begin{center}
\epsfig{file=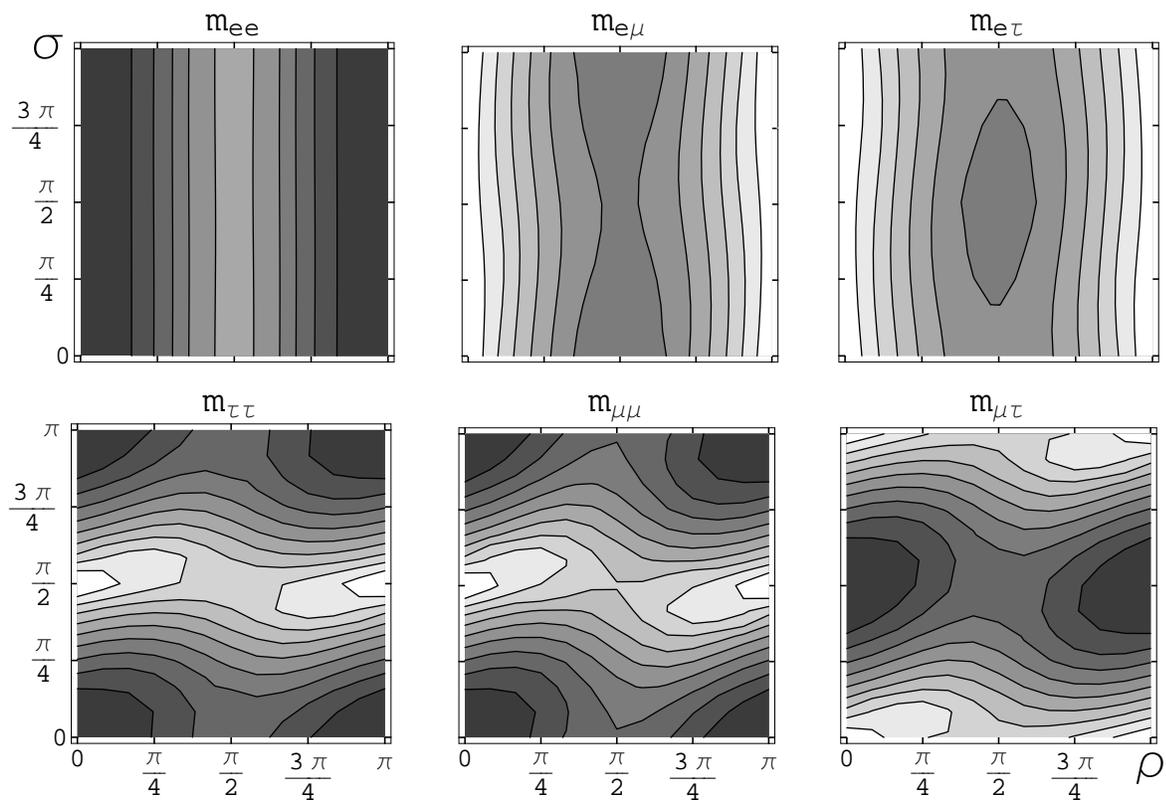,width=450pt,bbllx=70, bblly=455, bburx=520, bbury=765}
\end{center}
\caption{The same as in Fig.\ref{rs3}, but  for very small 1-3
mixing: $\theta_{13}=2^{\circ}$.}
\label{rs6}
\end{figure}

\begin{figure}
[!t]
\begin{center}
\epsfig{file=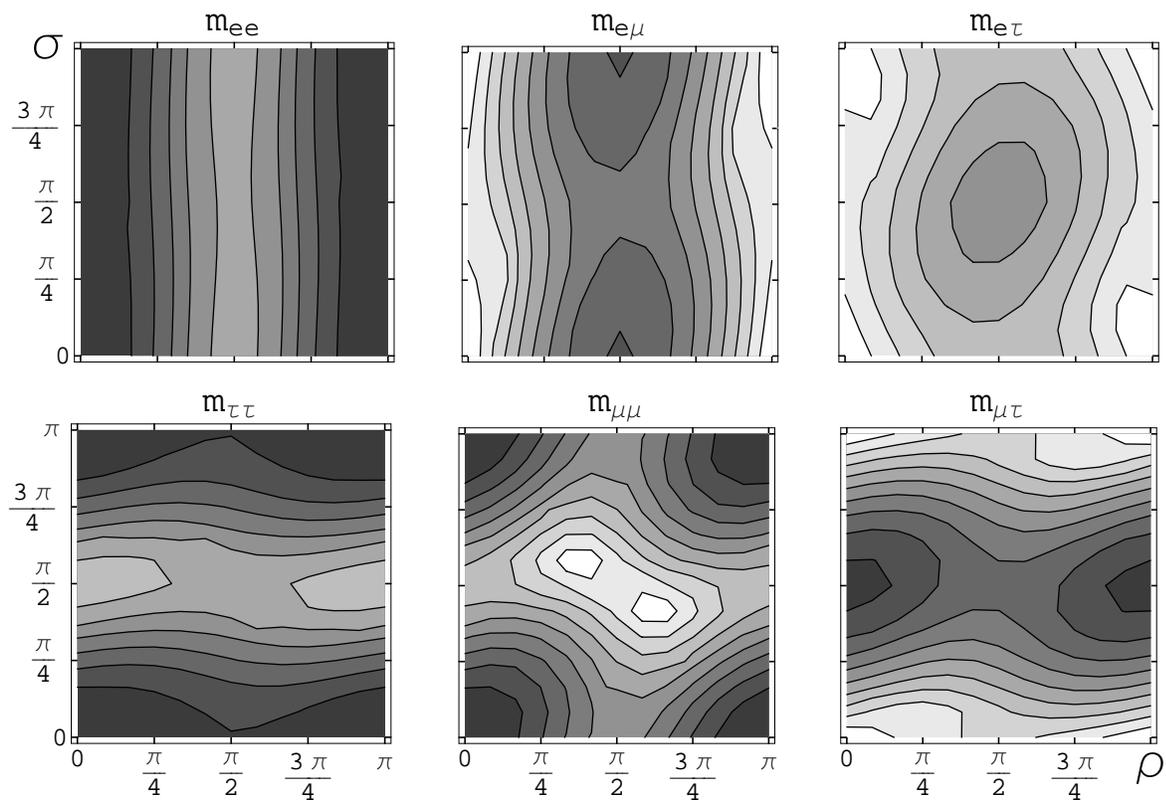,width=450pt,bbllx=70, bblly=455, bburx=520, bbury=765}
\end{center}
\caption{The same as in Fig.\ref{rs3}, but  for non-maximal 2-3
mixing: $\theta_{23}=35^{\circ}$.}
\label{rs4}
\end{figure}

\begin{figure}
[!b]
\begin{center}
\epsfig{file=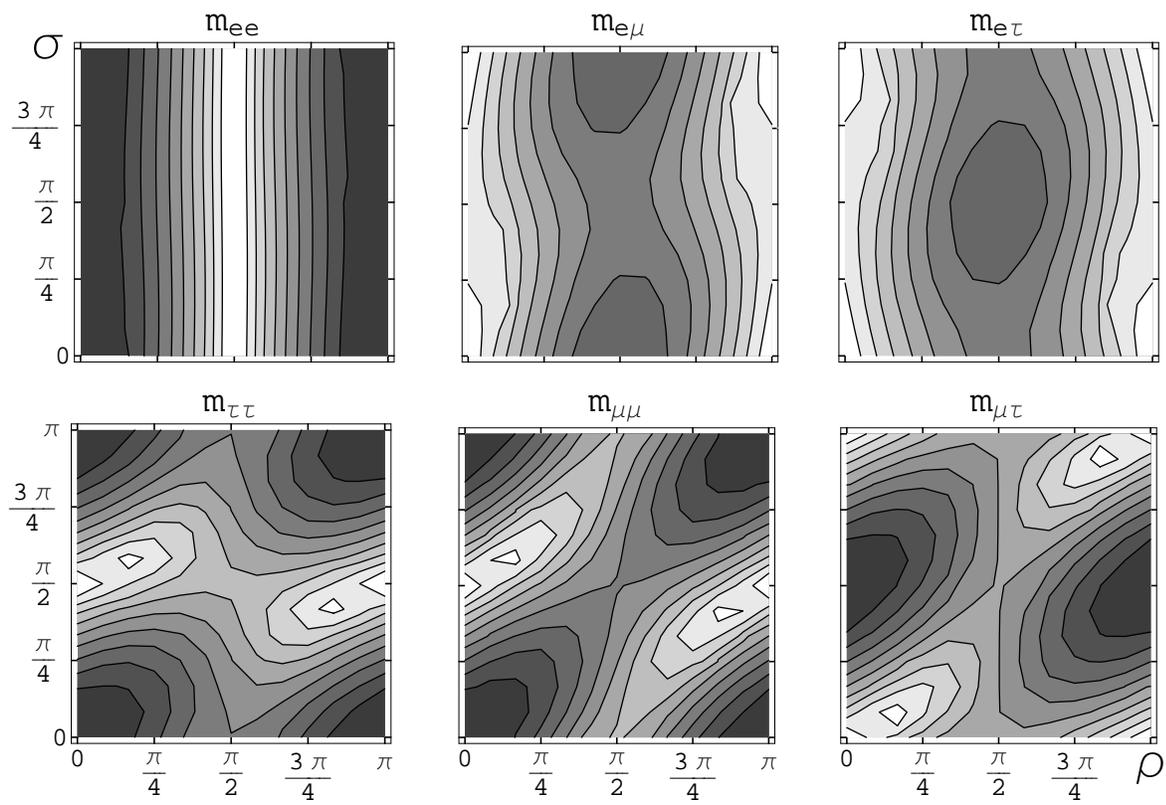,width=450pt,bbllx=70, bblly=455, bburx=520, bbury=765}
\end{center}
\caption{The same as in Fig.\ref{rs3}, but  for maximal 1-2
mixing: $\theta_{12}=45^{\circ}$.}
\label{rs7}
\end{figure}

In spite of large freedom related to the unknown  CP-phases,
$\sigma$, $\rho$, $\delta$,  scale $m_1$ and
$s_{13}$,  already
the present data give important restrictions
on  structure of the mass matrix.
The dependences of various matrix elements on phases are correlated.
These features can be seen in the $\rho-\sigma$ plots 
(Figs.\ref{rs1}-\ref{rs7}) which  show contours of constant values of
$m_{\alpha\beta}$ in the plane of the Majorana phases  $\rho$ and
$\sigma$.

Let us comment on properties of the $\rho-\sigma$ plots.

The periodicity in $\rho$ and $\sigma$  
implies that the opposite sides of the  plots must be identified. 
For example,
the case of equal CP parities  of $\nu_1,\,\nu_2$ and $\nu_3$ corresponds
to any of the four corners of  the plots.

In general, any pair  of values $(\rho,\sigma)$ in the
range $[0,\pi)\times[0,\pi)$ represents  a physically independent
situation (different mass matrices). However,
if $\delta=0$ or $\pi$, it follows from (\ref{ABC}) that
\beq
m_{\alpha\beta}(\sigma,\rho) = m_{\alpha\beta}(\pi-\sigma,\pi-\rho)\;.
\nonumber
\end{equation}
This reflection symmetry is present in Figs.\ref{rs1}-\ref{rs3}, 
Figs.\ref{rs6}-\ref{rs7}, but not in
Fig.\ref{rs5}, where $\delta=\pi/2$.

The phase $\rho$ is associated  with the
mass $m_1$, therefore, in the case of strong normal hierarchy, the dependence
of $m_{\alpha\beta}$  on $\rho$ disappears and the iso-mass contours
become parallel to the axis  $\rho$. In contrast, the contours for
$m_{ee}$ are nearly  parallel to $\sigma$ axis, since $m_{ee}$ depends on
$\sigma$
via $O(s_{13}^2)$ terms. There is a relative shift of $\pi/2$, along the
axis $\sigma$, between the patterns for $m_{e\mu}$ and $m_{e\tau}$.

The elements  $m_{\mu\mu}$ and $m_{\tau\tau}$
have have the same $\rho - \sigma$ pattern  in the limit of maximal
2-3 mixing and  zero $s_{13}$. The difference between them originates
from  deviation of $\theta_{23}$ from $45^{\circ}$ and
from the  terms (see (\ref{big}))
\beq
\pm \sin 2\theta_{23} s_{13} r e^{-i\delta} Y \;,
\label{assym}
\end{equation}
where the plus sign corresponds to
$m_{\tau\tau}$ and the minus sign to $m_{\mu\mu}$.
In the case of maximal 2-3 mixing, only the term (\ref{assym}) contributes
to the difference.
The pattern for $m_{\mu\tau}$ is complementary
to that for $m_{\mu\mu}$ and $m_{\tau\tau}$, in the
sense that regions  of large $m_{\mu\tau}$   correspond to
regions  of small  $m_{\mu\mu}$ and $m_{\tau\tau}$ and
{\it vice versa}.

Small values of the $\mu\tau$-block elements appear 
at the corners of the plots,
$\rho \approx 0,\pi$  as well as $\sigma \approx 0,\pi$, and in the
region $\sigma \sim \pi/2$. In the latter case, the  corresponding value
of $\rho$ depends on 2-3 mixing. For maximal mixing,  the regions of small
elements are at $\rho \sim 0,\pi$; with deviation  from maximal mixing,
the regions shift to the center of the plot and  merge at  $\rho \sim
\pi/2$ for large values of $\xi$.

Let us comment on specific features of Figs.~\ref{rs1} -\ref{rs7}.

In Fig.\ref{rs1} we show the plots for the non-degenerate spectrum.
There is a sharp separation of the $e$-row  and dominant
$\mu\tau$-block elements. Structuring within these two
groups is rather weak.

In Fig.\ref{rs2} we show the plots for spectrum with partial degeneracy.
Dependence of  elements on $\rho$ becomes stronger with increase of
$m_1$. The
$\mu\tau$-block elements have more profound structure. The elements
$m_{e\mu}$ and $m_{e\tau}$ are small 
in the regions near  the corners of the plots.

The plots for spectrum with strong  degeneracy are shown 
in Figs.~\ref{rs3} - \ref{rs7}.
Now the $e$-row elements depend strongly on $\rho$, whereas the dependence
on $\sigma$ is rather weak. 
With increase of $m_1$
the $\rho$-dependence becomes stronger
for  the $\mu\tau$-block elements (see (\ref{mutau-m})). 
The patterns for $m_{\mu\mu}$ and
$m_{\tau\tau}$ differ  due to order $s_{13}$ terms (\ref{assym}),
which also depend on $\delta$.
The contribution of the term (\ref{assym})
has minus sign for $m_{\mu\mu}$ and therefore
it  adds constructively
with the other $\rho$-dependent term (see (\ref{mutau-m})). For
$m_{\tau\tau}$, instead,  the contribution has an opposite sign,
therefore $\rho$-dependence remains weak.

In  Fig.\ref{rs5} we show the plots for $\delta = \pi/2$.
The difference between the plots for  $m_{\mu\mu}$
and $m_{\tau\tau}$ becomes smaller
in comparison with  the case $\delta = 0$: indeed,
for $\delta = \pi/2$, the term (\ref{assym}) has
pure imaginary coefficient and its contributions 
to $m_{\mu\mu}$ and  $m_{\tau\tau}$
become similar.
For $\delta = \pi$, the $\rho- \sigma$ plots for
$m_{\mu\mu}$ and $m_{\tau\tau}$
interchange as compared with those in Fig.\ref{rs3}.
The pattern for  $m_{\mu\tau}$ is almost unchanged.
In the first approximation, the effect of $\delta=\pi/2$ on the
$e$-row elements is reduced to a shift of
$\sigma$ by $\pi/4$  for $m_{e\mu}$ and $m_{e\tau}$
and by  $\pi/2$ for  $m_{ee}$.

In Fig.~\ref{rs6} we show the plots for small $s_{13}$.
With decrease of $s_{13}$, the dependence of $e$-row elements on $\sigma$
disappears, patterns for $m_{\mu\mu}$ and 
$m_{\tau\tau}$ become more similar,
their complementarity   to the pattern  for $m_{\mu\tau}$ becomes
sharp.

In Fig.~\ref{rs4} we show the plots for
non-maximal 2-3 mixing  ($\theta_{23} =  35^{\circ}$).
The pattern for $m_{ee}$ is unchanged
and the one for  $m_{\mu\tau}$ changes weakly.
In contrast, the  difference between the patterns for $m_{e\mu}$ and
$m_{e\tau}$ increases.
In particular,
$m_{e\mu}$ can be large for $\rho \approx\pi/2$ and  $\sigma \approx 0,\pi$.
Also difference  of the patterns for $m_{\mu\mu}$ and
$m_{\tau\tau}$ increases.
Dependence of $m_{\tau\tau}$
on phases becomes weaker and regions with very small values of $m_{\tau\tau}$
disappear. In contrast, for $m_{\mu\mu}$
the region of small values appears near the center of the plot:
$\rho \sim  \sigma \sim \pi/2$.
For $\theta_{23} >  45^{\circ}$ (not shown) the situation is opposite:
region of small  values at $\rho \sim  \sigma \sim \pi/2$
appears for $m_{\tau\tau}$. Also $m_{e\tau}$ becomes,
in general, larger than $m_{e\mu}$.

In Fig.\ref{rs7} we show the plots for maximal possible 
1-2 mixing.
The $\rho$ dependence becomes  strong for all the elements and especially
for $m_{ee}$. This element can be zero at $\rho \approx \pi/2$.

\subsection{Correlations of mass matrix elements. Extreme values}

The  $\rho - \sigma$ plots allow to systematically scan all possible
structures  of the mass matrices.
The pattern of the  $\rho - \sigma$ plots  themselves depends on 
the unknown parameters $m_1$, $\delta$, $s_{13}$, as well as on the
uncertainties of the known  
oscillation parameters. As follows from the figures,  the dependence of
the  plots on $m_1$ is very strong, 
whereas the dependences on $\delta$ and  $s_{13}$ are  relatively weak
(in view of the strong bound on $s_{13}$).

The $\rho - \sigma$ plots allow to see immediately the correlations
between the values of different matrix elements. 
Formally, the 6 independent
moduli of matrix elements depend on 5 free parameters:  
$$
m_{\alpha \beta} = m_{\alpha \beta} (m_1, \rho, \sigma, \delta, s_{13})
$$   
So, only  one relation should exist among the matrix elements.
Actually, the correlations are much stronger
due to relatively strong upper bound on $s_{13}$ and  the fact that the 
effect of $\delta$ is suppressed by a factor $s_{13}$.
  
In the physically interesting limits the number of free parameters
further decreases. Thus,  in the case  of strong mass hierarchy
($m_1 \rightarrow 0$) the  $m_1$ and, consequently,  $\rho$ dependences
disappear: $m_{\alpha\beta} = m_{\alpha\beta} (\sigma, \delta,
s_{13})$.
In the limit of strong mass degeneracy the structure of the mass matrix
does not depend on the absolute mass scale:
$m_{\alpha\beta} = m_1 f_{\alpha\beta} (\rho, \sigma, \delta,
s_{13})$, etc. . 

Each point in the  $\rho - \sigma$ diagram (obviously, the same point
should be taken in all six panels) corresponds to a mass matrix with
certain structure.  
A given set of $\rho - \sigma$ diagrams (which corresponds
to  fixed values of  $m_1$, $s_{13}$ and $\delta$)
shows 6 elements as functions of two parameters:
$\rho$ and  $\sigma$. Therefore, imposing
conditions  on two (or even one element) one
may reconstruct whole the matrix up to  certain discrete
ambiguity. {\it E.g.}, in the degenerate case,  
imposing condition that $m_{\tau\tau}$ is the heaviest
element, 
we find that $m_{ee}$ and
$m_{\mu \mu}$   should be equally large whereas three
other elements are small.

The $\rho - \sigma$ plots allow to find immediately maximal and minimal
values  of matrix elements. 
Using  Eq. (\ref{ABC}) it is easy to see that 
the maximal value of the individual matrix element is given by
\beq
m_{\alpha \beta}^{max} = \sum_i m_i |U_{\alpha i} U_{\beta i}|. 
\end{equation}
It does not depend on the Majorana phases. 
The minimal  value is zero or 
\beq
2\max_i(m_i |U_{\alpha i} U_{\beta i}|) - \sum_{i} m_i |U_{\alpha i}
U_{\beta i}|
\label{max}
\end{equation}
if the latter is above zero. 
The first term in (\ref{max}) is (two times) the largest  
among the contributions from the three mass eigenstates. 
These  statements have  been made for $m_{ee}$ element in \cite{viss2} and
generalized to other elements in \cite{NMF,rode}.

Extreme values of matrix elements depend on 
$m_{1}$ as well as  on values of oscillations parameters. 
In the case of strong mass hierarchy ($m_1=0$),
the present experimental results 
(see section \ref{angles}) allow for zero 
minimal values of the e-row elements (see also $\rho-\sigma$ plots). 
Maximal values of the elements are
$$
\begin{array}{rcl}
m_{ee}^{max} & = & c_{13}^2s_{12}^2m_2+s_{13}^2m_3 \;,\\
m_{e\mu}^{max} & = & c_{13}\left[ s_{12}
(c_{23}c_{12}+s_{23}s_{12}s_{13})m_2 + s_{23}s_{13}m_3 \right]
\end{array}
$$
and similarly for $m_{e\tau}^{max}$, with the substitution 
$s_{23}\leftrightarrow c_{23}$.

In contrast, the elements of $\mu\tau$-block have non-zero minimal values. 
In the limit $s_{13}=0$, minimal and  maximal values of these elements
are,
$$
m_{\mu\mu}^{max,min}=s_{23}^2m_3 \pm c_{23}^2c_{12}^2m_2
$$
and  $m_{\tau\tau}^{max,min} = 
m_{\mu\mu}^{max,min}(s_{23}\leftrightarrow
c_{23})$,  $m_{\mu\tau}^{max,min} = m_{\mu\mu}^{max,min} 
(s_{23}^2,c_{23}^2\leftrightarrow s_{23}c_{23})$.

In the case of strong degeneracy all elements, but $m_{ee}$, may have 
zero values. Taking into account recent data on solar neutrinos 
\cite{SKsolar,sno},  
we get $m_{ee}^{min} \approx 0.45 m_1$. Maximal values of the $\mu\tau$-block  
elements and   the $ee$-element  equal: 
\beq 
m_{ee}^{max} \approx m_{\mu \mu}^{max} \approx m_{\mu \tau}^{max} 
\approx m_{\tau \tau}^{max} \approx m_1. 
\end{equation}
In the limit of small $s_{13}$,  maximal values of the two other elements
are $m_{e\mu}^{max} \approx m_1 \sin 2\theta_{12} c_{23}$ and
$m_{e\tau}^{max} \approx  m_1 \sin 2\theta_{12} s_{23}$.

Due to correlations among the mass matrix elements imposed by 
experimental data as well as the sum rule condition (\ref{sum}), 
only some elements can take their maximal or minimal values
simultaneously. In particular,  according to the $\rho-\sigma$ plots of
Fig.~\ref{rs1},  in the hierarchical case only 
two $e$-row elements  can be zero (very small) 
simultaneously: $m_{ee}$ and  $m_{e\mu}$ or $m_{ee}$ and $m_{e\tau}$.  
In the case of partial degeneracy $m_{e\mu}$ and $m_{e\tau}$
can be very small simultaneously.   
In the case of strong degeneracy, we see, from Figs.~\ref{rs3}-\ref{rs7}, 
that there are two groups of elements  which can be  simultaneously 
very small:  1) $m_{e\mu}$,  $m_{e\tau}$,  $m_{\mu \mu}$,  
$m_{\tau \tau}$; 2)  $m_{e\mu}$,  $m_{e\tau}$, $m_{\mu \tau}$. 
Similarly, from  the  $\rho-\sigma$ plots one can get 
groups of elements which reach simultaneously their maxima.

\subsection{$\beta\beta_{0\nu}$-decay  and structure of the mass matrix
\label{betalimit}}

The $ee$-element, ${m}_{ee}$, is the only matrix element for which we have
immediate experimental access.
The $\rho - \sigma$ plots allow one
to find immediately the implications of the  results from
$\beta\beta_{0\nu}$-decay searches for the structure of the mass matrix
(in assumption that  the exchange of the light Majorana neutrinos is the
only mechanism of the decay).
For $m_{ee}$, the Majorana phase plots (using a different parameterization)  
have been considered in \cite{mina}. 

The  iso-mass contours of $m_{ee}$ are nearly parallel to
the axis $\sigma$. Weak dependence of $m_{ee}$ on $\sigma$ appears due to
term of the order $s_{13}^2$.
For  very small $s_{13}$  and (partially) degenerate spectrum, the
iso-mass contours are
determined by
\beq
m_{ee} = m_1\sqrt{1- \sin^2{2\theta_{12}}\sin^2{\rho}}.
\label{the}
\end{equation}

Suppose that experimental searches give the upper bound 
$m_{ee} < m_{ee}^{up}$. Then, according to  Figs.~\ref{rs3}-\ref{rs7}, 
there are two iso-mass contours  in the $\rho - \sigma$ plots,
which correspond to a given value $m_{ee}^{up}$  
($m_{ee}(\rho, \sigma) =  m_{ee}^{up}$) and  a given set of the other 
parameters ($m_1$, $s_{13}$, $\delta$, etc.): $\rho_1 = \rho_1(\sigma)$ 
$(\rho_1 < \pi/2)$ and   $\rho_2 = \rho_2(\sigma)$ 
$(\rho_1 > \pi/2)$. 
The upper experimental limit on $m_{ee}$ excludes the following regions
in the $\rho - \sigma$ plots (obviously for all the matrix elements):
\beq
\rho \in [ 0 , \rho_1(\sigma)]\;, ~~~~~~~~ \rho \in [\rho_2(\sigma), \pi]\;.
\end{equation}

The position and the shape of the contours 
$\rho_i(\sigma)$ (i = 1,2) 
depend  on $m_1$, $\theta_{12}$ and $s_{13}$.
Taking, {\it e.g.}, $m_1 = 0.5$ eV, $s_{13} = 0.1 $,
$\tan^2 \theta_{12} = 0.36$ and the bound
(\ref{meebeta}),  we find from  Fig.\ref{rs3}
 that the regions covered by the
three darkest strips are excluded. They
correspond  approximately to $\rho < \pi/4$ and $\rho > 3\pi/4$.
These regions are excluded for all the elements. In this particular
case, all corners of the plots  and sides with $\rho \approx 0, \pi$,
which correspond to hierarchical structure of the mass matrix, are
excluded. Clearly no constraint on the structure appears for
weaker bound, $m_{ee}^{up} >  0.5$ eV (which is allowed by the
uncertainty in  the nuclear matrix elements), 
or, more in general, for $m_{ee}^{up}>m_1$.\\

For small $s_{13}$,  the mass matrix can be written immediately in terms
of $m_{ee}$, using Eq.(\ref{pdmatrix}):
\beq
\tilde{m}=\left(
\begin{array}{ccc}
\tilde{m}_{ee} & c_{23} \sqrt{r^2 -\tilde{m}_{ee}^2} & s_{23}
\sqrt{r^2-\tilde{m}_{ee}^2}\\
\dots & |c_{23}^2 \tilde{m}_{ee} + s_{23}^2 e^{-2i\sigma_X}| &
s_{23} c_{23} |- \tilde{m}_{ee} + e^{-2i\sigma_X}|\\
\dots & \dots & |s_{23}^2 \tilde{m}_{ee} + c_{23}^2 e^{-2i\sigma_X}|
\end{array}
\right) \;.
\label{meema}
\end{equation}
Here $\tilde{m}_{ee}=x r\le r$.  This form shows how strongly the
determination of $\tilde{m}_{ee}$  can influence the structure of the mass
matrix. The $s_{13}$  corrections to (\ref{meema}),
can weakly modify the structure of the matrix.

Positive results of $\beta\beta_{0\nu}$-decay  searches will select
two strips in the $\rho - \sigma$  plot.

Substantial bounds on the structure of the mass matrix can be obtained when
future solar neutrino  experiments and KamLAND \cite{kamland} experiment
will stronger restrict the allowed range for $\theta_{12}$ and
also when future $\beta$ decay measurements (KATRIN \cite{katrin})
will strengthen the bound on the absolute mass scale.


\subsection{$\rho - \sigma$ plots for the phases of matrix elements 
\label{MEphases}}
 
The phases of  matrix elements are, in general, functions of
all the unknown physical parameters:
$\phi_{\alpha\beta} = \phi_{\alpha\beta} (m_1, \rho, \sigma, \delta,
s_{13})$. In the limits of strong mass hierarchy or/and small
$s_{13}$, the expressions are simplified and for some elements
the phases are zero. Also in certain situations the phases
of some  elements depend only on $\rho$ or on $\sigma$.
 
The values of phases correlate (or anticorrelate) with the absolute
values of the corresponding elements.
Strong change of phase occurs typically in the regions
of parameter space  where the
absolute value of the element is small. 
There are also correlations between phases of different
elements.
 
 
 
\begin{figure}
[t]
\begin{center}
\epsfig{file=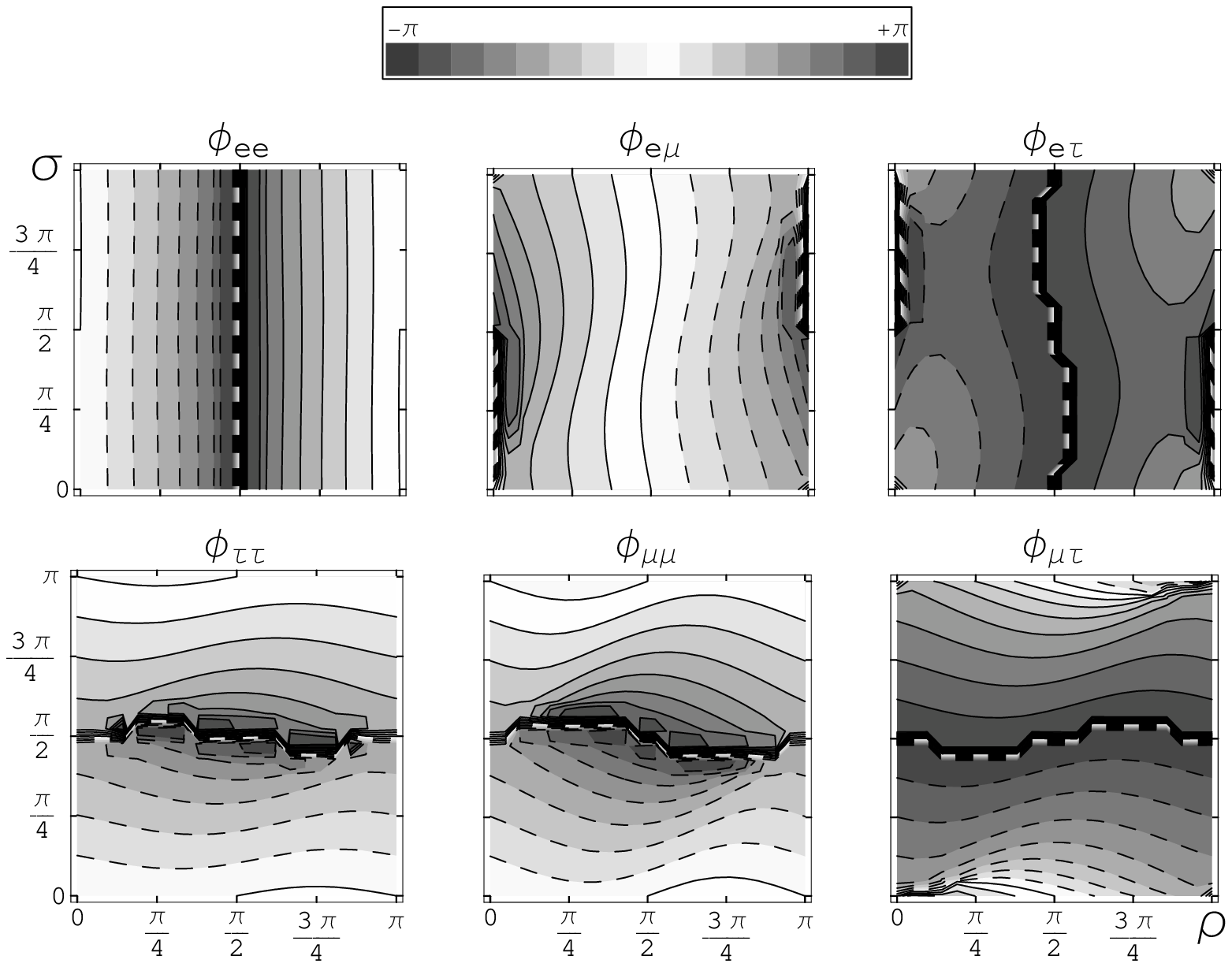,width=450pt,
bbllx=85, bblly=355, bburx=540, bbury=710}

\end{center}
\caption{The $\rho-\sigma$ plots of the phases of mass matrix elements in
the  case of degenerate spectrum. We use the same set of parameters as in
Fig.\ref{rs3}. 
Shown are the contours of constant phase
(iso-phase) $\phi_{\alpha,\beta}  = -\pi+ n \pi/8$ for
$n=0,...,7$ (dashed contours) and for $n=8,...,15$ (continuous contours).}
\label{phaseCD}
\end{figure}

In Fig.~\ref{phaseCD} we show the $\rho - \sigma$ plots for the phases 
of matrix elements in the case of degenerate spectrum
and the same choice of parameters as in Fig.~\ref{rs3}.
Notice that the pattern of $\rho-\sigma$ plots for  phases repeats 
partially the pattern for the absolute values. 
The phase $\phi_{ee}$ depends  strongly on $\rho$ 
and weakly on $\sigma$. 
The phases $\phi_{e\mu}$ and  $\phi_{e\tau}$  change
with $\rho$ and (weaker) with $\sigma$. The patterns are complementary to
some extent: at $\rho \sim \pi/2$,
$\phi_{e\mu}$ has minimum whereas  $\phi_{e\tau}$  maximum. 
The  phases of the $\mu\tau$-block
depend both on $\sigma$ and  (weaker) on $\rho$.
The patterns for  $\phi_{\mu\mu}$  and $\phi_{\tau\tau}$ are rather
similar.
Notice that maximal ($\pi$) values of these  phases are achieved at
$\sigma \sim \pi/2$ and minimal (zero) values are at
$\sigma \sim 0,  \pi$.

\section{CP phases and structure of the mass matrix 
\label{theo}}

Possible structures of the mass matrix can be classified
in the following way:

\begin{itemize}

\item Hierarchical matrices,  with certain dominant and sub-dominant
elements.

\item Matrices with flavor alignment  

\item Matrices with flavor disorder (flavor ``anarchy''). 

\item Matrices with certain  ordering of elements. In this case, the
elements $m_{\alpha\beta}$  have the same order of magnitude.

\item Democratic matrices, with  equal moduli of all the
elements:  $m_{\alpha\beta}\approx m_0$ for any choice of $\alpha,\beta$.

\end{itemize}

We will discuss these possibilities in order.

\subsection{Hierarchical mass matrices}

The regions of parameters which correspond to
a hierarchical structure of the mass matrix 
can be identified  as ``white" zones in the $\rho-\sigma$ plots, where
one or several elements have small values.
Notice that  the ``white" zones are  mainly at the corners 
or in the center of the plot, which corresponds to definite CP-parities 
or small CP-violating phases. So,  the most of  hierarchical structures
can be identified by  considering certain CP-parities.

A systematic search of possible hierarchical structures can be performed
in the following way. 
In the limit $s_{13}=0$, the elements of $e$-row equal:
\beq
\tilde{M}_{e\tau}
=-\tan{\theta_{23}}\tilde{M}_{e\mu}=-r s_{23}Y
=-r s_{23}s_{12}c_{12}(1-ke^{-2i\rho})~.
\label{erowlimit}
\end{equation}
Since $\tan{\theta_{23}}\sim 0.7 - 1.4$,  these elements can be either
both small or both large.

Let us consider first the case when  $m_{e\mu}$ and $m_{e\tau}$ do not
belong to the dominant structure, {\it i.e.}, $\tilde{M}_{e\mu}\approx
\tilde{M}_{e\tau}\approx 0$.  According to (\ref{erowlimit}), this implies
either $r\rightarrow 0$ or  $\rho\approx 0,\pi$. In the first
case we arrive at the  structure with dominant $\mu\tau$-block:
\beq
\tilde{M}=e^{-2i\sigma}\left(
\begin{array}{ccc}
0 & 0 & 0\\
\dots & s_{23}^2 & s_{23}c_{23}\\
\dots & \dots & c_{23}^2\\
\end{array}
\right)+{\cal O}(s_{13},r)\;,
\label{domblo}
\end{equation}
which holds for any value of the phases (see Fig.\ref{rs1}).
Weak ordering of elements  is possible in the $\mu\tau$-block. In the
second case, $\rho= 0,\pi$,  which corresponds to the same CP-parities of
$\nu_1$ and $\nu_2$, the ratio $r$ can  be of order $1$ and new structures 
appear. For $\rho= 0,\pi$, we get $X= Z= 1$ and  (see
(\ref{pdmatrix})):
\beq
\tilde{M}=\left(
\begin{array}{ccc}
r & 0 & 0\\
\dots & c_{23}^2r+s_{23}^2e^{-2i\sigma} & s_{23}c_{23}(-r+e^{-2i\sigma})\\
\dots & \dots & s_{23}^2r+c_{23}^2e^{-2i\sigma}\\
\end{array}
\right)+{\cal O}(s_{13})\;.
\label{tild}
\end{equation}
Such a possibility is realized near  the left and right borders of the
plots in Fig.\ref{rs2}. 
The determinant of the $\mu\tau$-block is given by
\beq
\tilde{M}_{\tau\tau}\tilde{M}_{\mu\mu} -\tilde{M}_{\mu\tau}^2\approx r
e^{-2i\sigma} \;.
\label{deter}
\nonumber
\end{equation}
So, with increase of $r$, it  deviates strongly  from zero.
In the first approximation,  we get mass matrix  with $4$ independent
dominant elements of the same order:  
$m_{ee}\sim m_{\mu\mu}\sim m_{\tau\tau} \sim m_{\mu\tau}$.

Hierarchy of elements  in the $\mu\tau$-block appears for special values
of the phase $\sigma$.  If, {\it e.g.}, $\tan{\theta_{23}}\le 1$, we can get
$M_{\mu\mu}\approx 0$ provided that
\beq
\sigma = \frac{\pi}{2}\;, ~~~~~~~~ r= \tan^2{\theta_{23}}\;.
\nonumber
\end{equation}
The mass matrix is then reduced to
\beq
\tilde{M} \approx \left(
\begin{array}{ccc}
\tan^2{\theta_{23}} & 0 & 0\\
\dots & 0 & -\tan{\theta_{23}}\\
\dots & \dots & -1+\tan^2{\theta_{23}}\\
\end{array}
\right)
=\left(
\begin{array}{ccc}
r & 0 & 0\\
\dots & 0 & -\sqrt{r}\\
\dots & \dots & -1+r\\
\end{array}
\right)
\;.
\label{matr-r}
\end{equation}
Let us underline that such a structure  is present in the case of partial
degeneracy only.

In the limit of complete  degeneracy, $r\rightarrow 1$, the
condition $M_{\mu\mu}\approx 0$ requires  $\tan{\theta_{23}}= 1$ and
therefore the matrix converges to
\beq
\tilde{m}=\left(
\begin{array}{ccc}
1 & {\cal O}(s_{13}) & {\cal O}(s_{13})\\
\dots & {\cal O}(s_{13}^2) & 1\\
\dots & \dots & {\cal O}(s_{13}^2)
\end{array}
\right)\;.
\label{oppoCP}
\end{equation}
This type of matrix has  been discussed previously, {\it e.g.} in
\cite{altfer}. If also $\delta=\pi/2 $, then $Z' = 0$ (see
(\ref{zprime})) and therefore order  $s_{13}$ terms
are zero (see (\ref{big}) and Fig.\ref{rs5}).

If $\tan{\theta_{23}}\ge 1$,  one can get $M_{\tau\tau}\approx 0$. This,
again, requires  $\sigma=\pi/2$ but $r=\cot^2{\theta_{23}}$ and,
in lowest order in $s_{13}$, the mass matrix has the form
\beq
\tilde{M}\approx\left(
\begin{array}{ccc}
\cot^2{\theta_{23}} & 0 & 0\\
\dots & -1+\cot^2{\theta_{23}} & -\cot{\theta_{23}}\\
\dots & \dots & 0\\
\end{array}
\right)
=\left(
\begin{array}{ccc}
r & 0 & 0\\
\dots & -1+r & -\sqrt{r}\\
\dots & \dots & 0\\
\end{array}
\right)
\;.
\label{matr-r2}
\end{equation}
It has the same limit (\ref{oppoCP})  in the case of completely degenerate
spectrum.

Notice that in the limit $s_{13}\rightarrow 0$, one should take into
account the deviations of $k$ from $1$.  This leads to appearance of terms
of order $\Delta m^2_{sol}/2m_1^2$ instead of zeros (see (\ref{soldev})).

According to   (\ref{tild}),  the off-diagonal elements  of
$\mu\tau$-block are zero for
$r=e^{-2i\sigma}$, that is  for $r=1$ and $\sigma=0,\pi$. In this case
$\tilde{M}_{ee}=\tilde{M}_{\mu\mu}=\tilde{M}_{\tau\tau}=1$ and
\beq
\tilde{m}=\left(
\begin{array}{ccc}
1 & {\cal O}(s_{13}) & {\cal O}(s_{13})\\
\dots & 1 & {\cal O}(s_{13}^2)\\
\dots & \dots & 1
\end{array}
\right)\;.
\label{diagm}
\end{equation}
So, the dominant structure reduces to the unit matrix
(as it has been described, {\it e.g.}, in \cite{altfer})
with  small off-diagonal corrections (Fig.\ref{masszero}a).
If also $\delta=0,\pi$, the
order $s_{13}$ terms are zero (see (\ref{big},\ref{zprime})).

Let us study possible  equalities of elements of the dominant
structure.
The conditions for the equality  $m_{\mu\mu} = m_{\tau\tau}$ follow from 
Eq.(\ref{mumta}).
All the elements of  $\mu\tau$-block  have the same absolute value
provided that 2-3 mixing is maximal and  $\sigma=\pi/4, 3\pi/4$. In
this
case:
\beq
\tilde{M}=\left(
\begin{array}{ccc}
r & 0 & 0\\
\dots & \frac 12 \sqrt{1+r^2} e^{\pm\phi} & \frac 12 \sqrt{1+r^2} e^{\mp\phi}\\
\dots & \dots & \frac 12 \sqrt{1+r^2} e^{\pm\phi}\\
\end{array}
\right)+{\cal O}(s_{13})\;,
\label{6e}
\end{equation}
where $\phi=\arctan{1/r}$.

For $r=1$, we get:
\beq
\tilde{m}=\left(
\begin{array}{ccc}
1 & {\cal O}(s_{13}) & {\cal O}(s_{13})\\
\dots & 1/\sqrt{2} & 1/\sqrt{2}\\
\dots & \dots & 1/\sqrt{2}
\end{array}
\right)\;.
\end{equation}
Order $s_{13}$ terms are zero if also $\delta=\sigma$ or $\sigma+\pi$.

Notice that mass matrices considered above depend on $r$ and $s_{23}$.
Dependence  on $\theta_{12}$ appears only via
$s_{13}$ and  $\Delta m^2_{sol}/m_1^2$  corrections.\\

Let us consider the case  where $M_{e\mu}$ and $M_{e\tau}$ belong to the
dominant structure. According  to (\ref{erowlimit}), this implies $r\sim
1$ and $\rho$ not to close to   $0,\pi$ (see regions with $\rho\sim \pi/2$
in Figs.~\ref{rs3}-\ref{rs7}). In this case, also the element $m_{ee}$ can
belong to the dominant structure. Indeed, minimal value of
$m_{ee}$ is achieved at $\rho=\pi/2$, so that:
\beq
\frac{m_{ee}^{min}}{m_{e\mu}}
=\frac{m_{ee}^{min}}{\tan{\theta_{23}}m_{e\tau}}
=\frac{\cos{2\theta_{12}}}{c_{23}\sin{2\theta_{12}}}\;.
\label{eratio}
\end{equation}
So, all the elements  of $e$-row  have comparable values, unless 1-2 mixing is
near maximal.  For $\theta_{12}\approx \pi/4$, 
the hierarchical structure
$m_{ee}\ll m_{e\mu},m_{e\tau}$ is realized (see Fig.\ref{rs7}).  

Let us consider the possibility of zeros in the $\mu\tau$-block. 
Now  the situation differs
from that of the case  $\rho= 0, \pi$ (see Eqs.(\ref{k1xyz},\ref{pdmatrix})).
Since $ x <1$,  we have $\tilde{m}_{\mu\tau} \neq 0$ and for maximal
mixing, all  elements of the 
$\mu\tau$-block differ from zero. Still,
for non maximal 2-3 mixing,  we can get $\tilde{m}_{\mu\mu}=0$
or $\tilde{m}_{\tau\tau}=0$. For instance,  if $\theta_{23}<45^{\circ}$,
$\tilde{m}_{\mu\mu}=0$  when $\sigma_X=\pi/2$ and
$\tan^2{\theta_{23}}= x r$.  
So, in the case of large $e$-row elements,  one  or two
of the diagonal elements can be zero: $m_{ee}$, for maximal 1-2 mixing
and/or
$m_{\mu\mu}$ ($m_{\tau\tau}$),  for special relation among
$\theta_{12},\sigma$ and $\theta_{23}$.
\noindent

Summarizing, the mass matrix has a hierarchical 
structure: 

(a) In the case of hierarchical mass spectrum:  
the $e$-row elements can be about 10 times smaller 
than the $\mu\tau$-block elements. 

(b) In the case of degenerate mass spectrum:  
the hierarchy determined by a factor $\sim 10$ or more 
appears in the regions near the corners of the $\rho - \sigma$ 
plots: 
$$
\rho \approx 0 - \pi/8, ~~~\sigma \approx 0 - \pi/32, 
$$ 
for the first corner,  and similar interval for the three other corners.  
In these cases the matrix equals approximately to the unit
matrix with small off-diagonal terms. 
Another possibility is 
$$
\rho \approx 0 - \pi/6, ~~~\sigma \approx (0.45 - 0.55)~\pi 
$$
and similar reflected region $\rho \rightarrow \pi - \rho$. 
In this case the mass matrix has a dominant structure with 
$\tilde{m}_{ee} \approx \tilde{m}_{\mu\tau}$, 
while all other elements are small. 

(c) For non-maximal 2-3 mixing:  the element $\tilde{m}_{\mu\mu}$ or 
$\tilde{m}_{\tau\tau}$ can be small for $\sigma\approx \pi/2$
and for a value of $\rho$ which depends  on the deviation $\xi$ of 2-3
mixing from maximal value. With increase  of $\xi$, the region of small
mass approaches the center of $\rho-\sigma$ plots ($\rho \sim \pi/2$).

\subsection{Flavor alignment and flavor disorder}

Does matrix show any  flavor ordering (alignment), that is, 
the correlation of the neutrino mass terms and the charged lepton masses?
To some extent, the lepton mixing matrix itself is the measure of the
flavor alignment, so that  
small mixing would imply strong alignment. The  observed large lepton
mixing means weak ordering  or absence of the flavor ordering. 
The question of flavor ordering can be 
studied in terms of mass matrix in flavor basis.
In this connection, 
let us consider  the possibility that masses decrease
with transition from the $\tau$-flavor to the $e$-flavor, that is,
\beq
m_{\tau\tau}\gtrsim m_{\mu\tau} \gtrsim m_{\mu\mu} \gtrsim m_{e\tau}
\gtrsim
m_{e\mu} \gtrsim m_{ee}\;.
\label{flaord}
\end{equation}
We will call this possibility  the {\it normal flavor ordering} or 
{\it alignment}. The ordering
with $m_{e\tau}\lesssim m_{\mu\mu}$ is also possible. Notice that,
according to (\ref{mumta}),  $m_{\tau\tau}>m_{\mu\mu}$ provided that
$\theta_{23}<45^{\circ}$. In contrast, one gets from (\ref{memta})
that $m_{e\tau}>m_{e\mu}$, if $\theta_{23}>45^{\circ}$.  So, in the
approximation
$s_{13}\approx 0$, the ``flavor ordering'' is impossible.
However, for near maximal 2-3 mixing, the differences 
($m_{\tau\tau}-m_{\mu\mu}$) and ($m_{e\tau}-m_{e\mu}$)  are so small that
corrections due to non-zero $s_{13}$ become important.
These corrections can produce flavor ordering, as  can be seen,
{\it e.g.}, in Fig.\ref{sigmam2}b, for the case 
of strong mass hierarchy, $m_1=0$, and in
Fig.\ref{m1sigma}e (shifting $e$-row lines), for the case $r \sim 1$.

There are other possibilities of the flavor ordering.  
Sets of parameters can be found for which 
the matrix has 

1) $\tau\tau$-alignment   (see, {\it e.g.}, Fig. \ref{sigmam2}a,  
$\sigma \approx 2.6$):
\beq
m_{\tau\tau} > m_{\mu \mu} \approx m_{\mu \tau}  > m_{e \tau} \approx
m_{e \mu} \approx m_{ee}\;;
\end{equation}

2) $e$-alignment (see Fig. \ref{sigma}d,  $\sigma \approx 2.4$),
when the masses are sensitive to $L_e$:
\beq
m_{\tau\tau} \approx  m_{\mu \mu} \approx m_{\mu \tau}  > m_{e \tau}
\approx
m_{e \mu} >  m_{ee} \;;
\end{equation}

3) other alignments, such as (Fig. \ref{sigma}b, $\sigma \approx 0$):
\beq
m_{\tau\tau} > m_{\mu \mu} \approx m_{\mu \tau}  > m_{e \tau} \approx
m_{e \mu} >  m_{ee}\;.
\end{equation} 

Although in many cases $m_{ee}$ can be the heaviest element, 
inverted flavor alignment (when the mass increases with 
change of flavor from $\tau$ to $e$) seems to be  impossible. \\

As follows from the Figs.~4 - 6, in a number of cases 
(partially degenerate, degenerate spectrum)
the matrix can show 
{\it flavor disorder}. That is, the  
matrix elements can take (relative)  values between 0 and 1 without
correlation with masses of the charge leptons.

\subsection{Mass matrices with specific ordering of elements}

For  $m_1\gtrsim \sqrt{\Delta m^2_{atm}}$ ($k\approx 1$),  in large part
of the phases  space,
all the elements of  the mass matrix are of the same order (see
Figs.\ref{rs3}-\ref{rs7}).
Values of free parameters can be chosen  in  such a way that any element
of the matrix can
be the smallest one or the  largest one. Also one can reach equalities
between some of the elements.  A number of configurations is possible,
with only a few restrictions   determined by relations among the
elements, discussed at the end of section \ref{secpd}.
Varying $r$, $x$, $\sigma_X$ and  $\theta_{23}$ (see (\ref{pdmatrix})), 
one can get equalities
among various elements of the  matrix. In particular,

1) $m_{ee}=m_{e\mu}$ for
\beq
x = \frac{c_{23}}{\sqrt{1+c_{23}^2}}\;.
\nonumber
\end{equation}

2) $m_{ee}=m_{e\tau}$ for $x$ given by a similar  expression with
the substitution $c_{23}\leftrightarrow s_{23}$.

3) All elements of the $e$-row are equal for maximal 2-3 mixing and
$x=1/\sqrt{3}$.

4) One can reach equality of the diagonal elements $m_{ee}=m_{\mu\mu}$ or
$m_{ee}=m_{\tau\tau}$ and  also $m_{ee}=m_{\mu\mu}=m_{\tau\tau}$; see,
{\it e.g.}, Fig.\ref{m1sigma}c.

5) The equality of  elements of the second diagonal,
$m_{e\tau}= m_{\mu\mu}= m_{\tau e}$, is possible, but in this case other
elements are not small: $m_{\tau\tau} \approx m_{\mu\mu}$, for example.

6) According to Fig.\ref{m1sigma}d, the following equalities can be 
satisfied:  
\beq
m_{ee}=m_{\mu\mu}=m_{\tau\tau}\approx 2m_{e\mu}=2m_{e\tau}=2m_{\mu\tau}
\nonumber
\end{equation}
for $\sigma\approx 0.7$.

7) For $\sigma\approx 1.2$ (Fig.\ref{m1sigma}d) we find
\beq
m_{ee}=m_{\mu\tau}\approx
2m_{\tau\tau}=2m_{e\mu}=2m_{e\tau}=2m_{\mu\mu}.
\nonumber
\end{equation}

However, it is not possible to get zero values of
all the diagonal elements.  Indeed,
$m_{ee}$ vanishes  for $r=0$ or $x=0$  (the latter corresponds to near
maximal
1-2 mixing). However, for $x =0$,  $m_{\mu\mu}$ and $m_{\tau\tau}$ are
non-zero: they belong to the dominant  structure. The only possibility
would be to consider inverted hierarchy  of the mass eigenvalues.

\subsection{Democratic mass matrix}

It is possible to have equal absolute
values for all the matrix  elements in the flavor basis.
To obtain such a ``democratic matrix"
one should satisfy five
equalities among  independent matrix elements $m_{\alpha\beta}$.
In  general, we have nine  parameters (three masses, three mixing angles
and three CP violating phases) and we should reproduce the solar as well as
atmospheric mass squared differences and mixing angles (4 relations) as 
well as
satisfy the CHOOZ bound. So,  in principle, the problem is non-trivial.
Let us present one realization of such a possibility.

The $e$-row elements  should be    as large as the $\mu\tau$-block
elements;  this  requires  $r\sim 1$ and $\rho\sim \pi/2$.
The  $\mu\tau$-block elements
are equal to each other only  for $\sigma \sim \pi/4,~ 3\pi/4$. Then, if
$s_{13}$ is very small,  also $\xi$ is required to be very small,
otherwise $m_{e\mu}$  differs inevitably from $m_{e\tau}$ and the
same is true for $m_{\mu\mu}$ and $m_{\tau\tau}$.

Taking  the limit $s_{13}=0, k=1$ (see (\ref{pdmatrix})),
we find from equality of the  $e$-row elements:
\beq
\xi=0\;, \qquad \sin^2{2\theta_{12}}\sin^2{\rho}=\frac 23\;.
\label{cond1}
\end{equation}
According to  second  equality in (\ref{cond1})  the solar  mixing
angle determines  the Majorana phase $\rho$. Equality of the
$\mu\tau$-block
elements leads to  the condition
\beq
\cos{2\sigma} c_{12}^2 + \cos{2(\sigma-\rho)} s_{12}^2 = 0 \;,
\label{cond2}
\end{equation}
which  fixes the value of $\sigma$.
If conditions  (\ref{cond1}) and  (\ref{cond2}) are satisfied,
it turns out that, taking $r=1$, the elements of  $e$-row and $\mu\tau$-block
are also equal. Moreover,  $\tilde{m}_{\alpha\beta}= \tilde{m}_0 = 1/\sqrt{3}$.
Notice that last  equality can be immediately obtained from
the mass squared conservation (section \ref{sumrule}). Indeed,
in the degeneracy case we have  
$S_0/m_3^2 \equiv \sum_{i} \tilde{m}_{i}^2 = 3$.
For  the democratic matrix the sum of all elements equals
$\sum_{\alpha,\beta} \tilde{m}_{\alpha \beta}^2
= 9 \tilde{m}_0^2$. According to the mass conservation, 
we have $9 \tilde{m}_0^2=S_0/m_3^2=3$,
or $\tilde{m}_0^2 = 1/3$.

\subsection{Bi-maximal mixing and its variations
\label{cons}}

The Fig.\ref{rs7} corresponds to bi-maximal mixing
($\theta_{12} = \theta_{23} = 45^{\circ}$). 
Notice that, in contrast  with pure bi-maximal mixing, $\theta_{13}$ is
non-zero here.  The limit $\theta_{13}\rightarrow 0$ leads to
disappearance of dependence of the $e$-row elements on $\sigma$ and to
equality of the patterns for $m_{\mu\mu}$ and $m_{\tau\tau}$.

According to Fig.\ref{rs7}, large variety  of mass matrix structures can
lead to bi-maximal mixing. In particular,  for $\rho=0,\pi$ and
$\sigma=0,\pi$ (corners of the plot), we get  the nearly diagonal matrix
(\ref{diagm}). For $\rho=0,\pi$ and $\sigma=\pi/2$,  the mass matrix has
the form (\ref{oppoCP}). For $\rho=\pi/2$, it  follows $x=0$. In this
case, neglecting ${\cal O}(s_{13})$ terms, for any value of $\sigma$ 
we get the matrix
\beq
\tilde m \approx \frac{1}{2}\left(
\begin{array}{ccc}
0 & \sqrt{2} & \sqrt{2} \\
\sqrt{2} & 1 & 1\\
\sqrt{2} & 1 & 1
\end{array}
\right)\;,
\end{equation}
discussed in the literature \cite{altfer}.

Apart from that, many other  structures  allowed, {\it e.g.} matrices
with nearly equal elements, etc., can lead to bi-maximal mixing.  

Notice that recent data on solar neutrinos strongly disfavor 
maximal 1-2 mixing \cite{globalfits}. Still mass matrix with 
bi-maximal mixing  
can be realized in the {\it symmetry basis}. In this case the observable 
non-maximal 1-2 mixing is the result of rotation of the charge lepton 
mixing matrix.    

\subsection{Parameterization of $M$
\label{flav}}

Let us consider the possibility to parameterize the mass matrix by
powers of a unique expansion parameter $\lambda\ll 1$:
\beq
\tilde{m}_{\alpha\beta}= c_{\alpha\beta} \lambda^{n_{\alpha\beta}} ,
\end{equation}
where $c_{\alpha\beta}$ are numbers of order 1.
In the flavor symmetry context, the exponents $n_{\alpha\beta}$
are related to the flavor charges of the  corresponding mass terms.
If $n_{\alpha\beta}=n_{\alpha}+n_{\beta}$,  where $n_{\alpha}$,
$n_{\beta}$ ($\alpha,\beta=e,\mu, \tau$) are numbers  associated with
corresponding
flavor states, factorization occurs:
\beq
\tilde{m}_{\alpha\beta} = c_{\alpha\beta}
\lambda^{n_{\alpha}}\lambda^{n_{\beta}}\;.
\end{equation}
In this case the smallness  of various mass terms is correlated:
$n_{\mu\mu}=2n_{\mu\tau}-n_{\tau\tau}$, $2n_{e\mu}=n_{ee}+n_{\mu\mu}$, etc.

Let us first  consider the case of spectrum with  mass hierarchy. 
As one can  see in Eq.(\ref{hmatrix}),  for maximal  2-3 mixing and
$\sigma\approx \pi/4,3\pi/4$, all
elements of the  dominant $\mu\tau$-block can be equal to each other.
Then, the elements of the $e$-row  should be suppressed by powers of
$\lambda$:
\beq
\tilde{m}_{e\beta}\propto \lambda^{n_{\beta}}\;,\qquad
\beta = e,\mu,\tau\;.
\end{equation}
As follows from our analysis, we can have all  the $e$-row elements to be
equal among themselves, simultaneously with equality of  $\mu\tau$-block
elements:
\begin{equation}
m \propto \left(
\begin{array}{ccc}
\lambda & \lambda & \lambda \\
\lambda  & 1 & 1\\ \lambda & 1 & 1
\end{array}
\right)	\;,
\label{matr-a}
\end{equation}
where (see (\ref{mean}))
\beq
\lambda \approx \sqrt{2(s_{13}^2 + r^2c_{12}^2s_{12}^2)}\;.
\nonumber
\end{equation}

For not too small $s_{13}$ further structuring is possible, when
$m_{ee}\propto \lambda^2$:
\begin{equation}
m \propto \left(
\begin{array}{ccc}
\lambda^2 & \lambda & \lambda \\
\lambda  & 1 & 1\\ \lambda & 1 & 1
\end{array}
\right)	\;.
\label{one}
\end{equation}
Such a situation is realized, {\it  e.g.},  in Fig.\ref{sigma}d (with
certain shift of the $e$-row lines due to $\delta$). In this
case $\lambda \approx s_{13}$. The matrix (\ref{one}) satisfies
the factorization condition. Other structuring of the $e-$row elements
is also possible, like $(\lambda^2,\lambda^2,\lambda)$  or
$(\lambda^2,\lambda,\lambda^2)$ with $\lambda\approx 0.3$
(see Fig.\ref{sigmam2}).

Mild hierarchy of elements  of  the $\mu\tau-$block is realized for
non-maximal 2-3 mixing or/and  non-trivial CP phases.
According to Fig.~\ref{sigmam2}a,~ \ref{sigmam2}b,  we may have
$m_{\tau\tau}\approx
m_{\mu\tau}> m_{\mu\mu}\approx m_{e\tau}>m_{e\mu}\approx m_{ee}$
which corresponds to parameterization:
\begin{equation}
m \propto \left(
\begin{array}{ccc}
\lambda^2 & \lambda^2 & \lambda \\
\lambda^2  & \lambda & 1\\ \lambda & 1 & 1
\end{array}
\right)	\;,
\label{matr-b}
\end{equation}
with $\lambda\approx 0.3$.
Also   $m_{\tau\tau}$ can be the
smallest element of the $\mu\tau-$block, instead of $m_{\mu\mu}$.

In the case of partial or complete degeneracy,  new dominant structures
appear and therefore new types of expansion is possible.
According to Fig.\ref{masszero}e and (\ref{6e}), the
mass matrix can have the following form:
\begin{equation}
m \propto \left(
\begin{array}{ccc}
1 & \lambda & \lambda \\
\lambda  & 1 & 1\\ \lambda & 1 & 1
\end{array}
\right)	\;,
\label{matr-c}
\end{equation}
with
\beq
\lambda \approx \frac{s_{13}}{r\sqrt{2}}\;.
\nonumber
\end{equation}
Two other  possibilities are (see Figs.~\ref{m1rho}f,~\ref{masszero}f):
\begin{equation}
m \propto \left(
\begin{array}{ccc}
1 & \lambda & \lambda \\
\lambda  & \lambda & 1\\
\lambda & 1 & \lambda
\end{array}
\right), \qquad
m \propto \left(
\begin{array}{ccc}
1 & \lambda & \lambda \\
\lambda  & \lambda^2 & 1\\
\lambda & 1 & \lambda^2
\end{array}
\right)	\;,
\label{matr-d}
\end{equation}
with  $\lambda \approx s_{13}/\sqrt{2}$, which  should be taken of  order
$0.1$ for the left matrix and  $0.2$ for the right matrix.

Notice that  value of $\lambda$
which appears in the matrices 
(\ref{matr-a})-(\ref{matr-d}) and,
therefore,  consistent with present data, can not be too small.
We find
\beq
\lambda\; \gtrsim 0.1 - 0.2 \;.
\label{lala}
\end{equation}
Values $\sim 0.3-0.4$ are also allowed.
The value of the parameter
(\ref{lala}) can be equal to $\sin\theta_c$, where $\theta_c$ is
the Cabibbo angle, used as an expansion parameter for  quark mass 
matrices.
In the flavor basis the structure of the charge lepton mass matrix is
characterized by  the two ratios:
$m_{\mu}/m_{\tau} = 0.059$ and $m_e/m_{\mu} = 0.0049$.
These ratios can also be reproduced as powers of
$\lambda$:
\beq
m_e : m_{\mu} : m_{\tau} \approx \lambda^6 : \lambda^2 : 1\;,
\nonumber
\end{equation}
with $\lambda\approx 0.24$.

In a large part of the parameter space,  the elements of the 
mass matrix have the
same order of magnitude, so that the  ratio of matrix elements is close
to $1$. In this case we can introduce  the ordering parameter
$\lambda_{ord}\sim {\cal O}(1)$. Typical  value of $\lambda_{ord}$ can be
determined, {\it e.g.}, by the possible spread of $\mu\tau$-block elements,
due to deviation of the 2-3 mixing from maximal value:
\beq
\lambda_{ord} \approx \tan{\theta_{23}} \approx 0.7\;.
\nonumber
\end{equation}
Another possible choice for $\lambda_{ord}$, in  the partial degeneracy
case, could be $r$.
We find the following structures in  Figs.\ref{sigma},\ref{sigmam2}
(omitting the subscript `$ord$'):
\beq
m \propto \left(
\begin{array}{ccc}
\lambda^4 & \lambda^3 & \lambda^2 \\
\lambda^3 & \lambda^2 & \lambda\\
\lambda^2 & \lambda & 1
\end{array}
\right)\;,\;\;\;\;\;
\;\;\; m \propto \left(
\begin{array}{ccc}
\lambda^6 & \lambda^4 & \lambda^3 \\
\lambda^4 & \lambda^2 & \lambda\\
\lambda^3 & \lambda & 1
\end{array}
\right) .
\label{two}
\end{equation}
These structures  require   rather large
$\theta_{13}$ to enhance the values of the $e$-row elements.

In the case of partial or complete  degeneracy, situation appears where all
elements are of the same order with  small spread, see, {\it e.g.},
Fig.\ref{m1sigma}f at $\sigma\approx 0.7$.  In this connection
one can consider the mass matrix as
small deviation from the democratic one:
\beq
M\propto M^D+\Delta M\;,
\nonumber
\end{equation}
where $|M^D_{\alpha\beta}|= 1$,  $\Delta M \sim {\cal O}(\lambda)$
and $\lambda$ is a small  parameter. Here $\lambda$ can be taken of order
$s_{13}$ or $\xi$ or $1-r$  (deviation from degeneracy). An interesting
possibility could be to take  for $\lambda$ the deviation of $\rho$ or
$\sigma$ from the values $0,\pi/2$,  which correspond to definite CP
parities.

\subsection{Remarks on the Symmetry basis \label{symmetry}}

As we have outlined in the introduction, to get further theoretical
inference,
one needs to find  the matrix in the symmetry basis and at the
symmetry scale.
In general, the  symmetry basis differs from the flavor basis and the mass
matrix of charged  leptons, $M_l$,  is non-diagonal there. The neutrino
mass
matrix in the symmetry  basis, $M_{\nu}$, is related to that in flavor
basis
as $M_{\nu}  = U_l^ T M U_l$,  where $U_l$ is the mixing matrix which
diagonalizes $M_l$.

The matrix $U_l$ is unknown and some additional assumptions are needed 
to fix its structure. Clearly  this introduces a further  ambiguity in
the analysis.  Here we mention two possibilities (two assumptions)  which
allow one to  immediately relate the matrices  in flavor basis  
and symmetry basis.  (The extensive discussion of this issue will be
given elsewhere~\cite{SF2}).

1) It  may happen that due to strong hierarchy of the
masses of the charged  leptons, the charged lepton mixing is rather small
and $U_l\approx \mathbb{I}$.  In this case, the structures of the mass
matrix $M$, discussed in this  paper, are not modified significantly under
transition to the symmetry basis.

2) Being
related to the ratio of masses of the $\mu$ and $\tau$ lepton,
the 2-3 angle, $\theta_{23}^l \sim \sqrt{m_{\mu}/m_{\tau}}$,  can be  the
only large angle
in $U_l$ (1-2 and 1-3 mixing  angles are very small, if they are connected
with the tiny electron mass). In this case, effect of
charged lepton mixing
on the neutrino mass matrix
is reduced to change of the neutrino 2-3 angle in the flavor basis:
$$
\theta_{23} = \theta_{23}^{sym} - \theta_{23}^l\;.
$$
Taking into account
this shift of the angle, one can use neutrino mass matrices
obtained in this paper as mass matrices in the symmetry
basis.
This shift can justify large deviations of the neutrino 2-3 mixing
from maximal value.

However, there are many models  in which charged lepton mass matrix is
strongly off-diagonal in symmetry basis; see,  {\it e.g.},
\cite{barr}\cite{barb}\cite{babubarr}.\\

Structures of the mass  matrix $M$ will not be modified substantially
due to running to high scales.  It was found \cite{renorm} that
renormalization of $M_{\alpha\beta}$  is smaller than $10^{-4}$ for the
Standard Model and about few percents for MSSM.

\section{Discussion and conclusions \label{seven}}

The  motivation of our study is to understand how far one can 
go in construction of the theory of neutrino mass  
using the bottom-up approach, that is, starting from experimental results. 
Neutrino mass matrix in flavor basis unifies information contained in 
masses and mixing angles measured in  experiment  and therefore can 
give deeper insight into the underlying physics. 

We have elaborated a method  which allows one to 
study dependences of the individual matrix elements and of  the 
structure  of the mass matrix as whole on the unknown yet parameters. 
In particular, we have performed a systematic and comprehensive study 
of dependences  of the neutrino mass matrix elements on the 
CP violating phases.  

We have introduced the $\rho - \sigma$ plots which 
show contours of constant mass in the 
plane of the Majorana phases   $\rho$ and $\sigma$.   
We used the $\rho - \sigma$ plots to analyze 
the possible structures of the mass matrix. 
Each point in the $\rho - \sigma$ plot represents a certain neutrino mass
matrix, so the $\rho - \sigma$ plots allow one to scan  all possible 
matrix structures.  

The $\rho - \sigma$ plots allow  to study in rather transparent and
straightforward way: 

-  influence of the phases on  magnitudes  of individual matrix elements.
In particular, one can find ranges in which
elements can change and their extremal values (minimal and maximal).

- correlations between values of different matrix elements. 
Taking a given element in some range one can see immediately
intervals in which other elements can change.
 
- correlations between the structure of the neutrino mass matrix and
the charged lepton masses.

- consequences  of  experimental measurements of  oscillation parameters
and $m_{ee}$  on the structure of the mass matrix. \\

Our results can be summarized in the following way.\\  

\noindent
1) The structure of the mass matrix changes significantly with $m_1$. 

For strongly  hierarchical mass spectrum ($m_1 \approx 0$) and
small $s_{13}$,  the mass matrix has a structure with the dominant
$\mu\tau$-block  and small $e$-row elements. The ratio of masses  of
these two groups can be as small as 0.1. 

The dominant structure becomes less profound for 
large $\Delta m^2_{sol}$, large $s_{13}$ and 
significant deviation from maximal 2-3 mixing. 
For $\Delta m^2_{sol} > 2 \cdot  10^{-4}$ eV$^2$, 
a separation of the elements in the dominant  $\mu\tau$-block  and 
sub-dominant $e$-row has no sense and one can 
consider certain non-hierarchical ordering of the elements. 
In particular, a configuration 
with nearly equal split among masses is possible.

For partially degenerate spectrum, 
the gap between the $\mu\tau$-block elements and $e$-row elements
disappears and all elements can be of the same order. 
Various equalities between the elements and orderings 
can be realized  depending on the CP violating phases. 

In the case of degenerate mass spectrum, the mass matrix can have 
a hierarchical structure with some elements (in particular,  from the 
$\mu\tau$-block) being much smaller than other elements. 
The hierarchical structures appear for specific ranges of
phases. 

In the case of complete  degeneracy, the structure of the mass matrix is
insensitive to the ordering  of mass eigenvalues. Therefore, our
conclusions are valid also for inverted ordering.\\

\noindent
2) The Majorana phases $\rho$ and $\sigma$ and the Dirac phase $\delta$ 
have different impact on the structure of mass matrix.  
This impact depends on values of  oscillation parameters and $m_1$. 

(a)  The Dirac phase $\delta$ is  associated with the small parameter
$s_{13}$.  The influence of this phase on the  $\mu\tau$-block elements 
is relatively weak for any type of spectrum (hierarchical or degenerate):
it is suppressed by factor 
$s_{13}$. In contrast, the elements of $e$-row can be substantially
influenced by $\delta$, especially in the case of hierarchical 
spectrum. In the first approximation 
$\delta$ enters $\tilde{m}_{e\mu}$ and $\tilde{m}_{e\tau}$  
in the combination  $(\delta - 2\sigma)$ and $\tilde{m}_{ee}$  -- in the 
combination $(2\delta - 2\sigma)$. So, 
the effect of $\delta$ is reduced
to the appropriate shifts of phase $\sigma$ for $\tilde{m}_{ee}$, 
$\tilde{m}_{e\mu}$ and $\tilde{m}_{e\tau}$.  In the 
$\rho - \sigma$ plot, for fixed pattern of the $\mu\tau$-block elements, 
the phase $\delta$ produces a shift of the patterns for $\tilde{m}_{e\mu}$
and $\tilde{m}_{e\tau}$, along the axis $\sigma$. 

Improvements of the upper bound on $s_{13}$ in future experiments 
will further suppress the influence of the Dirac  phase on the structure
of the mass matrix. 

(b) The phase  $\rho$ is associated with the mass eigenvalue 
$m_1$. So,  it has very small effect on  
the mass matrix in the case of hierarchical spectrum. 
The role of $\rho$ increases with $m_1$. The influence of this phase 
increases with the solar mixing angle.  
Therefore future measurements of $\theta_{12}$ in KamLAND
and  solar neutrino experiments will allow one to further restrict  the
effect of  $\rho$ on the structure of the mass matrix.

For the best fit value of $\theta_{12}$,  dependence of 
the $\mu\tau$-block elements on $\rho$ is not very strong. 
However,  existence  of hierarchical structure (zeros) 
in this block is related to specific values of $\rho$. 
There is a strong dependence  of the $e$-row 
elements on  $\rho$. 
Typically $\tilde{m}_{e\mu}$ and $\tilde{m}_{e\tau}$ have 
minima at $\rho \approx 0, \pi$ and they are maximal at 
$\rho \approx  \pi/2$.  
The $ee$-element depends  on $\rho$  most strongly. 
There is a chance to measure/restrict $\rho$ in the 
$\beta\beta_{0\nu}$-decay searches, provided that 
the absolute mass scale will be determined 
(further restricted) in the direct kinematic measurements.

(c) The phase $\sigma$ is associated with the heaviest mass 
eigenstate and, consequently, the $\sigma$-dependence is strong for all
the elements but $\tilde{m}_{ee}$. Variations  of  the $ee$-element
with  $\sigma$ are  suppressed by a factor $s_{13}^2$. 

The phase $\sigma$ enters the $e$-row elements, $\tilde{m}_{e\mu}$ and 
$\tilde{m}_{e\tau}$ with a factor $s_{13}$. In spite of this, 
in the case of hierarchical spectrum  variations 
of $\tilde{m}_{e\mu}$ and $\tilde{m}_{e\tau}$ with $\sigma$  
can be strong. With increase of $r$, the relative amplitude of variations 
of these elements with $\sigma$ decreases. 
In contrast, the dependence of $\mu\tau$-block elements
on $\sigma$ becomes stronger  with increase of  $r$. It can be enhanced, in
addition, if the 2-3 mixing is non-maximal. 
In the case of  degenerate spectrum, 
variations of the $\mu\tau$-block elements 
with 
$\sigma$ can be maximal, so that, at certain values of phases, 
a given element can be zero or the largest one. \\

\noindent
There are correlations among the dependences of the matrix 
elements on phases. In general, 
patterns of $\tilde{m}_{\mu\mu}$ and $\tilde{m}_{\tau\tau}$ 
are complementary to the pattern  of $\tilde{m}_{\mu\tau}$. The patterns
for $\tilde{m}_{e\mu}$  and $\tilde{m}_{e\tau}$ are shifted by $\Delta
\sigma = \pi/2$,  {\it etc.}.\\

\noindent
3) Using the dependences of the matrix elements on 
the unknown parameters  we have studied possible structures 
of mass matrices. 
 
The matrix may have hierarchical form  with various dominant 
structures and  small or zero  elements.  
The dominant structures can be identified
considering the limit $s_{13} \rightarrow 0$.
The terms of  order  $s_{13}$ give
small corrections to the dominant elements.
In contrast, the $s_{13}$-order terms can be important
or even give main contribution  to the sub-dominant
elements of the mass matrix.  
The phase $\delta$ does not determine the
dominant structure.
 
In the case of hierarchical mass spectrum the dominant structure is formed
by the $\mu \tau$-block  (see Eq.(\ref{domblo})). 
The $e$-row elements can be about 10 times smaller
than the $\mu\tau$-block elements. 
Properties  of this
block depend on the 2-3 mixing and on the phase $\sigma$.

In the case of degenerate mass spectrum 
the hierarchy determined by a factor $\sim 10$ or more
appears mainly in the left and right-hand sides of the $\rho - \sigma$
plots.

One arrives at two rather stable structures:  

(i) the matrix which equals approximately  the unit
matrix with small off-diagonal terms;

(ii) the matrix which has a dominant structure with
$\tilde{m}_{ee} \approx \tilde{m}_{\mu\tau}$,
while all other elements are small.

Apart from these known hierarchical matrices we have found 
several new structures with non-trivial values of CP violating phases.      
In particular, for non-maximal 2-3 mixing:  the element
$\tilde{m}_{\mu\mu}$ or $\tilde{m}_{\tau\tau}$ can be small for
$\sigma\approx \pi/2$ and for a value of $\rho$ which depends  on 
$\xi$. With increase  of $\xi$,
the region of small mass approaches the center of $\rho-\sigma$ plots
($\rho \sim \pi/2$).
  
Typically, CP violating phases   which differ substantially from 
from 0 , $\pi/2$ or $\pi$ lead to non-hierarchical matrices.

We have found that the matrix may have certain flavor ordering 
(alignment), when masses increase with change of the flavor from 
$e$ to $\tau$.  At the same time we find that the data can be reproduced 
by matrices with flavor disorder, when no correlation between the size of
the mass  terms and the flavor is observed. 
The democratic mass matrix is also  
possible.\\

\noindent   
4) Typical separations among the elements in the  hierarchical structures of
the neutrino mass matrix  are characterized by a factor 0.2 - 0.3.  
We have found that it is possible  to parameterize the
matrix by powers of a single  parameter $\lambda$ (whose origin can be
in the breaking of some flavor  symmetry at high energy). 
The value
$\lambda \approx 0.2-0.3$ is consistent with  the Cabibbo angle 
and also it can be related to the ratios of charge lepton masses. 

If 2-3 mixing is not maximal, one can introduce an ordering parameter 
$\lambda_{ord} \sim  \tan\theta_{23} \sim  0.6-0.7$. 
We find that the whole matrix can be parametrized in terms of powers of this
ordering parameter.\\

\noindent
5) The following results from forthcoming experiments will have crucial
impact on the structure of the neutrino mass matrix: 

- improvement of  bound  on (or determination of) the
deviation $\xi$ from maximal 2-3 mixing; 

- precise determination of the solar  oscillation parameters, 
$\Delta m_{sol}^2$ and $\theta_{12}$; 

- improvement of  bound  on (or determination of) $s_{13}$; 

- improvement of  bound  on (or determination of) $m_{ee}$;  

- direct kinematic measurements of the neutrino mass. \\

Is it possible to determine uniquely the mass matrix, 
at least in principle? The answer depends on future experimental results. 
Let us take the most optimistic situation: 
suppose that  the  neutrinoless $2\beta$ decay is discovered with 
$m_{ee} > 0.1$ eV and the direct measurements of neutrino mass 
give $m > 0.5$ eV with high precision. Let us assume also 
that mixing angles 
are measured with high accuracy. In this case,
the spectrum is strongly degenerate and one can use neutrinoless $2\beta$
decay data to determine the CP violating phases. 
The problem is that $m_{ee}$ depends both on $\rho$ and on 
$\sigma$, and moreover, the dependence on $\sigma$ is 
very weak being  suppressed by $s_{13}^2$. This means that 
$\rho$ can be measured with rather good accuracy, whereas no 
bound on $\sigma$ can be obtained: small variations of $\rho$ 
can imitate effect of $\sigma$ in the  whole possible range. 
The only exception is if the measured $m_{ee}$ 
is at maximal (or minimal) 
possible value predicted for a given (measured) absolute scale of the
mass. That would correspond to certain CP-parity of $\nu_1$ and 
$\delta - \sigma = 0$ or $\pi/2$. Then, measuring $\delta$ 
(in neutrino oscillation experiments) one can get $\sigma$. 
Clearly even this program looks very challenging. 
Other  experimental situations are even more difficult. 

The determination of $\sigma$ looks practically impossible,  
unless methods of direct  measurement or independent reconstruction of  at
least one another matrix element (apart from $m_{ee}$) will be  found,

The $\rho-\sigma$ plots give an idea 
of uncertainty in the structure of the mass matrix if $\sigma$ 
is unknown. 
If $2\beta$ decay searches give a positive result 
and direct measurements improve the bound on 
(or measure) $m_1$, 
we will be able to select a narrow vertical strip in the
$\rho - \sigma$ diagram. This will also restrict  
other elements, but significant  uncertainty will be left 
due to their dependence on the  phase $\sigma$.
In particular,  as follows from the figures in 
the case of degenerate spectrum, the  
structure of the $\mu\tau$-block will be largely unfixed.  

The hope is that even partial reconstruction of the mass matrix may give
important hint in favor of certain underlying theory.\\

\section*{Acknowledgements}
We would like to thank Francesco Vissani for useful discussion
and suggestions.




\appendix
\section*{Appendix: Analytic expressions for matrix elements}
\setcounter{equation}{0}
\renewcommand{\theequation}{A.\arabic{equation}}

For convenience of the reader we present here 
explicit expressions for the neutrino mass
matrix elements in flavor  basis, $M_{\alpha\beta}$
($\alpha,\beta=e,\mu,\tau$), as functions of the mass eigenvalues
$m_i\,$, of the mixing angles  $\theta_{ij}$ and of the CP violating
phases $\delta,\rho,\sigma$  (see (\ref{ABC})).

Denoting $c_{ij} \equiv
\cos \theta_{ij}$ and $s_{ij} \equiv \sin \theta_{ij}$,  we get the
following expressions for the matrix elements, as sums  of three terms
corresponding to the three mass eigenstates:

\begin{equation}
\begin{array}{rl}
M_{ee}= & c_{13}^2\,c_{12}^2\,m_1\,e^{-2i\rho}\;+\\
& c_{13}^2\,s_{12}^2\,m_2\;+\\
& s_{13}^2\,m_3\,e^{2i(\delta-\sigma)}\;;
\end{array}
\label{mee}
\end{equation}
\beq
\begin{array}{rl}
M_{e\mu}= & c_{13}\,c_{12}\,\left(-c_{23}\,s_{12} -
    s_{23}\,c_{12}\,s_{13}\,e^{-i\delta}\right)\,m_1\,e^{-2i\rho}\;+\\
& c_{13}\,s_{12}\,\left(c_{23}\,c_{12} -
    s_{23}\,s_{12}\,s_{13}\,e^{-i\delta}\right)\,m_2\;+\\
& c_{13}\,s_{23}\,s_{13}\,m_3\,e^{i(\delta-2\sigma)}\;;
\end{array}
\label{mem}
\end{equation}
\beq
\begin{array}{rl}
M_{e\tau}= & c_{13}\,c_{12}\,\left(s_{23}\,s_{12} -
	c_{23}\,c_{12}\,s_{13}\,e^{-i\delta}\right)\,m_1\,e^{-2i\rho}\;+\\
&  c_{13}\,s_{12}\,\left(-s_{23}\,c_{12} -
	c_{23}\,s_{12}\,s_{13}\,e^{-i\delta}\right)\,m_2\;+\\
&  c_{13}\,c_{23}\,s_{13}\,m_3\,e^{i(\delta-2\sigma)}\;;
\end{array}
\label{met}
\end{equation}
\beq
\begin{array}{rl}
M_{\mu\mu}= & \left(-c_{23}\,s_{12} - s_{23}\,c_{12}\,s_{13}\,e^{-i\delta}
	\right)^2\,m_1\,e^{-2i\rho}\;+\\
& \left(c_{23}\,c_{12} - s_{23}\,s_{12}\,s_{13}\,e^{-i\delta}\right)^2
	\,m_2\;+\\
& c_{13}^2\,s_{23}^2\,m_3\,e^{-2i\sigma}\;;
\end{array}
\label{mmu}
\end{equation}
\beq
\begin{array}{rl}
M_{\mu\tau}= & \left(s_{23}\,s_{12} - c_{23}\,c_{12}\,s_{13}\,e^{-i\delta}
	\right)
	\,\left(-c_{23}\,s_{12} - s_{23}\,c_{12}\,s_{13}\,e^{-i\delta}\right)\,
	m_1\,e^{-2i\rho}\;+\\
& 	\left(-s_{23}\,c_{12} - c_{23}\,s_{12}\,s_{13}\,e^{-i\delta}\right)\,
	\left(c_{23}\,c_{12} - s_{23}\,s_{12}\,s_{13}\,e^{-i\delta}\right)\,
	m_2\;+\\
&	c_{13}^2\,c_{23}\,s_{23}\,m_3\,e^{-2i\sigma}\;;
\end{array}
\label{mmt}
\end{equation}
\beq
\begin{array}{rl}
M_{\tau\tau}= & \left(s_{23}\,s_{12} - c_{23}\,c_{12}\,s_{13}\,e^{-i\delta}
	\right)^2\,m_1\,e^{-2i\rho}\;+\\
& \left(-s_{23}\,c_{12} - c_{23}\,s_{12}\,s_{13}\,e^{-i\delta}\right)^2
	\,m_2\;+\\
&  c_{13}^2\,c_{23}^2\,m_3\,e^{-2i\sigma}\;.
\end{array}
\label{mtt}
\end{equation}
\\

It is convenient to use also a representation  of the matrix elements as
series in powers of $s_{13}$.
Using the equations
(\ref{mee})-(\ref{mtt}) and the definitions (\ref{XYZ}), we get:
\beq
\begin{array}{l}
\tilde{M}_{ee}=rZ+s_{13}^2 Z'\;,\\
\tilde{M}_{e\mu}=c_{13}(c_{23} r Y+
s_{13}s_{23}e^{-i\delta} Z')\;,\\
\tilde{M}_{e\tau}=c_{13}(-s_{23}r Y+
s_{13}c_{23}e^{-i\delta} Z')\;,\\
\tilde{M}_{\mu\mu}= c_{23}^2 X r + s_{23}^2 e^{-2i\sigma}
-\sin{2\theta_{23}}s_{13} r e^{-i\delta}Y
-s_{23}^2 s_{13}^2 e^{-2i\delta} Z'\;,\\
\tilde{M}_{\tau\tau}= s_{23}^2 X r + c_{23}^2 e^{-2i\sigma}
+\sin{2\theta_{23}}s_{13} r e^{-i\delta}Y
-c_{23}^2 s_{13}^2 e^{-2i\delta} Z'\;,\\
\tilde{M}_{\mu\tau}= s_{23} c_{23} (-r X + e^{-2i\sigma})
-\cos{2\theta_{23}}s_{13} r e^{-i\delta}Y
-s_{23}c_{23} s_{13}^2 e^{-2i\delta} Z'\;,
\end{array}
\label{big}
\end{equation}
where
\beq
Z'\equiv e^{2i(\delta-\sigma)}-r Z\;.
\label{zprime}
\end{equation}

We will use also another form 
for the matrix elements, 
which is obtained
neglecting terms of the order $s_{13}^2$.
For the $\mu\tau$-block, we get:
\beq
\begin{array}{l}
\tilde{m}_{\mu\mu} \approx \left|
s_{23}^2 e^{-2i\sigma}
+ r c_{23}^2  (c_{12}^2 -  \epsilon_{\mu\mu})
+ k  r c_{23}^2 (s_{12}^2 + \epsilon_{\mu\mu}) e^{-2i\rho}
\right|\;, \\
\tilde{m}_{\tau\tau}
\approx \left|
c_{23}^2 e^{-2i\sigma} +
r s_{23}^2 (c_{12}^2  +  \epsilon_{\tau\tau}) +
k r s_{23}^2 (s_{12}^2  -  \epsilon_{\tau\tau})e^{-2i\rho}
\right|\;,\\
\tilde{m}_{\mu\tau} \approx s_{23}c_{23}
\left|- e^{-2i\sigma} +
r ( c_{12}^2 + \epsilon_{\mu\tau}) +
k r (s_{12}^2 -  \epsilon_{\mu\tau}) e^{-2i\rho}
\right|\;,
\end{array}
\label{mutau-m}
\end{equation}
where
\beq
\epsilon_{\alpha \beta}(\delta) =
\sin 2\theta_{12} s_{13} e^{-i\delta}\times
\left\{
\begin{array}{ll}
\tan\theta_{23}, & \epsilon_{\mu\mu}\\
\cot\theta_{23}, & \epsilon_{\tau\tau}\\
\cot 2\theta_{23},  & \epsilon_{\mu\tau}
\end{array}
\right. .
\label{corr}
\end{equation}
Notice that the three different terms in the Eqs.(\ref{mutau-m})
are contributions of the three masses $m_3$, $m_2$ and $m_1$.
For the elements of the $e$-row we get, up to an overall factor
$c_{13}$:
\beq
\begin{array}{l}
\tilde{m}_{e\mu} \approx \left|
r c_{12}s_{12} c_{23}
\left[1 - \epsilon_{e\mu} - k e^{-2i\rho}(1 + \epsilon_{e\mu}')\right]
+ s_{13}s_{23} e^{i(\delta - 2\sigma)} \right|,  \\
\tilde m_{e\tau} \approx \left|
r c_{12}s_{12} s_{23}
\left[1 + \epsilon_{e\tau} - k e^{-2i\rho}(1 - \epsilon_{e\tau}')\right]
-  s_{13}c_{23}e^{i(\delta-2\sigma)} \right|,
\end{array}
\label{em-m}
\end{equation}
where
\beq
\begin{array}{l}
(\epsilon_{e\mu},~\epsilon_{e\mu}') =
s_{13} e^{-i\delta} \tan\theta_{23} \times
(\tan\theta_{12},~\cot\theta_{12})\;,\\
(\epsilon_{e\tau},~\epsilon_{e\tau}') =
s_{13} e^{-i\delta} \cot\theta_{23} \times
(\tan\theta_{12},~\cot\theta_{12})\;.
\end{array}
\label{epsilon}
\end{equation}
The $ee$-element can be written  as
\beq
\tilde{m}_{ee} = \left|
c_{13}^2 (c_{12}^2 k  e^{-2i\rho} + s_{12}^2) r + s_{13}^2
e^{2i(\delta-\sigma)} \right|\;.
\label{ee-m}
\end{equation}
It does not depend on the angle $\theta_{23}$.


\end{document}